\documentclass[12pt]{article}
\pdfoutput=1
\usepackage{jheppub}

\usepackage{amsmath, amssymb, amsfonts, amsthm}
\usepackage{mathtools}
\usepackage{bm}
\usepackage{dsfont}
\usepackage{enumerate}
\usepackage{bbold}
\usepackage{physics}
\usepackage[table]{xcolor}
\usepackage{pgf}
\usepackage{simplewick}
\usepackage{tikz}
\usetikzlibrary{tikzmark}
\usetikzlibrary{calc}
\usepackage{array}

\numberwithin{equation}{section}

\newcommand{\be}{\begin{equation}}
\newcommand{\ee}{\end{equation}}
\newcommand{\bea}{\begin{aligned}}
\newcommand{\eea}{\end{aligned}}

\newcommand\mscript[1]{\mbox{\scriptsize\ensuremath{#1}}}
\newcommand{\sbmatrix}[1]{
{\mscript{\arraycolsep=0.3\arraycolsep\ensuremath{\begin{bmatrix}#1\end{bmatrix}}}}}
\newcommand{\fker}[6]{
\mathbb{F}_{#1 #2}\sbmatrix{#4 & #3 \\ #5 & #6}} 
\newcommand{\sker}[3]{\mathbb{S}_{#1 #2}[#3]}

\definecolor{myforestgreen}{RGB}{24, 150, 144}

\def\bbi{\mathbb{1}}
\def\Tr{{\mathrm{Tr}}}
\def\Z{{\mathbb Z}}
\def\R{{\mathbb R}}

\def\N{{\mathcal N}}
\def\e{\mathrm{e}}

\def\Im{\mathrm{Im}} 
\def\hat{\widehat}

\def\O{{\mathcal O}}

\def\bar{\overline}

\begin{document}
\tikzset{every picture/.style={line width=0.9pt}} 

\title{A non-rational Verlinde formula \\ from Virasoro TQFT}

\author[\dagger]{Boris Post,}
\affiliation[\dagger]{University of Amsterdam, Institute of Physics, ITFA, \\Science Park 904, Amsterdam, The Netherlands}
\emailAdd{b.p.post@uva.nl}
\author[\diamond]{Ioannis Tsiares}
\affiliation[\diamond]{Universit\'e Paris-Saclay, CNRS, CEA, Institut de Physique Th\'eorique, \\91191, Gif-sur-Yvette, France}
\emailAdd{ioannis.tsiares@ipht.fr}

\abstract{We use the Virasoro TQFT to derive an integral identity that we view as a non-rational generalization of the Verlinde formula for the Virasoro algebra with central charge $c\geq 25$. The identity expresses the Virasoro fusion kernel as an integral over a ratio of modular S-kernels on the (punctured) torus. In particular, it shows that the one-point S-kernel diagonalizes the Virasoro $6j$ symbol. After carefully studying the analytic properties of this `Virasoro-Verlinde formula', we present three applications. In boundary Liouville CFT, the formula ensures the open-closed duality of the boundary one-point function on the annulus. In pure 3d gravity, it provides an essential step in computing the partition function on hyperbolic 3-manifolds that fiber over the circle. Lastly, in AdS$_3$/CFT$_2$, the formula computes a three-boundary torus wormhole, which leads to a prediction for the statistical correlation between the density of states and two OPE coefficients in the dual large-$c$ CFT ensemble. We conclude by discussing the implications of our result for the fusion rules in generic non-rational 2d CFTs.  
}
\maketitle

\section{Introduction}\label{sec:Intro}
In 1988, Erik Verlinde conjectured a by now famous formula, valid in two-dimensional rational conformal field theory (RCFT), which relates the fusion coefficients of the theory to its modular S-matrix \cite{Verlinde:1988sn}:
\be\label{eq:original_formula} 
N_{ij}\text{}^k = \sum_{h\in \mathcal{P}}\,\frac{S_{i h}S_{jh}S^*_{hk}}{S_{\bbi h}}\,.
\ee 
Here the sum runs over a finite set of primary operators of the chiral algebra. The Verlinde formula is often phrased as the statement that \emph{the modular S-matrix diagonalizes the fusion rules} of an RCFT. A very striking feature of this formula is that it relates local data of the CFT to the global features of the CFT on the torus. Another important consequence is that it makes manifest the fact that the fusion rules form an algebra, which is both commutative and associative.

Soon after Verlinde's discovery, Moore and Seiberg  proved the formula \eqref{eq:original_formula}, by using the consistency conditions imposed on any 2d CFT so as to define it on an arbitrary (punctured) Riemann surface \cite{Moore:1988uz, Moore:1988qv}. The modern formulation of these consistency conditions uses the theory of modular tensor categories \cite{Fuchs:2002cm,Huang:2004dg}. For further applications and reviews of the Verlinde formula, we refer to \cite{Witten:1988hf,Cardy:1989ir,Dijkgraaf:1988tf,DiFrancesco:639405,blumenhagen2009}.

A natural question is whether a version of the Verlinde formula \eqref{eq:original_formula} continues to hold for non-rational conformal field theories.\footnote{This question was partially addressed in
\cite{Jego:2006ta} for Liouville theory (restricted to degenerate representations) and two examples of non-compact WZW coset models. The possibility was also briefly considered in the outlook of \cite{Teschner:2008qh}, but not pursued in later works. Our results can be seen as a further development of the ideas in those works.} For now, we restrict ourselves to non-rational CFT's with only Virasoro symmetry (i.e.\! with no extended chiral algebra) and central charge $c\geq 25$. However, in this case one immediately runs into a problem: the space of allowed unitary highest-weight representations is non-compact, meaning that instead of a finite sum over primaries the right-hand side of \eqref{eq:original_formula} would be replaced by a continuous integral:
\begin{equation}\label{eq:naive_guess}
N_{P_1P_2}^{P_3} \stackrel{?}{=}\int_0^\infty \dd P\, \frac{\sker{P_1}{P}{\bbi}\sker{P_2}{P}{\bbi}\mathbb{S}_{PP_3}^*[\bbi]}{\sker{\bbi}{P}{\bbi}}\,.
\end{equation}
For non-degenerate primaries $\O_{1,2,3}$ with conformal weights $\Delta_i,\bar\Delta_i$ parametrized in terms of Liouville momenta $P_i$, $\bar P_i$ the modular S-kernels $\sker{P_i}{P}{\bbi}$ and $\sker{\bbi}{P}{\bbi}$ are known explicitly, and it turns out that the integral in the naive guess \eqref{eq:naive_guess} is \emph{divergent}. This is of course incompatible with the definition of a fusion coefficient $N_{P_1P_2}^{P_3}$ as the dimension of the space of 3-point conformal blocks on the sphere, which, for the Virasoro algebra, should be either zero or one \cite{Ribault:2014hia}. 

In order to resolve this problem, we use the recently constructed Virasoro TQFT \cite{Collier_2023,Collier:2024mgv} to derive a new formula that expresses the \emph{Virasoro fusion kernel}, with a particular repetition of its arguments, as an integral over a ratio of S-kernels: 
\be\label{eq:main_formula}
\fker{P_1}{P_3}{P_3}{P_0}{P_1}{P_2} = \int_0^\infty \dd P\, \frac{\sker{P_1}{P}{P_0}\sker{P_2}{P}{\bbi}\mathbb{S}_{PP_3}^*[P_0]}{\sker{\bbi}{P}{\bbi}}\,.
\ee 
Importantly, the integrand contains the \emph{one-point S-kernel}, which implements the modular S-transform of the torus one-point conformal block \cite{Eberhardt:2023mrq}. The presence of the extra Liouville parameter $P_0$ (which labels the conformal weight of the operator $\O_0$ inserted on the punctured torus) makes the integral in \eqref{eq:main_formula} convergent. This motivates us to define the left-hand side of \eqref{eq:main_formula} as a `regularized fusion density',
\be\label{eq:fusion_density}
\N_{P_0}[P_1,P_2,P_3] \coloneqq \fker{P_1}{P_3}{P_3}{P_0}{P_1}{P_2} ,
\ee 
which is now a well-defined meromorphic function of the (left-moving\footnote{Since the conformal blocks of the Virasoro algebra factorize into left- and right-movers, the crossing kernels are also holomorphically factorized. Hence equations \eqref{eq:main_formula} and \eqref{eq:fusion_density} both have a right-moving counterpart that depends on the  anti-holomorphic coordinates $\bar P_i$.}) Liouville momenta of $\O_{1,2,3}$. The external momentum $P_0$ plays the role of a regulator, in the sense that the limit $\O_0\to \bbi$ is divergent (due to the $P^{-2}$ scaling of the ratio \eqref{eq:naive_guess} for small $P$) but the analytic properties of the crossing kernels allow us to extract non-trivial information about the fusion rules \emph{before} taking the regulator to the identity.

As an example, consider the analytic continuation of the regularized fusion density \eqref{eq:fusion_density} to the $\langle 2, 1\rangle$ degenerate representation $P_2 = P_{\langle 2,1\rangle} \equiv ib + \frac{i}{2b}$, where $b$ parametrizes the central charge. A careful analysis of the Virasoro fusion kernel then shows that  
\begin{equation}
   \lim_{P_0\to\bbi} \N_{P_0}[P_1,P_{\langle 2,1\rangle},P_3] = \delta(P_1-(P_3+\tfrac{ib}{2}))\,+\, \delta(P_1-(P_3-\tfrac{ib}{2})),
\end{equation}
which indeed takes the form of a density $\sum_{k}N_{ij}\text{}^k \,\delta(P-P_k)$, where $N_{ij}\text{}^k$ are the known fusion coefficients of the generalized Virasoro minimal model. More generally, using the known closed-form expressions for the fusion kernel found by Ponsot and Teschner \cite{Ponsot:1999uf,Ponsot:2000mt}, we derive a wealth of analytic properties of the fusion density $\N_{P_0}$, including reality, positivity, degenerate and non-degenerate selection rules, commutativity and associativity. These properties will be established in section \ref{sec:properties}. 

There is a very interesting rewriting of the main formula \eqref{eq:main_formula} that makes contact with the quantum $6j$ symbol of the Virasoro algebra. If we conveniently normalize the one-point S-kernels (which we denote by a hat: see  \eqref{eq:S_normalization}), then equation \eqref{eq:main_formula} can be written in terms of the Racah-Wigner quantum $6j$ symbol as:
\begin{equation}\label{eq:intro_6j}
    \begin{Bmatrix}
  P_3 & P_3 & P_0 \\
  P_1 & P_1 & P_2 
 \end{Bmatrix}_{6j} = \int_0^\infty \dd\mu(P) \,\frac{\hat{\mathbb{S}}_{P_1P}[P_0]\mathbb{S}_{P_2P}[\bbi]\hat{\mathbb{S}}_{PP_3}^*[P_0]}{\mathbb{S}_{\bbi P}[\bbi]} \,.
\end{equation}
We can rephrase this formula as the statement that \emph{the one-point S-kernel $\widehat{\mathbb{S}}_{PP'}[P_0]$ diagonalizes the Virasoro 6j symbol}, with `eigenvalues' given by the ratio of S-kernels: 
\begin{equation}\label{eq:ratio}
    \frac{\mathbb{S}_{P_2P}[\bbi]}{\mathbb{S}_{\bbi P}[\bbi]} = \frac{\cos(4\pi PP_2)}{2\sinh(2\pi bP)\sinh(2\pi b^{-1}P)}.
\end{equation}
This resonates with the interpretation of the standard Verlinde formula as stating that the modular S-matrix diagonalizes the fusion rules in RCFT. Note that the left-hand side of \eqref{eq:intro_6j} is not the most general $6j$ symbol, since the external operators $\O_{1,3}$ are pairwise identical. It is proportional to the fusion kernel $\fker{P_0}{P_2}{P_3}{P_1}{P_1}{P_3}$, which governs the crossing transformation between the $s$- and $t$-channel conformal block expansions of the sphere four-point function $\langle \O_1 \O_1\O_3\O_3\rangle$.

The rewriting \eqref{eq:intro_6j} reveals an underlying quantum group structure of the Virasoro-Verlinde formula. Since the work of Ponsot and Teschner \cite{Ponsot:1999uf,Ponsot:2000mt}, it has been known that the representation theory of the Virasoro algebra, for the continuous series $P\in \R_+$, coincides with the representation theory of the modular double of the quantum group $\mathcal{U}_q(\mathfrak{sl}(2,\R))$. In particular, the Virasoro $6j$ symbol is equal to the Racah-Wigner coefficient of the modular double. Moreover, the integration measure 
\begin{equation}
    \dd\mu(P) \coloneqq \dd P\,\rho_0(P)
\end{equation}
where $\rho_0(P) = 4\sqrt{2}\sinh(2\pi bP)\sinh(2\pi b^{-1}P)$, coincides with the Plancherel measure on the space of representations of $\mathcal{U}_q(\mathfrak{sl}(2,\R))$. Since the $6j$ symbol plays a prominent role in Liouville theory \cite{Teschner:2012em,Teschner:2013tqy}, 3D gravity \cite{Jackson:2014nla, Mertens:2022aou,Collier:2024mgv} and the tensor-matrix model of \cite{Belin:2023efa,Jafferis:2024jkb}, the new identity \eqref{eq:intro_6j} will likely prove to be useful in these contexts. 

Indeed, in section \ref{sec:III} we give three applications of the Virasoro-Verlinde formula in a variety of theories governed by Virasoro symmetry. These applications --- in Liouville boundary CFT, quantum topology and AdS$_3$/CFT$_2$ --- are summarized pictorially in figure \ref{fig:intro_summary}. The results of this section can be formulated as follows: 

\begin{figure}
    \centering
    \begin{tikzpicture}[x=0.75pt,y=0.75pt,yscale=-1.2,xscale=1.2]
\draw  [line width=0.75]  (138.75,154.83) .. controls (138.75,135.37) and (154.25,119.58) .. (173.38,119.58) .. controls (192.5,119.58) and (208,135.37) .. (208,154.83) .. controls (208,174.3) and (192.5,190.08) .. (173.38,190.08) .. controls (154.25,190.08) and (138.75,174.3) .. (138.75,154.83) -- cycle ;
\draw  [line width=0.75]  (157.99,154.83) .. controls (157.99,146.18) and (164.88,139.17) .. (173.38,139.17) .. controls (181.87,139.17) and (188.76,146.18) .. (188.76,154.83) .. controls (188.76,163.49) and (181.87,170.5) .. (173.38,170.5) .. controls (164.88,170.5) and (157.99,163.49) .. (157.99,154.83) -- cycle ;
\draw  [line width=0.75]  (135.05,158.51) -- (142.54,151.1)(135.07,151.03) -- (142.52,158.57) ;
\draw  [line width=0.75]  (241.33,154.33) .. controls (241.33,135) and (244.53,119.33) .. (248.47,119.33) .. controls (252.41,119.33) and (255.6,135) .. (255.6,154.33) .. controls (255.6,173.66) and (252.41,189.33) .. (248.47,189.33) .. controls (244.53,189.33) and (241.33,173.66) .. (241.33,154.33) -- cycle ;
\draw [line width=0.75]    (248.47,119.33) -- (312.66,119.33) ;
\draw  [draw opacity=0][line width=0.75]  (312.14,119.39) .. controls (312.27,119.36) and (312.4,119.34) .. (312.53,119.34) .. controls (316.47,119.34) and (319.67,135.01) .. (319.67,154.34) .. controls (319.67,173.38) and (316.57,188.87) .. (312.71,189.33) -- (312.53,154.34) -- cycle ; \draw  [line width=0.75]  (312.14,119.39) .. controls (312.27,119.36) and (312.4,119.34) .. (312.53,119.34) .. controls (316.47,119.34) and (319.67,135.01) .. (319.67,154.34) .. controls (319.67,173.38) and (316.57,188.87) .. (312.71,189.33) ;  
\draw [line width=0.75]    (248.47,189.33) -- (312.66,189.33) ;
\draw  [draw opacity=0][dash pattern={on 1.5pt off 1.5pt on 1.5pt off 1.5pt}][line width=0.75]  (310.86,188.37) .. controls (307.73,184.68) and (305.4,170.85) .. (305.4,154.34) .. controls (305.4,135.01) and (308.59,119.34) .. (312.53,119.34) .. controls (313.04,119.34) and (313.53,119.59) .. (314,120.08) -- (312.53,154.34) -- cycle ; \draw  [dash pattern={on 1.5pt off 1.5pt on 1.5pt off 1.5pt}][line width=0.75]  (310.86,188.37) .. controls (307.73,184.68) and (305.4,170.85) .. (305.4,154.34) .. controls (305.4,135.01) and (308.59,119.34) .. (312.53,119.34) .. controls (313.04,119.34) and (313.53,119.59) .. (314,120.08) ;  
\draw  [line width=0.75]  (252.91,158.5) -- (258.26,151.1)(252.93,151.03) -- (258.24,158.57) ;
\draw [draw opacity=0][line width=0.75]    (374.59,128.65) .. controls (376.51,108.67) and (409.15,118.83) .. (386.93,147.53) ;
\draw [draw opacity=0][line width=0.75]    (374.59,135.2) .. controls (375.14,146.02) and (380.52,148.66) .. (388.58,155.58) .. controls (396.64,162.51) and (398.25,171.85) .. (387,184.56) ;
\draw [draw opacity=0][line width=0.75]    (397.9,132.68) .. controls (423.14,143.25) and (410.66,183.59) .. (397.7,188.15) .. controls (384.74,192.71) and (364.72,177.74) .. (380.35,155.33) ;
\draw [draw opacity=0][line width=0.75]    (381.19,188.37) .. controls (368.85,193.47) and (356.82,173.45) .. (356.9,157.49) .. controls (356.99,141.52) and (366.21,128.38) .. (390.97,131.48) ;
\draw [line width=0.75]    (374.92,128.81) .. controls (376.84,108.83) and (409.48,118.99) .. (387.27,147.69) ;
\draw [line width=0.75]    (374.92,135.35) .. controls (375.47,146.18) and (380.85,148.82) .. (388.91,155.74) .. controls (396.97,162.66) and (398.58,172.01) .. (387.33,184.72) ;
\draw [line width=0.75]    (398.24,132.84) .. controls (423.47,143.41) and (410.99,183.74) .. (398.03,188.31) .. controls (385.07,192.87) and (365.05,177.89) .. (380.68,155.49) ;
\draw [line width=0.75]    (381.19,188.37) .. controls (368.85,193.47) and (356.82,173.45) .. (356.9,157.49) .. controls (356.99,141.52) and (366.21,128.38) .. (390.97,131.48) ;
\draw  [fill={rgb, 255:red, 255; green, 255; blue, 255 }  ,fill opacity=1 ] (449.5,152.67) .. controls (449.5,132.78) and (475.39,116.67) .. (507.33,116.67) .. controls (539.27,116.67) and (565.16,132.78) .. (565.16,152.67) .. controls (565.16,172.55) and (539.27,188.67) .. (507.33,188.67) .. controls (475.39,188.67) and (449.5,172.55) .. (449.5,152.67) -- cycle ;
\draw  [fill={rgb, 255:red, 128; green, 128; blue, 128 }  ,fill opacity=0.3 ] (458.59,152.67) .. controls (458.59,137.62) and (480.42,125.43) .. (507.33,125.43) .. controls (534.25,125.43) and (556.07,137.62) .. (556.07,152.67) .. controls (556.07,167.71) and (534.25,179.9) .. (507.33,179.9) .. controls (480.42,179.9) and (458.59,167.71) .. (458.59,152.67) -- cycle ;
\draw  [fill={rgb, 255:red, 255; green, 255; blue, 255 }  ,fill opacity=1 ] (466.16,152.67) .. controls (466.16,141.23) and (484.59,131.95) .. (507.33,131.95) .. controls (530.07,131.95) and (548.5,141.23) .. (548.5,152.67) .. controls (548.5,164.11) and (530.07,173.38) .. (507.33,173.38) .. controls (484.59,173.38) and (466.16,164.11) .. (466.16,152.67) -- cycle ;
\draw  [fill={rgb, 255:red, 155; green, 155; blue, 155 }  ,fill opacity=0.3 ] (476.34,152.67) .. controls (476.34,144.7) and (490.54,138.24) .. (508.06,138.24) .. controls (525.57,138.24) and (539.77,144.7) .. (539.77,152.67) .. controls (539.77,160.64) and (525.57,167.1) .. (508.06,167.1) .. controls (490.54,167.1) and (476.34,160.64) .. (476.34,152.67) -- cycle ;
\draw  [fill={rgb, 255:red, 255; green, 255; blue, 255 }  ,fill opacity=1 ] (483.91,152.67) .. controls (483.91,147.87) and (494.72,143.97) .. (508.06,143.97) .. controls (521.4,143.97) and (532.21,147.87) .. (532.21,152.67) .. controls (532.21,157.47) and (521.4,161.36) .. (508.06,161.36) .. controls (494.72,161.36) and (483.91,157.47) .. (483.91,152.67) -- cycle ;
\draw  [draw opacity=0] (522.1,151.88) .. controls (519.66,154.83) and (514.39,156.87) .. (508.28,156.87) .. controls (501.58,156.87) and (495.89,154.42) .. (493.83,151.01) -- (508.28,148.18) -- cycle ; \draw   (522.1,151.88) .. controls (519.66,154.83) and (514.39,156.87) .. (508.28,156.87) .. controls (501.58,156.87) and (495.89,154.42) .. (493.83,151.01) ;  
\draw  [draw opacity=0] (496,153.15) .. controls (497.58,150.73) and (502.29,148.98) .. (507.84,148.98) .. controls (513.62,148.98) and (518.47,150.87) .. (519.85,153.44) -- (507.84,154.93) -- cycle ; \draw   (496,153.15) .. controls (497.58,150.73) and (502.29,148.98) .. (507.84,148.98) .. controls (513.62,148.98) and (518.47,150.87) .. (519.85,153.44) ;  
\draw  [draw opacity=0] (458.59,152.44) .. controls (458.62,150.41) and (460.3,148.78) .. (462.37,148.78) .. controls (464.32,148.78) and (465.93,150.23) .. (466.13,152.09) -- (462.37,152.49) -- cycle ; \draw   (458.59,152.44) .. controls (458.62,150.41) and (460.3,148.78) .. (462.37,148.78) .. controls (464.32,148.78) and (465.93,150.23) .. (466.13,152.09) ;  
\draw  [draw opacity=0] (476.35,152.56) .. controls (476.4,150.41) and (478.07,148.69) .. (480.13,148.69) .. controls (482.17,148.69) and (483.84,150.4) .. (483.91,152.53) -- (480.13,152.67) -- cycle ; \draw   (476.35,152.56) .. controls (476.4,150.41) and (478.07,148.69) .. (480.13,148.69) .. controls (482.17,148.69) and (483.84,150.4) .. (483.91,152.53) ;  
\draw  [draw opacity=0] (532.21,152.54) .. controls (532.27,150.4) and (533.94,148.69) .. (535.99,148.69) .. controls (538.08,148.69) and (539.77,150.47) .. (539.77,152.67) .. controls (539.77,152.72) and (539.77,152.77) .. (539.77,152.82) -- (535.99,152.67) -- cycle ; \draw   (532.21,152.54) .. controls (532.27,150.4) and (533.94,148.69) .. (535.99,148.69) .. controls (538.08,148.69) and (539.77,150.47) .. (539.77,152.67) .. controls (539.77,152.72) and (539.77,152.77) .. (539.77,152.82) ;  
\draw  [draw opacity=0] (548.5,152.68) .. controls (548.5,152.68) and (548.5,152.67) .. (548.5,152.67) .. controls (548.5,150.47) and (550.2,148.69) .. (552.29,148.69) .. controls (554.34,148.69) and (556.02,150.42) .. (556.07,152.57) -- (552.29,152.67) -- cycle ; \draw   (548.5,152.68) .. controls (548.5,152.68) and (548.5,152.67) .. (548.5,152.67) .. controls (548.5,150.47) and (550.2,148.69) .. (552.29,148.69) .. controls (554.34,148.69) and (556.02,150.42) .. (556.07,152.57) ;  
\draw (218.44,152.59) node [anchor=north west][inner sep=0.75pt]    {$=$};
\draw (117.83,185.4) node [anchor=north west][inner sep=0.75pt]    {$a)$};
\draw (334.97,185.03) node [anchor=north west][inner sep=0.75pt]    {$b)$};
\draw (428.63,184.37) node [anchor=north west][inner sep=0.75pt]    {$c)$};
\end{tikzpicture}
    \caption{Three applications of the Virasoro-Verlinde formula: $a)$ open-closed duality of the annulus one-point function in boundary Liouville CFT; $b)$ gravitational path integrals on knot and link complements in $S^3$ that fiber over the circle; $c)$ a three-boundary torus wormhole in AdS$_3$, with new implications for the statistical properties of universal CFT$_2$ data.}
    \label{fig:intro_summary}
\end{figure}
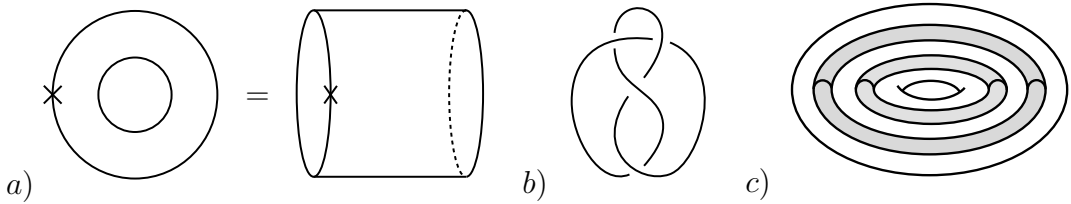

\begin{itemize}
    \item In section \ref{sec:open_closed}, we show that the Virasoro-Verlinde formula proves the open-closed duality of the boundary one-point function on the annulus in boundary Liouville CFT, thanks to the relation between the non-perturbative Liouville BCFT structure constants and the Virasoro crossing kernels.
    \item In section \ref{sec:whitehead}, we demonstrate that the Virasoro-Verlinde formula provides the relation between two distinct ways of evaluating the Virasoro TQFT partition function (as a so-called `Heegaard splitting' and as a `mapping torus') for finite volume hyperbolic 3-manifolds that admit a fibration over the circle. We make this explicit for two well-known knot and link complements, the figure-8 knot and the Whitehead link.
    \item In section \ref{sec:3boundary}, we find that the Virasoro-Verlinde formula precisely computes the gravitational path integral on a three-boundary torus wormhole with a conical defect in the bulk. This gives a bulk interpretation to the regulator $P_0$  in \eqref{eq:main_formula} as parametrizing the mass of a matter Wilson line that stabilizes the wormhole. Assuming a `coarse-grained' AdS$_3$/CFT$_2$ dictionary between 3d gravity (with defects) and the statistical ensemble of universal large-$c$ CFT data of \cite{Chandra:2022bqq}, our computation gives a new prediction for the connected contribution to the `mixed' expectation value $\overline{\rho C^2}$ between the density of primary states and two OPE coefficients. 
\end{itemize}
After these applications, in section \ref{sec:universal} we turn to the question of whether the Virasoro-Verlinde formula defines a particular fusion product within the space of unitary highest-weight representations of the Virasoro algebra. We find a neat connection with a fusion product previously considered in \cite{Teschner:2008qh} which we discuss in detail. For `generic' non-rational 2d CFTs with only Virasoro symmetry -- comprising a \textit{discrete} set of left-right representations -- we raise the issue of the existence of a `generalized Virasoro-Verlinde' formula that appropriately captures the discreteness of the representations. This leads us to conjecture a novel formula relating the modular-invariant density of states on $S^1$ with the associated fusion numbers, which, as we explain, can be used to derive novel bootstrap constraints on the density of states.

\subsubsection*{Organization of the paper}

This article is organized as follows. In sections \ref{sec:RCFT_recap} and \ref{sec:VTQFT_proof}, we prove the main formula \eqref{eq:main_formula} using the correspondence between Virasoro crossing kernels in $c\geq 25$ 2d CFT and the 3d Virasoro TQFT. An alternative proof that uses only the 2d CFT consistency conditions is given in appendix \ref{sec:CFTproof}. After introducing the regularized fusion density $\N_{P_0}$, we derive many of its properties in section \ref{sec:properties}. In section \ref{sec:III}, we give three applications of the formula, which were summarized above. Then, in section \ref{sec:universal}, we comment on the implications of our formula for general CFTs with only Virasoro symmetry, leading us to propose a new bootstrap condition on the spectral density of an non-rational CFT. We end with a discussion and outlook. In appendix \ref{app:prelim}, we collect the necessary background on crossing kernels and Virasoro TQFT.

\section{A Verlinde formula for the Virasoro algebra}\label{sec:II}

In this section, we give the proof of the main formula \eqref{eq:main_formula} using the Virasoro TQFT. In appendix \ref{sec:CFTproof}, we also provide a separate CFT derivation that only makes use of the (non-rational) Moore-Seiberg consistency conditions. After establishing the Virasoro-Verlinde formula in section \ref{sec:VTQFT_proof}, we will describe the properties of the fusion density in section \ref{sec:properties}. However, before we go into the Virasoro TQFT, let us briefly recall how the rational Verlinde formula can be proven using 3d TQFT techniques.\footnote{We thank Lorenz Eberhardt for explaining the argument. More details are found in sec.\! 4 of \cite{Witten:1988hf}.}

\subsection{Rational CFT recap}\label{sec:RCFT_recap}
In the standard set-up where the Verlinde formula applies, we have a rational CFT$_2$ with some chiral algebra $\mathcal{A}$ and finite-dimensional Hilbert spaces of conformal blocks $\mathcal{H}_{g,n}$, labeled by the data of a punctured Riemann surface $\Sigma_{g,n}$ in a given pair-of-pants decomposition $\mathcal{C}$. Crossing transformations between the channels $\mathcal{C}$ and $\mathcal{C}'$ are finite-dimensional matrices, and the fusion coefficients are $N_{ij}\text{}^{k}= \mathrm{dim}(\mathcal{H}_{0,3})$. 

There is a dual description of rational CFT's in terms of 3d Chern-Simons theory with compact gauge group \cite{Witten:1988hf,Moore:1989yh}. So, to set up the derivation of the Verlinde formula in RCFT, let us consider the Chern-Simons partition function on the following link, embedded in the 3-sphere, consisting of three Wilson loops:\vspace{1mm}
\begin{equation}\label{eq:ring}
    \begin{tikzpicture}[x=0.75pt,y=0.75pt,yscale=-1.2,xscale=1.2,baseline={([yshift=-0.5ex]current bounding box.center)}]
\draw  [draw opacity=0]  (203.87,107.6) .. controls (203.87,107.65) and (203.88,107.7) .. (203.88,107.75) .. controls (203.88,118.8) and (199.87,127.75) .. (194.94,127.75) .. controls (190,127.75) and (186,118.8) .. (186,107.75) .. controls (186,96.7) and (190,87.75) .. (194.94,87.75) .. controls (198.9,87.75) and (202.27,93.53) .. (203.44,101.54) -- (194.94,107.75) -- cycle ; \draw    (203.87,107.6) .. controls (203.87,107.65) and (203.88,107.7) .. (203.88,107.75) .. controls (203.88,118.8) and (199.87,127.75) .. (194.94,127.75) .. controls (190,127.75) and (186,118.8) .. (186,107.75) .. controls (186,96.7) and (190,87.75) .. (194.94,87.75) .. controls (198.9,87.75) and (202.27,93.53) .. (203.44,101.54) ;  
\draw  [draw opacity=0]  (189.14,113.06) .. controls (195.31,106.48) and (204.08,102.38) .. (213.81,102.38) .. controls (215.74,102.38) and (217.63,102.54) .. (219.46,102.85) -- (213.81,136.19) -- cycle ; \draw    (189.14,113.06) .. controls (195.31,106.48) and (204.08,102.38) .. (213.81,102.38) .. controls (215.74,102.38) and (217.63,102.54) .. (219.46,102.85) ;  
\draw  [draw opacity=0]  (225.92,104.61) .. controls (238.61,109.48) and (247.63,121.78) .. (247.63,136.19) .. controls (247.63,154.86) and (232.49,170) .. (213.81,170) .. controls (195.14,170) and (180,154.86) .. (180,136.19) .. controls (180,130.32) and (181.49,124.81) .. (184.12,120) -- (213.81,136.19) -- cycle ; \draw    (225.92,104.61) .. controls (238.61,109.48) and (247.63,121.78) .. (247.63,136.19) .. controls (247.63,154.86) and (232.49,170) .. (213.81,170) .. controls (195.14,170) and (180,154.86) .. (180,136.19) .. controls (180,130.32) and (181.49,124.81) .. (184.12,120) ;  
\draw  [draw opacity=0]  (238.84,117.34) .. controls (237.34,123.68) and (234.37,128) .. (230.94,128) .. controls (226,128) and (222,119.05) .. (222,108) .. controls (222,96.95) and (226,88) .. (230.94,88) .. controls (235.87,88) and (239.88,96.95) .. (239.88,108) .. controls (239.88,108.8) and (239.85,109.6) .. (239.81,110.38) -- (230.94,108) -- cycle ; \draw    (238.84,117.34) .. controls (237.34,123.68) and (234.37,128) .. (230.94,128) .. controls (226,128) and (222,119.05) .. (222,108) .. controls (222,96.95) and (226,88) .. (230.94,88) .. controls (235.87,88) and (239.88,96.95) .. (239.88,108) .. controls (239.88,108.8) and (239.85,109.6) .. (239.81,110.38) ;  
\draw (176.74,90.37) node [anchor=north west][inner sep=0.75pt]  [font=\small]  {$i$};
\draw (242.8,91.26) node [anchor=north west][inner sep=0.75pt]  [font=\small]  {$j$};
\draw (251.49,130.14) node [anchor=north west][inner sep=0.75pt]  [font=\small]  {$k$};
\end{tikzpicture}\,\,.\vspace{2mm}
\end{equation}
The labels $i,j$ and $k$ specify the representation of the gauge group that the Wilson loops are in. The main idea of the argument, that we will sketch, is that the Chern-Simons partition function of the link complement in $S^3$ can be computed in two different ways:
\begin{enumerate}[1)]
    \item First unlink the $i$ and $j$ loops from the $k$ loop, using the TQFT identity:\footnote{This formula is valid in any 3d TQFT. It is shown in appendix \ref{app:TQFT_identities} that a similar formula holds for the Virasoro TQFT, where the ratio of S-kernels is given by \eqref{eq:ratio}.}
    \begin{equation}\label{eq:TQFT_rule_1}
    \begin{tikzpicture}[x=0.75pt,y=0.75pt,yscale=-1.3,xscale=1.3,baseline={([yshift=-.5ex]current bounding box.center)}] 
\draw   (100.1,105) -- (119.11,105) ;
\draw  [draw opacity=0] (146.06,109.65) .. controls (145.08,121.23) and (140.5,130) .. (135,130) .. controls (128.79,130) and (123.75,118.81) .. (123.75,105) .. controls (123.75,91.19) and (128.79,80) .. (135,80) .. controls (140.68,80) and (145.38,89.35) .. (146.14,101.5) -- (135,105) -- cycle ; \draw   (146.06,109.65) .. controls (145.08,121.23) and (140.5,130) .. (135,130) .. controls (128.79,130) and (123.75,118.81) .. (123.75,105) .. controls (123.75,91.19) and (128.79,80) .. (135,80) .. controls (140.68,80) and (145.38,89.35) .. (146.14,101.5) ; 
\draw    (128.11,105) -- (168.75,105) ;
\draw (155,91.4) node [anchor=north west][inner sep=0.75pt]  [font=\small]  {$k$};
\draw (115,82.4) node [anchor=north west][inner sep=0.75pt]  [font=\small]  {$j$};
\end{tikzpicture} \;=\, \frac{S_{jk}}{S_{\bbi k}}\,\,
\begin{tikzpicture}[x=0.75pt,y=0.75pt,yscale=-1.2,xscale=1.2,baseline={([yshift=-1.7ex]current bounding box.center)}]
\draw    (99.9,105) -- (168.75,105) ;
\draw (126.2,92.2) node [anchor=north west][inner sep=0.75pt]  [font=\small]  {$k$};
\end{tikzpicture}\,\,.
\end{equation}
Here $S_{jk}$ is the modular S-matrix of the RCFT and $\bbi$ denotes the identity module. Then, after unlinking $i$ and $j$, we are left with a single Wilson loop $Z_{\text{CS}}\left[\begin{tikzpicture}[x=0.75pt,y=0.75pt,yscale=-0.5,xscale=0.5,baseline={([yshift=-0.5ex]current bounding box.center)}]
\draw   (55.5,74.5) .. controls (55.5,60.69) and (66.69,49.5) .. (80.5,49.5) .. controls (94.31,49.5) and (105.5,60.69) .. (105.5,74.5) .. controls (105.5,88.31) and (94.31,99.5) .. (80.5,99.5) .. controls (66.69,99.5) and (55.5,88.31) .. (55.5,74.5) -- cycle ;
\draw (84,62.9) node [anchor=north west][inner sep=0.75pt]    {{\tiny $k$}};
\end{tikzpicture}\right]$. 
    \item First fuse the loops $i$ and $j$ into a new Wilson loop $h$. The allowed representations for $h$ are determined by the fusion coefficient $N_{ij}\text{}^h$:
    \begin{equation}
        \begin{tikzpicture}[x=0.75pt,y=0.75pt,yscale=-1,xscale=1,baseline={([yshift=-0.5ex]current bounding box.center)}]
\draw  [draw opacity=0]  (223.87,127.6) .. controls (223.87,127.65) and (223.88,127.7) .. (223.88,127.75) .. controls (223.88,138.8) and (219.87,147.75) .. (214.94,147.75) .. controls (210,147.75) and (206,138.8) .. (206,127.75) .. controls (206,116.7) and (210,107.75) .. (214.94,107.75) .. controls (218.9,107.75) and (222.27,113.53) .. (223.44,121.54) -- (214.94,127.75) -- cycle ; \draw    (223.87,127.6) .. controls (223.87,127.65) and (223.88,127.7) .. (223.88,127.75) .. controls (223.88,138.8) and (219.87,147.75) .. (214.94,147.75) .. controls (210,147.75) and (206,138.8) .. (206,127.75) .. controls (206,116.7) and (210,107.75) .. (214.94,107.75) .. controls (218.9,107.75) and (222.27,113.53) .. (223.44,121.54) ;  
\draw  [draw opacity=0]  (209.14,133.06) .. controls (215.31,126.48) and (224.08,122.38) .. (233.81,122.38) .. controls (235.74,122.38) and (237.63,122.54) .. (239.46,122.85) -- (233.81,156.19) -- cycle ; \draw    (209.14,133.06) .. controls (215.31,126.48) and (224.08,122.38) .. (233.81,122.38) .. controls (235.74,122.38) and (237.63,122.54) .. (239.46,122.85) ;  
\draw  [draw opacity=0]  (245.92,124.61) .. controls (258.61,129.48) and (267.63,141.78) .. (267.63,156.19) .. controls (267.63,174.86) and (252.49,190) .. (233.81,190) .. controls (215.14,190) and (200,174.86) .. (200,156.19) .. controls (200,150.32) and (201.49,144.81) .. (204.12,140) -- (233.81,156.19) -- cycle ; \draw    (245.92,124.61) .. controls (258.61,129.48) and (267.63,141.78) .. (267.63,156.19) .. controls (267.63,174.86) and (252.49,190) .. (233.81,190) .. controls (215.14,190) and (200,174.86) .. (200,156.19) .. controls (200,150.32) and (201.49,144.81) .. (204.12,140) ;  
\draw  [draw opacity=0]  (258.84,137.34) .. controls (257.34,143.68) and (254.37,148) .. (250.94,148) .. controls (246,148) and (242,139.05) .. (242,128) .. controls (242,116.95) and (246,108) .. (250.94,108) .. controls (255.87,108) and (259.88,116.95) .. (259.88,128) .. controls (259.88,128.8) and (259.85,129.6) .. (259.81,130.38) -- (250.94,128) -- cycle ; \draw    (258.84,137.34) .. controls (257.34,143.68) and (254.37,148) .. (250.94,148) .. controls (246,148) and (242,139.05) .. (242,128) .. controls (242,116.95) and (246,108) .. (250.94,108) .. controls (255.87,108) and (259.88,116.95) .. (259.88,128) .. controls (259.88,128.8) and (259.85,129.6) .. (259.81,130.38) ;  
\draw [color={rgb, 255:red, 208; green, 2; blue, 27 }  ,draw opacity=1 ]   (220.6,104.2) .. controls (227.77,94.93) and (237.86,97.75) .. (244.02,101.81) ;
\draw [shift={(246.4,103.6)}, rotate = 220.6] [fill={rgb, 255:red, 208; green, 2; blue, 27 }  ,fill opacity=1 ][line width=0.08]  [draw opacity=0] (5.36,-2.57) -- (0,0) -- (5.36,2.57) -- cycle    ;
\draw  [draw opacity=0]  (428.47,128.43) .. controls (427.59,137.42) and (424.02,144.15) .. (419.74,144.15) .. controls (414.8,144.15) and (410.8,135.2) .. (410.8,124.15) .. controls (410.8,113.1) and (414.8,104.15) .. (419.74,104.15) .. controls (423.97,104.15) and (427.52,110.74) .. (428.44,119.58) -- (419.74,124.15) -- cycle ; \draw    (428.47,128.43) .. controls (427.59,137.42) and (424.02,144.15) .. (419.74,144.15) .. controls (414.8,144.15) and (410.8,135.2) .. (410.8,124.15) .. controls (410.8,113.1) and (414.8,104.15) .. (419.74,104.15) .. controls (423.97,104.15) and (427.52,110.74) .. (428.44,119.58) ;  
\draw  [draw opacity=0]  (415.02,122.66) .. controls (416.46,122.47) and (417.92,122.38) .. (419.41,122.38) .. controls (421.34,122.38) and (423.23,122.54) .. (425.06,122.85) -- (419.41,156.19) -- cycle ; \draw    (415.02,122.66) .. controls (416.46,122.47) and (417.92,122.38) .. (419.41,122.38) .. controls (421.34,122.38) and (423.23,122.54) .. (425.06,122.85) ;  
\draw  [draw opacity=0]  (423.96,122.68) .. controls (440.48,124.9) and (453.23,139.06) .. (453.23,156.19) .. controls (453.23,174.86) and (438.09,190) .. (419.41,190) .. controls (400.74,190) and (385.6,174.86) .. (385.6,156.19) .. controls (385.6,141.71) and (394.7,129.35) .. (407.5,124.53) -- (419.41,156.19) -- cycle ; \draw    (423.96,122.68) .. controls (440.48,124.9) and (453.23,139.06) .. (453.23,156.19) .. controls (453.23,174.86) and (438.09,190) .. (419.41,190) .. controls (400.74,190) and (385.6,174.86) .. (385.6,156.19) .. controls (385.6,141.71) and (394.7,129.35) .. (407.5,124.53) ;  
\draw (196.74,110.37) node [anchor=north west][inner sep=0.75pt]  [font=\small]  {$i$};
\draw (262.8,111.26) node [anchor=north west][inner sep=0.75pt]  [font=\small]  {$j$};
\draw (271.49,150.14) node [anchor=north west][inner sep=0.75pt]  [font=\small]  {$k$};
\draw (293.8,143.6) node [anchor=north west][inner sep=0.75pt]    {$=$};
\draw (317.8,141.8) node [anchor=north west][inner sep=0.75pt]    {$\sum_{h}$};
\draw (432.34,103.17) node [anchor=north west][inner sep=0.75pt]  [font=\small]  {$h$};
\draw (457.09,150.14) node [anchor=north west][inner sep=0.75pt]  [font=\small]  {$k$};
\draw (346.6,141.8) node [anchor=north west][inner sep=0.75pt]    {$N_{ij}\text{}^h$};
\end{tikzpicture}
    \end{equation}
    
    Then use the TQFT formula \eqref{eq:TQFT_rule_1} to unlink $h$ from $k$. What's left is again a single Wilson loop inside $S^3$, labeled by $k$. 
\end{enumerate}
The two ways of computing the link should give the same answer for the Chern-Simons partition function. Therefore, consistency of the TQFT demands that:
\begin{equation}
    \frac{S_{ik}S_{jk}}{S_{\bbi k}S_{\bbi k}} = \sum_{h} \, N_{ij}\text{}^h\, \frac{S_{hk}}{S_{\bbi k}}\,.
\end{equation}
This is basically Verlinde's formula. We can bring it in the equivalent form \eqref{eq:original_formula} by dividing out a common factor $S_{\bbi k}$ and acting with the inverse S-matrix $(S^{-1})_{hk} = S^*_{kh}$.

A simple generalization of this argument is to consider a ring labeled by $k$, as in \eqref{eq:ring}, but now with multiple Wilson loops $i_1,\dots i_{n-1}$ linked to it. There are again two ways to compute the Chern-Simons partition function of this configuration: either we first unlink all loops $i_m$ using $n-1$ applications of \eqref{eq:TQFT_rule_1}, or we first fuse all loops with each other successively. The latter simply computes the dimension of $\mathcal{H}_{0,n}$.

\subsection{Virasoro TQFT proof}\label{sec:VTQFT_proof}

 Inspired by the above strategy, we explore to what extent the derivation should be modified for non-rational CFT's with only Virasoro symmetry and $c\geq 25$. In this case, the Hilbert space of Virasoro conformal blocks $\mathcal{H}_{g,n}$ is infinite-dimensional \cite{Verlinde:1989ua}. The Virasoro conformal blocks are labeled by continuous conformal weights, for which we use the conventional Liouville parametrization:
\begin{equation}\label{eq:Liouv_par}
    \Delta = \frac{c-1}{24} + P^2, \quad \bar \Delta = \frac{c-1}{24} + \bar{P}^2, \quad  c = 1 + 6 Q^2  , \quad Q = b + b^{-1} \ .
\end{equation}
Here $\Delta,\bar \Delta$ label the conformal weights of primaries with respect to the Virasoro algebra $\mathrm{Vir}\times \overline{\mathrm{Vir}}$. The blocks factorize holomorphically as $\mathcal{F}^{\,\mathcal{C}}_{g,n}(P_i;P_{\O_i};\Omega)\,\overline{\mathcal{F}}^{\,\mathcal{C}}_{g,n}(\bar{P}_i;\bar{P}_{\O_i};\bar\Omega)$, where $\{P_i\}_{i=1}^{3g-3+n}$ are a set of `internal momenta' associated to a given channel $\mathcal{C}$ of the block (corresponding to a choice of pair-of-pants decomposition of the Riemann surface), and $\{P_{\O_i}\}_{i=1}^n$ label the conformal weights of the external operators $\O_i$ inserted at the punctures. Finally, $\Omega$ denotes the set moduli of the Riemann surface.

A complete basis of non-degenerate Virasoro conformal blocks is parametrized by positive real values of $P_i$ (so $\Delta_i\geq \frac{c-1}{24}$). Degenerate conformal blocks, such as the vacuum block, exist as analytic continuations in $P_i$ to discrete points on the imaginary axis.\footnote{For the degenerate blocks one has to subtract any null-states by hand, as is well-understood \cite{Eberhardt:2023mrq}.} To make the space of $c\geq 25$ Virasoro conformal blocks into a Hilbert space, an inner product  was constructed in \cite{Collier_2023}; see also our appendix \ref{app:prelim}.

Conformal blocks in one channel $\mathcal{C}$ can be mapped to conformal blocks in a different channel $\mathcal{C}'$ by so-called \emph{crossing transformations}. For conformal blocks of the Virasoro algebra at $c\geq 25$, these transformations are represented as integral kernels $\mathbb{K}$, integrated over real internal momenta:
\begin{equation}
    \mathcal{F}^{\,\mathcal{C}}_{g,n}(P_i;P_{\O_k};\Omega) = \int_0^\infty \prod_{i} \dd P'_i \,\mathbb{K}_{\{P_i\},\{P_i'\}}\Big[\{P_{\O_k}\}_{k=1}^n\Big] \mathcal{F}^{\,\mathcal{C}'}_{g,n}(P'_i;P_{\O_k};\Omega').
\end{equation}
An excellent review of Virasoro crossing kernels was recently given in \cite{Eberhardt:2023mrq}. We also give a brief overview of these kernels in appendix \ref{app:prelim}. A complete set of generators for the Virasoro crossing kernels is given by the following three basic transformations: 
\begin{align}
    &\text{Sphere four-point fusion kernel:} &&\hspace{-1cm}\fker{P}{P'}{P_1}{P_2}{P_3}{P_4}, \label{eq:F}\\[0.2em] & \text{Torus one-point S-kernel:} &&\hspace{-1cm}\sker{P}{P'}{P_0},\label{eq:S}\\[0.5em]  &\text{Sphere three-point braiding phase:} &&\hspace{-1cm}\mathbb{B}_{P}^{P_2P_3} \delta(P-P').\label{eq:B}
\end{align}
These generators satisfy a non-rational version of the Moore-Seiberg construction \cite{Teschner:2005bz}.  Even though the conformal blocks are not known analytically for general $g,n$, the above generators of the Moore-Seiberg group \emph{are} known in closed form. The solutions were originally derived in \cite{Ponsot:1999uf,Ponsot:2000mt}, and a short review of their functional form is provided in appendix \ref{app:kernels}. Many more of their properties are collected in \cite{Eberhardt:2023mrq}. 
 
The existence of a non-rational analog of the Moore-Seiberg construction for the Virasoro algebra at $c\geq 25$ allows one to define a consistent 3d topological quantum field theory, which was named the Virasoro TQFT in \cite{Collier_2023,Collier:2024mgv}. There, it was used as an intermediate step in the calculation of exact partition functions in pure AdS$_3$ gravity on hyperbolic 3-manifolds. The Virasoro TQFT allows for efficient computations of multiboundary wormholes in 3d gravity, which capture higher statistical moments of OPE data \cite{deBoer:2024kat}. In this article, we will use the Virasoro TQFT as a powerful framework in which the infinite number of integral identities relating the generators \eqref{eq:F}, \eqref{eq:S} and \eqref{eq:B} are encoded in topological properties of 3-manifolds. This will allow us to easily derive the non-rational analog of the Verlinde formula. The connection to 3d gravity and multiboundary wormholes will be made in section \ref{sec:III}.

\paragraph{Two ways of computing a 3d partition function.} Let us now turn to the TQFT derivation of the Virasoro-Verlinde formula \eqref{eq:main_formula}. We would like to mimic the argument sketched in section \ref{sec:RCFT_recap}. However, a direct translation of that argument to the non-rational case does not work, for the following reason. An important technicality of the Virasoro TQFT is that the inner product for $\mathcal{H}_{g,n}$ is only well-defined for stable surfaces, which means that we should slice bulk 3-manifolds along surfaces of negative Euler characteristic. In particular, the sphere partition function is ill-defined.

So in order to use the Virasoro TQFT properly, our main idea is to slice the 3-manifold depicted in \eqref{eq:ring} along a four-punctured sphere. The Hilbert space of four-punctured sphere conformal blocks $\mathcal{H}_{0,4}$ has a well-defined inner product \eqref{eq:inner_product} and a complete basis of states \cite{Collier_2023}. Specifically, we compute the Virasoro TQFT partition function on a ball with boundary $\Sigma_{0,4}$, with the following configuration of bulk Wilson lines\footnote{The term Wilson line is still appropriate thanks to the connection between Virasoro TQFT and the quantization of $PSL(2,\R)$ Chern-Simons theory, or more precisely, its Teichm\"uller component \cite{Collier_2023}.}: 
\begin{equation}\label{eq:four_point_loop}
 M =\begin{tikzpicture}[x=0.75pt,y=0.75pt,yscale=-1.3,xscale=1.3,baseline={([yshift=-0.5ex]current bounding box.center)}]
\draw  [dash pattern={on 20pt off 4pt on 50.5pt off 0.75pt}]  (120,130) -- (190,130) ;
\draw [shift={(190,130)}, rotate = 0] [color={rgb, 255:red, 0; green, 0; blue, 0 }  ][fill={rgb, 255:red, 0; green, 0; blue, 0 }  ]      (0, 0) circle [x radius= 1.34, y radius= 1.34]   ;
\draw [shift={(120,130)}, rotate = 0] [color={rgb, 255:red, 0; green, 0; blue, 0 }  ][fill={rgb, 255:red, 0; green, 0; blue, 0 }  ]      (0, 0) circle [x radius= 1.34, y radius= 1.34]   ;
\draw  [dash pattern={on 20pt off 4pt on 50pt off 0.75pt}]  (120,160) -- (190,160) ;
\draw [shift={(190,160)}, rotate = 0] [color={rgb, 255:red, 0; green, 0; blue, 0 }  ][fill={rgb, 255:red, 0; green, 0; blue, 0 }  ]      (0, 0) circle [x radius= 1.34, y radius= 1.34]   ;
\draw [shift={(120,160)}, rotate = 0] [color={rgb, 255:red, 0; green, 0; blue, 0 }  ][fill={rgb, 255:red, 0; green, 0; blue, 0 }  ]      (0, 0) circle [x radius= 1.34, y radius= 1.34]   ;
\draw  [draw opacity=0] (165.27,163.22) .. controls (162.58,167.43) and (158.97,170) .. (155,170) .. controls (146.72,170) and (140,158.81) .. (140,145) .. controls (140,131.19) and (146.72,120) .. (155,120) .. controls (158.9,120) and (162.45,122.48) .. (165.11,126.54) -- (155,145) -- cycle ; 
\draw  [color={rgb, 255:red, 208; green, 2; blue, 27 }  ,draw opacity=1 ] (165.27,163.22) .. controls (162.58,167.43) and (158.97,170) .. (155,170) .. controls (146.72,170) and (140,158.81) .. (140,145) .. controls (140,131.19) and (146.72,120) .. (155,120) .. controls (158.9,120) and (162.45,122.48) .. (165.11,126.54) ;  
\draw  [draw opacity=0] (168.3,133.44) .. controls (169.39,136.9) and (170,140.83) .. (170,145) .. controls (170,149.77) and (169.2,154.24) .. (167.8,158.03) -- (155,145) -- cycle ; \draw  [color={rgb, 255:red, 208; green, 2; blue, 27 }  ,draw opacity=1 ] (168.3,133.44) .. controls (169.39,136.9) and (170,140.83) .. (170,145) .. controls (170,149.77) and (169.2,154.24) .. (167.8,158.03) ; 
\draw  [dash pattern={on 1.5pt off 1.5pt on 1.5pt off 1.5pt}] (109.93,145) .. controls (109.93,120.11) and (130.11,99.93) .. (155,99.93) .. controls (179.89,99.93) and (200.07,120.11) .. (200.07,145) .. controls (200.07,169.89) and (179.89,190.07) .. (155,190.07) .. controls (130.11,190.07) and (109.93,169.89) .. (109.93,145) -- cycle ;
\draw (121,116.4) node [anchor=north west][inner sep=0.75pt]  [font=\small]  {$P_{1}$};
\draw (121,146.4) node [anchor=north west][inner sep=0.75pt]  [font=\small]  {$P_{3}$};
\draw (149.86,105.97) node [anchor=north west][inner sep=0.75pt]  [font=\small]  {$P_2$};
\end{tikzpicture}\vspace{1mm}
\end{equation}
The Wilson lines are labeled by continuous Liouville momenta $P_i$ which parametrize the conformal weight of a non-degenerate Virasoro primary through \eqref{eq:Liouv_par}. 

By the usual rules of 3d TQFT, the path integral on the manifold $M$ defines a state 
    $Z_{\text{Vir}}(M) \in \mathcal{H}_{0,4}.$  
We can evaluate this partition function in two distinct ways:
\begin{enumerate}[1)]
    \item First unlink the red $P_2$ loop from the parallel Wilson lines $P_1$ and $P_3$ using the Virasoro TQFT identities \eqref{eq:linked_lines} and \eqref{eq:unlinking2}, which involve the one-point S kernel:
    \begin{equation}
    \begin{split}
       & Z_{\text{Vir}}\!\left( \begin{tikzpicture}[x=0.75pt,y=0.75pt,yscale=-1,xscale=1,baseline={([yshift=-1.2ex]current bounding box.center)}]
\draw  [dash pattern={on 15pt off 3.75pt on 37.5pt off 0.75pt}]  (101.25,110) -- (171.25,110) ;
\draw [shift={(101.25,110)}, rotate = 0] [color={rgb, 255:red, 0; green, 0; blue, 0 }  ][fill={rgb, 255:red, 0; green, 0; blue, 0 }  ]      (0, 0) circle [x radius= 1.34, y radius= 1.34]   ;
\draw [shift={(171.25,110)}, rotate = 0] [color={rgb, 255:red, 0; green, 0; blue, 0 }  ][fill={rgb, 255:red, 0; green, 0; blue, 0 }  ]      (0, 0) circle [x radius= 1.34, y radius= 1.34]   ;
\draw  [dash pattern={on 15pt off 3.75pt on 37.5pt off 0.75pt}]  (100,140) -- (170,140) ; 
\draw [shift={(100,140)}, rotate = 0] [color={rgb, 255:red, 0; green, 0; blue, 0 }  ][fill={rgb, 255:red, 0; green, 0; blue, 0 }  ]      (0, 0) circle [x radius= 1.34, y radius= 1.34]   ;
\draw [shift={(170,140)}, rotate = 0] [color={rgb, 255:red, 0; green, 0; blue, 0 }  ][fill={rgb, 255:red, 0; green, 0; blue, 0 }  ]      (0, 0) circle [x radius= 1.34, y radius= 1.34]   ;
\draw  [draw opacity=0] (145.27,143.22) .. controls (142.58,147.43) and (138.97,150) .. (135,150) .. controls (126.72,150) and (120,138.81) .. (120,125) .. controls (120,111.19) and (126.72,100) .. (135,100) .. controls (138.9,100) and (142.45,102.48) .. (145.11,106.54) -- (135,125) -- cycle ; \draw  [color={rgb, 255:red, 208; green, 2; blue, 27 }  ,draw opacity=1 ] (145.27,143.22) .. controls (142.58,147.43) and (138.97,150) .. (135,150) .. controls (126.72,150) and (120,138.81) .. (120,125) .. controls (120,111.19) and (126.72,100) .. (135,100) .. controls (138.9,100) and (142.45,102.48) .. (145.11,106.54) ;  
\draw  [draw opacity=0] (148.3,113.44) .. controls (149.39,116.9) and (150,120.83) .. (150,125) .. controls (150,129.77) and (149.2,134.24) .. (147.8,138.03) -- (135,125) -- cycle ; \draw  [color={rgb, 255:red, 208; green, 2; blue, 27 }  ,draw opacity=1 ] (148.3,113.44) .. controls (149.39,116.9) and (150,120.83) .. (150,125) .. controls (150,129.77) and (149.2,134.24) .. (147.8,138.03) ; 
\draw (101,96.4) node [anchor=north west][inner sep=0.75pt]  [font=\scriptsize]  {$1$};
\draw (101,126.4) node [anchor=north west][inner sep=0.75pt]  [font=\scriptsize]  {$3$};
\draw (131,88.4) node [anchor=north west][inner sep=0.75pt]  [font=\scriptsize]  {$2$};
\end{tikzpicture}\right)  =\\&\int_0^\infty \dd P_0 \dd P_0'\,\fker{\bbi}{P_0}{P_1}{P_1}{P_1}{P_1}\frac{\mathbb{S}_{P_1P_2}[P_0]}{\sker{\bbi}{P_2}{\bbi}}\fker{\bbi}{P_0'}{P_2}{P_2}{P_2}{P_2}\frac{\mathbb{S}^*_{P_2P_3}[P_0']}{\sker{\bbi}{P_3}{\bbi}}\,Z_{\text{Vir}}\!\left(\,\,\begin{tikzpicture}[x=0.75pt,y=0.75pt,yscale=-1,xscale=1,baseline={([yshift=-0.7ex]current bounding box.center)}]
\draw    (232.29,186.59) -- (232.29,203.87) ;
\draw    (232.29,223.07) -- (232.29,240.36) ;
\draw    (200,186.27) -- (270,186.27) ;
\draw [shift={(270,186.27)}, rotate = 0] [color={rgb, 255:red, 0; green, 0; blue, 0 }  ][fill={rgb, 255:red, 0; green, 0; blue, 0 }  ]      (0, 0) circle [x radius= 1.34, y radius= 1.34]   ;
\draw [shift={(200,186.27)}, rotate = 0] [color={rgb, 255:red, 0; green, 0; blue, 0 }  ][fill={rgb, 255:red, 0; green, 0; blue, 0 }  ]      (0, 0) circle [x radius= 1.34, y radius= 1.34]   ;
\draw    (200.03,240.18) -- (270.03,240.18) ;
\draw [shift={(270.03,240.18)}, rotate = 0] [color={rgb, 255:red, 0; green, 0; blue, 0 }  ][fill={rgb, 255:red, 0; green, 0; blue, 0 }  ]      (0, 0) circle [x radius= 1.34, y radius= 1.34]   ;
\draw [shift={(200.03,240.18)}, rotate = 0] [color={rgb, 255:red, 0; green, 0; blue, 0 }  ][fill={rgb, 255:red, 0; green, 0; blue, 0 }  ]      (0, 0) circle [x radius= 1.34, y radius= 1.34]   ;
\draw  [color={rgb, 255:red, 208; green, 2; blue, 27 }  ,draw opacity=1 ] (222.69,213.47) .. controls (222.69,208.17) and (226.98,203.87) .. (232.29,203.87) .. controls (237.59,203.87) and (241.89,208.17) .. (241.89,213.47) .. controls (241.89,218.77) and (237.59,223.07) .. (232.29,223.07) .. controls (226.98,223.07) and (222.69,218.77) .. (222.69,213.47) -- cycle ;
\draw (201,174.34) node [anchor=north west][inner sep=0.75pt]  [font=\scriptsize]  {$1$};
\draw (201.03,226.58) node [anchor=north west][inner sep=0.75pt]  [font=\scriptsize]  {$3$};
\draw (234.29,225.47) node [anchor=north west][inner sep=0.75pt]  [font=\scriptsize]  {$P_{0} '$};
\draw (244.26,208.28) node [anchor=north west][inner sep=0.75pt]  [font=\scriptsize]  {$2$};
\draw (234.29,189.99) node [anchor=north west][inner sep=0.75pt]  [font=\scriptsize]  {$P_{0}$};
\end{tikzpicture}\,\,\right).
\end{split}
    \end{equation}
   These unlinking formulas are derived in appendix \ref{app:TQFT_identities} and have not appeared in the literature before. 
   Next, remove the red `Wilson bubble', using the Virasoro TQFT identity \eqref{eq:WilsonBubble}. This gives:
    \begin{equation}\label{eq:part_fn_1}
        Z_{\text{Vir}}(M)= \int_0^\infty \dd P_0\, \fker{\bbi}{P_0}{P_1}{P_1}{P_1}{P_1}\, \frac{\mathbb{S}_{P_1P_2}[P_0]\,\mathbb{S}^*_{P_2P_3}[P_0]}{\sker{\bbi}{P_2}{\bbi}\,\sker{\bbi}{P_3}{\bbi}} \,Z_{\text{Vir}}\!\left(\, \begin{tikzpicture}[x=0.75pt,y=0.75pt,yscale=-1,xscale=1,baseline={([yshift=-1ex]current bounding box.center)}]
\draw    (235,181.42) -- (235.12,230) ;
\draw    (200,181.42) -- (270,181.42) ;
\draw [shift={(270,181.42)}, rotate = 0] [color={rgb, 255:red, 0; green, 0; blue, 0 }  ][fill={rgb, 255:red, 0; green, 0; blue, 0 }  ]      (0, 0) circle [x radius= 1.34, y radius= 1.34]   ;
\draw [shift={(200,181.42)}, rotate = 0] [color={rgb, 255:red, 0; green, 0; blue, 0 }  ][fill={rgb, 255:red, 0; green, 0; blue, 0 }  ]      (0, 0) circle [x radius= 1.34, y radius= 1.34]   ;
\draw    (200.12,230) -- (270.12,230) ;
\draw [shift={(270.12,230)}, rotate = 0] [color={rgb, 255:red, 0; green, 0; blue, 0 }  ][fill={rgb, 255:red, 0; green, 0; blue, 0 }  ]      (0, 0) circle [x radius= 1.34, y radius= 1.34]   ;
\draw [shift={(200.12,230)}, rotate = 0] [color={rgb, 255:red, 0; green, 0; blue, 0 }  ][fill={rgb, 255:red, 0; green, 0; blue, 0 }  ]      (0, 0) circle [x radius= 1.34, y radius= 1.34]   ;
\draw (201,169.49) node [anchor=north west][inner sep=0.75pt]  [font=\scriptsize]  {$1$};
\draw (201.12,216.4) node [anchor=north west][inner sep=0.75pt]  [font=\scriptsize]  {$3$};
\draw (238.45,199.63) node [anchor=north west][inner sep=0.75pt]  [font=\small]  {$P_0$};
\end{tikzpicture}\,\right).
    \end{equation}
    Here we used the relation \eqref{eq:C0} between $\mathbb{F}$ and $C_0$ to simplify the integrand. So we have written $Z_{\text{Vir}}(M)$ as a superposition of sphere four-point blocks in the $t$-channel, with an exchanged heavy operator of weight $\Delta_0$. 
\item First fuse $P_1$ and $P_3$, by inserting the identity line and applying an identity fusion move: 
\begin{equation}
Z_\text{Vir}\left(
    \begin{tikzpicture}[x=0.75pt,y=0.75pt,yscale=-1,xscale=1,baseline={([yshift=-1ex]current bounding box.center)}] 
\draw  [dash pattern={on 15pt off 3.75pt on 37.5pt off 0.75pt}]  (101.25,110) -- (171.25,110) ;
\draw [shift={(171.24,110)}, rotate = 0] [color={rgb, 255:red, 0; green, 0; blue, 0 }  ][fill={rgb, 255:red, 0; green, 0; blue, 0 }  ]      (0, 0) circle [x radius= 1.34, y radius= 1.34]   ;
\draw [shift={(101.25,110)}, rotate = 0] [color={rgb, 255:red, 0; green, 0; blue, 0 }  ][fill={rgb, 255:red, 0; green, 0; blue, 0 }  ]      (0, 0) circle [x radius= 1.34, y radius= 1.34]   ;
\draw  [dash pattern={on 15pt off 3.75pt on 37.5pt off 0.75pt}]  (100,140) -- (170,140) ;
\draw [shift={(170,140)}, rotate = 0] [color={rgb, 255:red, 0; green, 0; blue, 0 }  ][fill={rgb, 255:red, 0; green, 0; blue, 0 }  ]      (0, 0) circle [x radius= 1.34, y radius= 1.34]   ;
\draw [shift={(100,140)}, rotate = 0] [color={rgb, 255:red, 0; green, 0; blue, 0 }  ][fill={rgb, 255:red, 0; green, 0; blue, 0 }  ]      (0, 0) circle [x radius= 1.34, y radius= 1.34]   ;
\draw  [draw opacity=0] (145.27,143.22) .. controls (142.58,147.43) and (138.97,150) .. (135,150) .. controls (126.72,150) and (120,138.81) .. (120,125) .. controls (120,111.19) and (126.72,100) .. (135,100) .. controls (138.9,100) and (142.45,102.48) .. (145.11,106.54) -- (135,125) -- cycle ; \draw  [color={rgb, 255:red, 208; green, 2; blue, 27 }  ,draw opacity=1 ] (145.27,143.22) .. controls (142.58,147.43) and (138.97,150) .. (135,150) .. controls (126.72,150) and (120,138.81) .. (120,125) .. controls (120,111.19) and (126.72,100) .. (135,100) .. controls (138.9,100) and (142.45,102.48) .. (145.11,106.54) ;  
\draw  [draw opacity=0] (148.3,113.44) .. controls (149.39,116.9) and (150,120.83) .. (150,125) .. controls (150,129.77) and (149.2,134.24) .. (147.8,138.03) -- (135,125) -- cycle ; \draw  [color={rgb, 255:red, 208; green, 2; blue, 27 }  ,draw opacity=1 ] (148.3,113.44) .. controls (149.39,116.9) and (150,120.83) .. (150,125) .. controls (150,129.77) and (149.2,134.24) .. (147.8,138.03) ; 
\draw  [dash pattern={on 1.5pt off 1.5pt on 1.5pt off 1.5pt}]  (160,110) -- (160,140) ;
\draw (101,96.4) node [anchor=north west][inner sep=0.75pt]  [font=\scriptsize]  {$1$};
\draw (101,126.4) node [anchor=north west][inner sep=0.75pt]  [font=\scriptsize]  {$3$};
\draw (131,88.4) node [anchor=north west][inner sep=0.75pt]  [font=\scriptsize]  {$2$};
\draw (167,118.4) node [anchor=north west][inner sep=0.75pt]  [font=\scriptsize]  {$\bbi$};
\end{tikzpicture}\right) = \int_0^\infty \dd P \,\fker{\bbi}{P}{P_3}{P_1}{P_1}{P_3} \,Z_{\text{Vir}}\left(\begin{tikzpicture}[x=0.75pt,y=0.75pt,yscale=-1,xscale=1,baseline={([yshift=-1ex]current bounding box.center)}]
\draw  [dash pattern={on 15pt off 3.75pt on 37.5pt off 0.75pt}]  (101.25,110) -- (150,110) ;
\draw [shift={(101.25,110)}, rotate = 0] [color={rgb, 255:red, 0; green, 0; blue, 0 }  ][fill={rgb, 255:red, 0; green, 0; blue, 0 }  ]      (0, 0) circle [x radius= 1.34, y radius= 1.34]   ;
\draw  [dash pattern={on 15pt off 3.75pt on 37.5pt off 0.75pt}]  (100,140) -- (150,140) ;
\draw [shift={(100,140)}, rotate = 0] [color={rgb, 255:red, 0; green, 0; blue, 0 }  ][fill={rgb, 255:red, 0; green, 0; blue, 0 }  ]      (0, 0) circle [x radius= 1.34, y radius= 1.34]   ;
\draw  [draw opacity=0] (145.27,143.22) .. controls (142.58,147.43) and (138.97,150) .. (135,150) .. controls (126.72,150) and (120,138.81) .. (120,125) .. controls (120,111.19) and (126.72,100) .. (135,100) .. controls (138.9,100) and (142.45,102.48) .. (145.11,106.54) -- (135,125) -- cycle ; \draw  [color={rgb, 255:red, 208; green, 2; blue, 27 }  ,draw opacity=1 ] (145.27,143.22) .. controls (142.58,147.43) and (138.97,150) .. (135,150) .. controls (126.72,150) and (120,138.81) .. (120,125) .. controls (120,111.19) and (126.72,100) .. (135,100) .. controls (138.9,100) and (142.45,102.48) .. (145.11,106.54) ;  
\draw  [draw opacity=0] (148.3,113.44) .. controls (149.39,116.9) and (150,120.83) .. (150,125) .. controls (150,129.77) and (149.2,134.24) .. (147.8,138.03) -- (135,125) -- cycle ; \draw  [color={rgb, 255:red, 208; green, 2; blue, 27 }  ,draw opacity=1 ] (148.3,113.44) .. controls (149.39,116.9) and (150,120.83) .. (150,125) .. controls (150,129.77) and (149.2,134.24) .. (147.8,138.03) ; 
\draw    (160.22,125.33) -- (180.22,125.33) ;
\draw    (150,110) -- (160.22,125.33) ;
\draw    (150,140) -- (160.22,125.33) ;
\draw    (190,110) -- (180.22,125.33) ;
\draw [shift={(190,110)}, rotate = 0] [color={rgb, 255:red, 0; green, 0; blue, 0 }  ][fill={rgb, 255:red, 0; green, 0; blue, 0 }  ]      (0, 0) circle [x radius= 1.34, y radius= 1.34]   ;
\draw    (180.22,125.33) -- (190,140) ;
\draw [shift={(190,140)}, rotate = 0] [color={rgb, 255:red, 0; green, 0; blue, 0 }  ][fill={rgb, 255:red, 0; green, 0; blue, 0 }  ]      (0, 0) circle [x radius= 1.34, y radius= 1.34]   ;
\draw (101,96.4) node [anchor=north west][inner sep=0.75pt]  [font=\scriptsize]  {$1$};
\draw (101,126.4) node [anchor=north west][inner sep=0.75pt]  [font=\scriptsize]  {$3$};
\draw (131,88.4) node [anchor=north west][inner sep=0.75pt]  [font=\scriptsize]  {$2$};
\draw (163,125.96) node [anchor=north west][inner sep=0.75pt]  [font=\scriptsize]  {$P$};
\draw (193,102.4) node [anchor=north west][inner sep=0.75pt]  [font=\scriptsize]  {$1$};
\draw (193,132.4) node [anchor=north west][inner sep=0.75pt]  [font=\scriptsize]  {$3$};
\end{tikzpicture} \right) \,.
\end{equation}
The operation of adding an identity line can always be done `for free', as it corresponds to cutting the 3-manifold locally along a four-punctured sphere surrounding a neighbourhood of the endpoints of the identity line. Next,
apply the TQFT rule for the Verlinde loop operator \eqref{eq:TQFTrule} to remove the $P_2$ loop: 
\begin{equation}
    \begin{tikzpicture}[x=0.75pt,y=0.75pt,yscale=-1,xscale=1,baseline={([yshift=-1ex]current bounding box.center)}]
\draw  [dash pattern={on 15pt off 3.75pt on 37.5pt off 0.75pt}]  (101.25,110) -- (150,110) ;
\draw [shift={(101.25,110)}, rotate = 0] [color={rgb, 255:red, 0; green, 0; blue, 0 }  ][fill={rgb, 255:red, 0; green, 0; blue, 0 }  ]      (0, 0) circle [x radius= 1.34, y radius= 1.34]   ;
\draw  [dash pattern={on 15pt off 3.75pt on 37.5pt off 0.75pt}]  (100,140) -- (150,140) ;
\draw [shift={(100,140)}, rotate = 0] [color={rgb, 255:red, 0; green, 0; blue, 0 }  ][fill={rgb, 255:red, 0; green, 0; blue, 0 }  ]      (0, 0) circle [x radius= 1.34, y radius= 1.34]   ;
\draw  [draw opacity=0] (145.27,143.22) .. controls (142.58,147.43) and (138.97,150) .. (135,150) .. controls (126.72,150) and (120,138.81) .. (120,125) .. controls (120,111.19) and (126.72,100) .. (135,100) .. controls (138.9,100) and (142.45,102.48) .. (145.11,106.54) -- (135,125) -- cycle ; \draw  [color={rgb, 255:red, 208; green, 2; blue, 27 }  ,draw opacity=1 ] (145.27,143.22) .. controls (142.58,147.43) and (138.97,150) .. (135,150) .. controls (126.72,150) and (120,138.81) .. (120,125) .. controls (120,111.19) and (126.72,100) .. (135,100) .. controls (138.9,100) and (142.45,102.48) .. (145.11,106.54) ;  
\draw  [draw opacity=0] (148.3,113.44) .. controls (149.39,116.9) and (150,120.83) .. (150,125) .. controls (150,129.77) and (149.2,134.24) .. (147.8,138.03) -- (135,125) -- cycle ; \draw  [color={rgb, 255:red, 208; green, 2; blue, 27 }  ,draw opacity=1 ] (148.3,113.44) .. controls (149.39,116.9) and (150,120.83) .. (150,125) .. controls (150,129.77) and (149.2,134.24) .. (147.8,138.03) ; 
\draw    (160.22,125.33) -- (180.22,125.33) ;
\draw    (150,110) -- (160.22,125.33) ;
\draw    (150,140) -- (160.22,125.33) ;
\draw    (190,110) -- (180.22,125.33) ;
\draw [shift={(190,110)}, rotate = 0] [color={rgb, 255:red, 0; green, 0; blue, 0 }  ][fill={rgb, 255:red, 0; green, 0; blue, 0 }  ]      (0, 0) circle [x radius= 1.34, y radius= 1.34]   ;
\draw    (180.22,125.33) -- (190,140) ;
\draw [shift={(190,140)}, rotate = 0] [color={rgb, 255:red, 0; green, 0; blue, 0 }  ][fill={rgb, 255:red, 0; green, 0; blue, 0 }  ]      (0, 0) circle [x radius= 1.34, y radius= 1.34]   ;
\draw [color={rgb, 255:red, 208; green, 2; blue, 27 }  ,draw opacity=1 ] [dash pattern={on 1.5pt off 1.5pt on 1.5pt off 1.5pt}]  (145.56,101.22) .. controls (151.85,99.94) and (160.23,111.55) .. (166.08,117.62) ;
\draw [shift={(168.22,119.67)}, rotate = 220.24] [fill={rgb, 255:red, 208; green, 2; blue, 27 }  ,fill opacity=1 ][line width=0.08]  [draw opacity=0] (3.57,-1.72) -- (0,0) -- (3.57,1.72) -- cycle    ;
\draw (101,96.4) node [anchor=north west][inner sep=0.75pt]  [font=\scriptsize]  {$1$};
\draw (101,126.4) node [anchor=north west][inner sep=0.75pt]  [font=\scriptsize]  {$3$};
\draw (131,88.4) node [anchor=north west][inner sep=0.75pt]  [font=\scriptsize]  {$2$};
\draw (163,125.96) node [anchor=north west][inner sep=0.75pt]  [font=\scriptsize]  {$P$};
\draw (193,102.4) node [anchor=north west][inner sep=0.75pt]  [font=\scriptsize]  {$1$};
\draw (193,132.4) node [anchor=north west][inner sep=0.75pt]  [font=\scriptsize]  {$3$};
\end{tikzpicture} \,\, =\,\, \begin{tikzpicture}[x=0.75pt,y=0.75pt,yscale=-1,xscale=1,baseline={([yshift=-1ex]current bounding box.center)}]
\draw    (182.5,145) -- (200.22,145) ;
\draw [shift={(160,130)}, rotate = 0] [color={rgb, 255:red, 0; green, 0; blue, 0 }  ][fill={rgb, 255:red, 0; green, 0; blue, 0 }  ]      (0, 0) circle [x radius= 1.34, y radius= 1.34]   ;
\draw    (160,130) -- (170.22,145) ;
\draw    (160,160) -- (170.22,145) ;
\draw [shift={(160,160)}, rotate = 0] [color={rgb, 255:red, 0; green, 0; blue, 0 }  ][fill={rgb, 255:red, 0; green, 0; blue, 0 }  ]      (0, 0) circle [x radius= 1.34, y radius= 1.34]   ;
\draw    (210,130) -- (200.22,145) ;
\draw [shift={(210,130)}, rotate = 0] [color={rgb, 255:red, 0; green, 0; blue, 0 }  ][fill={rgb, 255:red, 0; green, 0; blue, 0 }  ]      (0, 0) circle [x radius= 1.34, y radius= 1.34]   ;
\draw    (200.22,145) -- (210,160) ;
\draw [shift={(210,160)}, rotate = 0] [color={rgb, 255:red, 0; green, 0; blue, 0 }  ][fill={rgb, 255:red, 0; green, 0; blue, 0 }  ]      (0, 0) circle [x radius= 1.34, y radius= 1.34]   ;
\draw  [draw opacity=0] (190.15,149.58) .. controls (189.75,161.36) and (187.69,170.33) .. (185.22,170.33) .. controls (182.46,170.33) and (180.22,159.14) .. (180.22,145.33) .. controls (180.22,131.53) and (182.46,120.33) .. (185.22,120.33) .. controls (187.68,120.33) and (189.72,129.2) .. (190.14,140.88) -- (185.22,145.33) -- cycle ; \draw  [color={rgb, 255:red, 208; green, 2; blue, 27 }  ,draw opacity=1 ] (190.15,149.58) .. controls (189.75,161.36) and (187.69,170.33) .. (185.22,170.33) .. controls (182.46,170.33) and (180.22,159.14) .. (180.22,145.33) .. controls (180.22,131.53) and (182.46,120.33) .. (185.22,120.33) .. controls (187.68,120.33) and (189.72,129.2) .. (190.14,140.88) ;  
\draw    (170.22,145.33) -- (178.21,145.36) ;
\draw (148.6,122.6) node [anchor=north west][inner sep=0.75pt]  [font=\scriptsize]  {$1$};
\draw (148.8,153) node [anchor=north west][inner sep=0.75pt]  [font=\scriptsize]  {$3$};
\draw (168.5,146.15) node [anchor=north west][inner sep=0.75pt]  [font=\scriptsize]  {$P$};
\draw (211,122.4) node [anchor=north west][inner sep=0.75pt]  [font=\scriptsize]  {$1$};
\draw (211,152.4) node [anchor=north west][inner sep=0.75pt]  [font=\scriptsize]  {$3$};
\draw (174.14,113.26) node [anchor=north west][inner sep=0.75pt]  [font=\scriptsize]  {$2$};
\end{tikzpicture}\,\, = \,\, \frac{\mathbb{S}_{P_2P}[\bbi]}{\mathbb{S}_{\bbi P}[\bbi]} \,\,\begin{tikzpicture}[x=0.75pt,y=0.75pt,yscale=-1,xscale=1,baseline={([yshift=-1ex]current bounding box.center)}]
\draw    (170.22,145.33) -- (200.22,145.33) ;
\draw    (160,130) -- (170.22,145.33) ;
\draw [shift={(160,130)}, rotate = 0] [color={rgb, 255:red, 0; green, 0; blue, 0 }  ][fill={rgb, 255:red, 0; green, 0; blue, 0 }  ]      (0, 0) circle [x radius= 1.34, y radius= 1.34]   ;
\draw    (160,160) -- (170.22,145.33) ;
\draw [shift={(160,160)}, rotate = 0] [color={rgb, 255:red, 0; green, 0; blue, 0 }  ][fill={rgb, 255:red, 0; green, 0; blue, 0 }  ]      (0, 0) circle [x radius= 1.34, y radius= 1.34]   ;
\draw    (210,130) -- (200.22,145.33) ;
\draw [shift={(210,130)}, rotate = 0] [color={rgb, 255:red, 0; green, 0; blue, 0 }  ][fill={rgb, 255:red, 0; green, 0; blue, 0 }  ]      (0, 0) circle [x radius= 1.34, y radius= 1.34]   ;
\draw    (200.22,145.33) -- (210,160) ;
\draw [shift={(210,160)}, rotate = 0] [color={rgb, 255:red, 0; green, 0; blue, 0 }  ][fill={rgb, 255:red, 0; green, 0; blue, 0 }  ]      (0, 0) circle [x radius= 1.34, y radius= 1.34]   ;
\draw (148.6,122.6) node [anchor=north west][inner sep=0.75pt]  [font=\scriptsize]  {$1$};
\draw (148.8,153) node [anchor=north west][inner sep=0.75pt]  [font=\scriptsize]  {$3$};
\draw (177.94,147.04) node [anchor=north west][inner sep=0.75pt]  [font=\scriptsize]  {$P$};
\draw (213,122.4) node [anchor=north west][inner sep=0.75pt]  [font=\scriptsize]  {$1$};
\draw (213,152.4) node [anchor=north west][inner sep=0.75pt]  [font=\scriptsize]  {$3$};
\end{tikzpicture}\,.
\end{equation}
The ratio of identity S-kernels that arises has the simple form \eqref{eq:ratio}. Finally, apply another fusion move to bring the conformal block back to the $t$-channel. This gives the following expression for the partition function:
\begin{equation}\label{eq:part_fn_2}
   Z_{\text{Vir}}(M) = \int_0^\infty \dd P \dd P_0\, \fker{\bbi}{P}{P_3}{P_1}{P_1}{P_3} \frac{\mathbb{S}_{P_2P}[\bbi]}{\sker{\bbi}{P}{\bbi}} \fker{P}{P_0}{P_1}{P_1}{P_3}{P_3}\,Z_{\text{Vir}}\!\left(\,\begin{tikzpicture}[x=0.75pt,y=0.75pt,yscale=-1,xscale=1,baseline={([yshift=-1ex]current bounding box.center)}]
\draw    (235,181.42) -- (235.12,230) ;
\draw    (200,181.42) -- (270,181.42) ;
\draw [shift={(270,181.42)}, rotate = 0] [color={rgb, 255:red, 0; green, 0; blue, 0 }  ][fill={rgb, 255:red, 0; green, 0; blue, 0 }  ]      (0, 0) circle [x radius= 1.34, y radius= 1.34]   ;
\draw [shift={(200,181.42)}, rotate = 0] [color={rgb, 255:red, 0; green, 0; blue, 0 }  ][fill={rgb, 255:red, 0; green, 0; blue, 0 }  ]      (0, 0) circle [x radius= 1.34, y radius= 1.34]   ;
\draw    (200.12,230) -- (270.12,230) ;
\draw [shift={(270.12,230)}, rotate = 0] [color={rgb, 255:red, 0; green, 0; blue, 0 }  ][fill={rgb, 255:red, 0; green, 0; blue, 0 }  ]      (0, 0) circle [x radius= 1.34, y radius= 1.34]   ;
\draw [shift={(200.12,230)}, rotate = 0] [color={rgb, 255:red, 0; green, 0; blue, 0 }  ][fill={rgb, 255:red, 0; green, 0; blue, 0 }  ]      (0, 0) circle [x radius= 1.34, y radius= 1.34]   ;
\draw (201,169.49) node [anchor=north west][inner sep=0.75pt]  [font=\scriptsize]  {$1$};
\draw (201.12,216.4) node [anchor=north west][inner sep=0.75pt]  [font=\scriptsize]  {$3$};
\draw (238.45,199.63) node [anchor=north west][inner sep=0.75pt]  [font=\small]  {$P_0$};
\end{tikzpicture}\,\right).
\end{equation}
\end{enumerate}
Consistency of the Virasoro TQFT demands that the two ways of calculating the partition function $Z_\text{Vir}(M)$ should be equal. Remark that we expressed both  \eqref{eq:part_fn_1} and \eqref{eq:part_fn_2} as superpositions of the four-point sphere conformal block in the $t$-channel, with exchanged momentum $P_0$. Using the fact that the $t$-channel conformal blocks with $P_0\in \mathbb{R}_+$ form a complete orthogonal basis for the Hilbert space $\mathcal{H}_{0,4}$, we conclude that the expansion coefficients should be equal:\vspace{2mm}
\begin{equation}\label{eq:intermediate}
    \int_0^\infty \dd P \,\fker{\bbi}{P}{P_3}{P_1}{P_1}{P_3} \frac{\mathbb{S}_{P_2P}[\bbi]}{\sker{\bbi}{P}{\bbi}} \fker{P}{P_0}{P_1}{P_1}{P_3}{P_3} = \fker{\bbi}{P_0}{P_1}{P_1}{P_1}{P_1}\, \frac{\mathbb{S}_{P_1P_2}[P_0]\mathbb{S}^*_{P_2P_3}[P_0]}{\sker{\bbi}{P_2}{\bbi}\sker{\bbi}{P_3}{\bbi}}.\vspace{2mm}
\end{equation}

This is essentially the Virasoro-Verlinde formula \eqref{eq:main_formula} that we wanted to prove. We can massage the equation into the desired form, by acting on both sides with the modular S kernel $\sker{P_2'}{P_2}{\bbi}$ (that is, taking the inverse Fourier transform of \eqref{eq:intermediate}) and using the permutation symmetry of the $C_0$ formula:
\begin{equation}
    \frac{\fker{\bbi}{P}{P_3}{P_1}{P_1}{P_3}}{\sker{\bbi}{P}{\bbi}} = C_0(P,P_1,P_3) = C_0(P_3,P,P_1) = \frac{\fker{\bbi}{P_3}{P_1}{P}{P}{P_1}}{\sker{\bbi}{P_3}{\bbi}}.
\end{equation}
Rearranging the F kernels to one side of the equation, and relabeling dummy variables, we arrive at the following equation:
\begin{equation}\label{eq:main_formula2}
    \frac{\fker{\bbi}{P_3}{P_2}{P_1}{P_1}{P_2}\fker{P_2}{P_0}{P_1}{P_1}{P_3}{P_3}}{\fker{\bbi}{P_0}{P_1}{P_1}{P_1}{P_1}} = \int_0^\infty \dd P \,\frac{\sker{P_1}{P}{P_0}\sker{P_2}{P}{\bbi}\mathbb{S}_{PP_3}^*[P_0]}{\sker{\bbi}{P}{\bbi}}.
\end{equation}
Finally, we can use the tetrahedral symmetry of the fusion kernel (see equation (2.57) in \cite{Eberhardt:2023mrq}) to write the left-hand side in terms of the fusion density \eqref{eq:fusion_density}:
\begin{equation}\label{eq:fusion_density2}
    \N_{P_0}[P_1,P_2,P_3] \coloneqq \fker{P_1}{P_3}{P_3}{P_0}{P_1}{P_2} =  \frac{\fker{\bbi}{P_3}{P_2}{P_1}{P_1}{P_2}\fker{P_2}{P_0}{P_1}{P_1}{P_3}{P_3}}{\fker{\bbi}{P_0}{P_1}{P_1}{P_1}{P_1}}\,.
\end{equation}
This concludes the Virasoro TQFT proof. 

\subsection{Properties of the fusion density}\label{sec:properties}
We will now explore the properties of the fusion density $\N_{P_0}$. As we saw above, it is expressed as a fusion kernel \eqref{eq:fusion_density2}, but it is not \emph{a priori} clear that it satisfies the properties expected of a fusion coefficient $N_{ij}\text{}^k$. 
In this section, we derive many of the relevant properties of $\N_{P_0}[P_1,P_2,P_3]$ and compare them to general properties expected of fusion coefficients. This will explain why it deserves the name fusion density.

\paragraph{Reality.}
In general, a fusion coefficient $N_{ij}\text{}^k$ is a positive integer if $k$ appears in the fusion of $i$ and $j$, and zero otherwise. We are proposing a density of fusion coefficients, so we do not expect to get an integer, but at least we expect the fusion density to be \emph{real} and \emph{non-negative}. Let us first check the reality of $\mathcal{N}_{P_0}$.

Recall that the fusion density is defined through the fusion kernel $\mathbb{F}$, see \eqref{eq:fusion_density2}. For real values of its arguments $P_i$, and $c\geq 25$, the fusion kernel is known to take real values \cite{Ponsot:1999uf}. So at least for the heavy states, the fusion density is real. This can also be seen directly from the formula \eqref{eq:main_formula}, since
\begin{equation}
    \big(\N_{P_0}[P_1,P_2,P_3]\big)^* = \int_0^\infty \dd P \frac{\mathbb{S}_{P_1P}^*[P_0]\sker{P_2}{P}{\bbi}\mathbb{S}_{PP_3}[P_0]}{\sker{\bbi}{P}{\bbi}} =  \N_{P_0}[P_1,P_2,P_3],
\end{equation}
where in the last equality we used that for all $P_a,P_b\in\R$ \cite{Eberhardt:2023mrq}:
\begin{equation}
    \mathbb{S}_{P_aP_b}^*[P_0] = \e^{-\pi i \Delta_0}\mathbb{S}_{P_aP_b}[P_0].
\end{equation}
Viewed as a meromorphic function of complex variables $P_i\in \mathbb{C}$, $\N_{P_0}$ is not everywhere real-valued in the complex plane, but we will check later in this section that the fusion density \emph{is} real at the degenerate points $P_{\langle m,n\rangle} = \frac{ibm}{2}+ \frac{in}{2b}$.

\paragraph{Positivity.} To show that the fusion density is non-negative, we need to work a little harder. We can use the closed-form solution of the fusion kernel (see equation \eqref{eq:F_kernel_integral}) to give an explicit integral expression for the fusion density:\footnote{To arrive at this formula, we evaluated \eqref{eq:F_kernel_integral} on $P_4 = P_0, P_s = P_1, P_t = P_3$. Then we performed the change of variables $z = \tfrac{Q}{4} + ix-\tfrac{i}{2}(P_2-P_0)$, and used the fact that the double sine function satisfies $S_b(z) = 1/S_b(Q-z)$. The change of variables is allowed since we assume $0\leq \mathrm{Im}(P_0) < \frac{Q}{2}$, so the contour does not cross any poles.}
\begin{multline}\label{eq:fusion_density_explicit}
    \mathcal{N}_{P_0}[P_1,P_2,P_3] = \fker{P_1}{P_3}{P_3}{P_0}{P_1}{P_2} = \fker{P_1}{P_3}{P_1}{P_2}{P_3}{P_0} =\\[0.8em]  A^{P_0}_{P_1P_3} \int_{-\infty}^\infty \dd x \,S_b\big(\tfrac{Q}{4} + \tfrac{i}{2}P_0  \pm i(x+\tfrac{P_2}{2})\pm iP_1\big)S_b\big(\tfrac{Q}{4} - \tfrac{i}{2}P_0  \pm i(x-\tfrac{P_2}{2})\pm iP_3\big).
\end{multline}
Here $S_b(z)$ is the double sine function \cite{Teschner:2012em,Teschner:2013tqy}, and we used the standard notation that $\pm$ denotes a product over all choices of sign. The normalization factor is given by a ratio of Barnes double gamma functions:
\begin{equation}\label{eq:prefactor}
    A^{P_0}_{P_1P_3} = \frac{\Gamma_b(Q\pm 2i P_1)}{\Gamma_b(\pm 2i P_3)} \frac{\Gamma_b(\frac{Q}{2}+iP_0\pm 2i P_3)}{\Gamma_b(\frac{Q}{2}+iP_0\pm 2i P_1)}. 
\end{equation}
Now, assume $P_0 = ip_0$ is pure imaginary with $p_0\in[0,\frac{Q}{2})$, and also choose $P_{1,2,3}\in \mathbb{R}$. Then the integrand can be written as an absolute value square:
\begin{multline}
    \mathcal{N}_{ip_0}[P_1,P_2,P_3] =\\[0.8em]  A^{ip_0}_{P_1P_3} \int_{-\infty}^\infty \dd x\, \Big|S_b\big(\tfrac{Q}{4} - \tfrac{p_0}{2}  + i(x+\tfrac{P_2}{2})\pm iP_1\big)S_b\big(\tfrac{Q}{4} + \tfrac{p_0}{2}  + i(x-\tfrac{P_2}{2})\pm iP_3\big)\Big|^2.
\end{multline}
Here we used that the double sine function behaves under complex conjugation as \cite{Eberhardt:2023mrq}:
\begin{equation}
    \overline{S_b(z)} = S_b(\bar z).
\end{equation}
 Since the integrand is manifestly non-negative, the integral is non-negative too. 
 
 The prefactor is also real and non-negative when $\O_0$ is light and $\O_{1,3}$ are heavy:
\begin{equation}
    A^{ip_0}_{P_1P_3} \in \mathbb{R}_+ \quad \forall\, p_0 \in [0,\tfrac{Q}{2}), \,P_1,P_3\in \R.
\end{equation}
This follows from the following reality property of the double gamma function:
\begin{equation}
    \overline{\Gamma_b(x +iP)} = \Gamma_b(x-iP) \quad \forall \,x,P\in \R_{+}.
\end{equation}
This property is easily derived from the integral formula of the $\Gamma_b$ function, written for example in equation (2.19) of \cite{Collier_2023}. We conclude that the fusion density is real and non-negative for $\O_0$ light and $\O_{1,2,3}$ heavy:
\begin{equation}
    \N_{P_0}[P_1,P_2,P_3] \geq 0.
\end{equation}

\paragraph{Identity selection rule.} In any CFT, the fusion coefficients should obey the selection rule $N_{i\bbi}\text{}^k = \delta_{ik}$, which follows from the same property of the OPE coefficient $C_{i\bbi}\text{}^k$. The fusion density satisfies the continuous analog of this selection rule when $P_2\rightarrow \bbi$:
\begin{equation}\label{eq:selection_rule}
    \N_{P_0}[P_1,\bbi,P_3] = \fker{P_1}{P_3}{P_3}{P_0}{P_1}{\bbi} = \delta(P_1-P_3). 
\end{equation}
for arbitrary $P_0\in \mathbb{C}$. This follows from an analysis of the fusion kernel (see e.g.\! equation (3.76) in \cite{Eberhardt:2023mrq}) or more directly by using the RHS of the Virasoro-Verlinde formula \eqref{eq:main_formula}:
\begin{equation}\label{eq:identityP2}
    \fker{P_1}{P_3}{P_3}{P_0}{P_1}{\bbi} = \int_0^\infty \dd P\,\mathbb{S}_{P_1P}[P_0]\mathbb{S}^*_{PP_3}[P_0] = \delta(P_1-P_3).
\end{equation}
In the second equality we used the idempotency of the one-point S-kernel.

The other two limits, $P_1\rightarrow \bbi$ or $P_3\rightarrow \bbi$, are not well-defined for arbitrary $P_0$ as can be seen from the RHS of \eqref{eq:main_formula} (i.e. the modular kernel with a non-trivial external operator $P_0$ cannot have the identity propagating in one of the internal channels). To study those, we need to send $P_1 (P_3)\rightarrow \bbi$ in a coordinated way together with $P_0\rightarrow \bbi$. Taking this limit carefully we get $\lim_{P_0,P_1\rightarrow\bbi}\mathbb{S}_{P_1P}[P_0]=\rho_0(P)$, as expected.\footnote{For more details about this limit see e.g.\! appendix A of \cite{Collier_2020}.} Then 
\be
\lim_{P_0,P_1\rightarrow\bbi}\N_{P_0}[P_1,P_2,P_3] = \int_0^{\infty} \dd P \ \mathbb{S}_{P_2P}[\bbi]\mathbb{S}_{PP_3}[\bbi] = \delta(P_2-P_3) .
\ee
A similar analysis holds for the other limit $P_3\rightarrow \bbi$.
\paragraph{Degenerate representations.} Instead of taking $P_2 \to \bbi$, we can also consider $P_2$ to be a different non-identity degenerate representation. The irreducible degenerate representations of the Virasoro algebra occur at the discrete imaginary values 
\begin{equation}
    P_{\langle m,n\rangle} = \frac{ibm}{2}+ \frac{in}{2b}, \quad m,n \in \mathbb{Z}_{>0}.  
\end{equation}
Even though we derived equation \eqref{eq:main_formula} for real momenta only, both sides of the Virasoro-Verlinde formula are meromorphic functions of their complex arguments $P_i$. We can therefore evaluate our  formula on $P_2 = P_{\langle m,n\rangle}$:
\begin{equation}\label{eq:degenerate_Verlinde_formula}
       \fker{P_1}{P_3}{P_3}{P_0}{P_1}{\langle m,n\rangle} =  \int_0^\infty \dd P \,\frac{\mathbb{S}_{P_1P}[P_0]\mathbb{S}_{\langle m,n\rangle P}[\bbi]\mathbb{S}^*_{PP_3}[P_0]}{\mathbb{S}_{\bbi P}[\bbi]}.
\end{equation}
For this equation to hold true, one has to use the degenerate crossing kernels (which appear in the generalized minimal models, as well as Liouville theory with only degenerate vertex operators). The degenerate S-kernel is given in \eqref{eq:degenerateS}, and the ratio 
\begin{equation}\label{eq:degenerate_ratio}
    \frac{\mathbb{S}_{\langle m,n\rangle P}[\bbi]}{ \mathbb{S}_{\bbi P}[\bbi]} = \frac{\sinh(2\pi b^{-1}  mP)\sinh(2\pi b \, n  P)}{\sinh(2\pi b^{-1}  P)\sinh(2\pi b  P)}
\end{equation}
 is now regular at $P=0$. So we can safely set the `regulator' to $P_0\to \bbi$. 

As an example, let us take $\langle m,n\rangle = \langle 2,1\rangle$. In that case, the BPZ equation fixes the degenerate 4-point conformal block in terms of the Gauss hypergeometric function \cite{Belavin:1984vu}. The connection identity satisfied by the hypergeometric $_2 F_1$ allows one to read off the $\langle 2,1\rangle$ degenerate fusion kernel \cite{Eberhardt:2023mrq}:
\begin{equation}
\begin{split}
    \fker{P-\frac{ib}{2},}{P_3}{P}{P_0}{P_1}{\langle 2,1\rangle} &= \frac{\Gamma(1-2b iP)\Gamma(2biP_1)}{\Gamma(\frac{1}{2}+b iP_1\pm b i P_0-b i P)}\delta(P_3-(P_1+\tfrac{ib}{2}))\\[1em] &\quad+ \frac{\Gamma(1-2b iP)\Gamma(-2biP_1)}{\Gamma(\frac{1}{2}-b iP_1\pm b i P_0-b i P)}\delta(P_3-(P_1-\tfrac{ib}{2})).
\end{split}
\end{equation}
So the degenerate fusion kernel `clicks' at two places if we integrate over $P_3$. Satisfyingly, the ratios of gamma functions become equal to 1 in the limit that the regulator is removed (i.e. $P_0 = \frac{iQ}{2}$), if we simultaneously take $P = P_1 +\tfrac{ib}{2}$. So the fusion density is indeed a sum of delta functions, whose coefficients are precisely the integer fusion coefficients of the generalized minimal model: 
\begin{equation}
    \N_\bbi[P_1,P_{\langle 2,1\rangle},P_3] = \delta(P_3-(P_1+\tfrac{ib}{2}))\,+\, \delta(P_3-(P_1-\tfrac{ib}{2})).
\end{equation}
This reflects the well-known fusion rule $[V_{\langle 2,1\rangle}]\cdot [V_\alpha] = [V_{\alpha - b/2}]+[V_{\alpha+b/2}]$ between the $\langle 2,1\rangle $ degenerate representation of the Virasoro algebra and a general non-degenerate one. From the point of view of the RHS of \eqref{eq:degenerate_Verlinde_formula} the above result can also be obtained by a careful analysis of the poles that pinch the $P$ contour whenever $P_3-P_1 = \pm \frac{ib}{2}$ in the limit that $P_0\rightarrow \bbi$.

More generally, we expect the fusion density at degenerate momenta to be a sum over delta functions whose expansion coefficients are the fusion rules of the generalized Virasoro minimal model:
\begin{equation}\label{eq:degenerate_fusion_rules}
    \mathcal{N}_\bbi[P_1,P_{\langle m,n\rangle},P_3] = \sum_{h\in \mathcal{S}} N_{h_1,\langle m,n\rangle}\text{}^{h}\, \delta(h_3-h). 
\end{equation}
To show this, one has to use a recursive formula that determines the degenerate fusion kernel (see equation (3.15) in \cite{Eberhardt:2023mrq}). Starting from the basic case $\langle 2,1\rangle$, the recursion relation determines the degenerate fusion kernel on the left-hand side of \eqref{eq:degenerate_Verlinde_formula} for any $m,n >0$ as a sum over delta functions. Evaluating each kernel on $\Delta_0 = \bbi$ then gives the degenerate fusion coefficients \eqref{eq:degenerate_fusion_rules} in a similar way as above. 

A different way of showing the equality \eqref{eq:degenerate_fusion_rules} is to invert the degenerate Virasoro-Verlinde formula  \eqref{eq:degenerate_Verlinde_formula} by acting on both sides with the one-point S-kernel:
\begin{equation}\label{eq:inverted}
       \int_0^\infty \dd P_1 \,\sker{P}{P_1}{P_0}\,\fker{P_1}{P_3}{P_3}{P_0}{P_1}{\langle m,n\rangle} =  \frac{\mathbb{S}_{\langle m,n\rangle P}[\bbi]\mathbb{S}^*_{PP_3}[P_0]}{\mathbb{S}_{\bbi P}[\bbi]}.
\end{equation}
By computing the boundary Liouville partition function on the cylinder with mixed FZZT and ZZ boundary conditions, it was shown in \cite{Jego:2006ta} that the ratio on the right-hand side of \eqref{eq:inverted}, when we take $\Delta_0= \bbi$, can be rewritten as:
\begin{equation}\label{eq:troost}
    \frac{\mathbb{S}_{\langle m,n\rangle P}[\bbi]\mathbb{S}_{PP_3}[\bbi]}{\mathbb{S}_{\bbi P}[\bbi]} = \sum_{m',n'\in \mathbb{Z}} \int_0^\infty \dd P_1 \, \sker{P,}{P_1+P_{\langle m,n\rangle}}{\bbi}\, N_{n,m,P_3}^{n',m',P_1}
\end{equation}
where $N_{n,m,P_3}^{n',m',P_1}$ are the degenerate fusion rules of Liouville theory (which coincide with the fusion rules of the generalized minimal model): 
\begin{equation}
    N_{n,m,P_3}^{n',m',P_1} = \begin{cases}
			1 & \text{if }\,\scalebox{0.7}{$\begin{cases}
			    1 - m \leq m' \leq m-1, &m+m' \text{ is odd} \\
                1-n \leq n' \leq n-1, &n+n' \text{ is odd} \\ \text{and } P_1=P_3.
			\end{cases}$}\\
            0 & \text{else}.
		 \end{cases}
\end{equation}
Comparing \eqref{eq:inverted} with \eqref{eq:troost} and changing variables $P_1 \to P_1 + P_{\langle m,n\rangle}$ in the integral, we arrive at the desired result \eqref{eq:degenerate_fusion_rules}.

\paragraph{Non-degenerate representations.} Another interesting aspect of the fusion density concerns its behavior in the limit $P_0\to \bbi$ for \emph{non-degenerate} representations. This property will be crucial for making the connection to holographic non-rational CFTs dual to pure AdS$_3$ gravity, which we study in section \ref{sec:3boundary}, since representations with $P\in \mathbb{R}_+$ correspond to black hole states.

If we take the naive limit that the regulator $P_0$ goes to the identity, we would obtain:
\begin{equation}\label{eq:identity_limit}
    \lim_{P_0 \to \bbi}\N_{P_0}[P_1,P_2,P_3] =  \lim_{P_0 \to \bbi} \fker{P_1}{P_3}{P_3}{P_0}{P_1}{P_2}  \to \infty.
\end{equation}
This can be seen in multiple ways. Firstly, from the Virasoro-Verlinde formula \eqref{eq:main_formula}, we see that at $\Delta_0 =\bbi$ the $P$-integrand scales as $P^{-2}$ for small $P$, so we get a divergence of the integral at $P=0$. Another way to see the divergence is from the following selection rule satisfied by the fusion kernel \cite{Eberhardt:2023mrq}:
\begin{equation}
    \fker{P_1}{P_3}{P_3'}{\bbi}{P_1}{P_2} = \delta(P_3-P_3').
\end{equation}
Clearly, when $P_3 = P_3'$ we get a divergence. We want to view the appearance of the infinite factor $\delta(0)$ as the continuous analog of a fusion rule of the form $N_{P_1P_2}^{P_3}=1$, which would simply say that the fusion of three heavy states is allowed for any value of $(P_1,P_2,P_3)$ above the black hole threshold. This is indeed the case for Liouville CFT, and we will comment on the implication for generic non-rational CFT in section \ref{sec:universal}.

We can make this intuition more precise by isolating the divergence in the $P_0\to \bbi$ limit. To do so, we swap the internal indices of the fusion kernel using \cite{Eberhardt:2023mrq}:
\begin{equation}
    \frac{\fker{P_2}{P_0}{P_1}{P_1}{P_3}{P_3}}{\rho_0(P_0)C_0(P_0,P_1,P_1)C_0(P_0,P_3,P_3)} = \frac{\fker{P_0}{P_2}{P_3}{P_1}{P_1}{P_3}}{\rho_0(P_2)C_0(P_2,P_3,P_1)C_0(P_2,P_3,P_1)}.
\end{equation}
Together with the permutation symmetry of $C_0$ and the relation between $C_0$ and the identity fusion kernel \eqref{eq:C0}, this allows us to rewrite the fusion density as:
\begin{align}\label{eq:fusion_dens}
    \N_{P_0}[P_1,P_2,P_3] &= \frac{\fker{\bbi}{P_3}{P_1}{P_2}{P_2}{P_1}\fker{P_2}{P_0}{P_1}{P_1}{P_3}{P_3}}{\fker{\bbi}{P_0}{P_1}{P_1}{P_1}{P_1}} =\rho_0(P_3)C_0(P_0,P_3,P_3)\frac{\fker{P_0}{P_2}{P_3}{P_1}{P_1}{P_3}}{\fker{\bbi}{P_2}{P_3}{P_1}{P_1}{P_3}}.
\end{align}

Now it becomes clear what the $P_0$ divergence is in the limit $P_0 \to \bbi$: it is just a simple pole from the $C_0$ function at $C_0(\frac{iQ}{2},P_3,P_3)$. To compute the residue of the pole, we parametrize $P_0 = \frac{iQ}{2}-i\epsilon$, and expand the ratio of double gamma functions in \eqref{eq:C_0_function} around $\epsilon = 0$. This gives:
\begin{equation}
    C_0(i(\tfrac{Q}{2}-\epsilon),P_3,P_3) = \frac{1}{\pi \rho_0(P_3)} \epsilon^{-1}+ O(\epsilon^0).
\end{equation}
So if we define $z = i\pi P_0$, then the residue of the fusion density at $P_0 = \bbi$ is one:
\begin{equation}
\mathrm{Res}_{z = -\frac{\pi Q}{2}}\, \N_{\frac{z}{i\pi}}[P_1,P_2,P_3] = 1.
\end{equation}
This makes precise in what sense $N_{P_1P_2}^{P_3}=1$: the fusion density knows about the fusion coefficients through residues at its poles. We will use this in section \ref{sec:universal} to discuss universal properties of CFT$_2$ and their relation to the fusion density. 

\paragraph{Quantum group symmetry and the Virasoro $6j$ symbol.}
A very interesting rewriting of our formula \eqref{eq:main_formula} makes the connection to the Virasoro $6j$ symbol manifest. First, let us define the normalized S-kernel:
\begin{equation}\label{eq:S_normalization}
    \hat{\mathbb{S}}_{P'P}[P_0] := \sqrt{\frac{C_0(P',P',P_0)}{C_0(P,P,P_0)}}\frac{\mathbb{S}_{P'P}[P_0]}{\rho_0(P)}.
\end{equation}
This normalization of the one-point S-kernel, which also appears in quantum Teichm\"uller theory \cite{Teschner:2012em, Teschner:2013tqy}, is sometimes referred to as the `Racah-Wigner normalization'. It is symmetric under exchanging $P$ and $P'$:
\begin{equation}
    \hat{\mathbb{S}}_{P'P}[P_0] = \hat{\mathbb{S}}_{PP'}[P_0],
\end{equation}
thanks to the formula for swapping the internal indices of the one-point S-kernel (see e.g.\! equation (2.66) in \cite{Eberhardt:2023mrq}). In this symmetric normalization, the Virasoro-Verlinde formula can be written in the following elegant form:
\begin{equation}\label{eq:VV_6j}
    \int_0^\infty \dd\mu(P) \,\frac{\hat{\mathbb{S}}_{P_1P}[P_0]\mathbb{S}_{P_2P}[\bbi]\hat{\mathbb{S}}_{PP_3}^*[P_0]}{\mathbb{S}_{\bbi P}[\bbi]} = \begin{Bmatrix}
  P_3 & P_3 & P_0 \\
  P_1 & P_1 & P_2 
 \end{Bmatrix}_{6j}\,.
\end{equation}

Here $\dd\mu(P) \coloneqq \dd P\rho_0(P)$, and the right-hand side is the Virasoro $6j$ symbol, with pairwise identical external operators $P_1$ and $P_3$. It is related to the fusion kernel through the Racah-Wigner normalization (see eq.\! \eqref{eq:6j_fusion}) and hence it governs the crossing transformation between the $s$- and $t$-channel four-point blocks:\footnote{In the Racah-Wigner normalization, conformal blocks are normalized by factors of $\sqrt{C_0(P_i,P_j,P_k)}$ at each trivalent vertex $(ijk)$.} 
\begin{equation}
\begin{tikzpicture}[x=0.75pt,y=0.75pt,yscale=-0.7,xscale=0.7,baseline={([yshift=-.5ex]current bounding box.center)}]
\draw  [draw opacity=0][line width=1.5]  (266.32,194.22) .. controls (255.36,193.97) and (246.5,170.41) .. (246.5,141.38) .. controls (246.5,112.37) and (255.35,88.82) .. (266.32,88.55) -- (266.5,141.38) -- cycle ; \draw  [color={rgb, 255:red, 208; green, 2; blue, 27 }  ,draw opacity=1 ][line width=1.5]  (266.32,194.22) .. controls (255.36,193.97) and (246.5,170.41) .. (246.5,141.38) .. controls (246.5,112.37) and (255.35,88.82) .. (266.32,88.55) ;  
\draw  [draw opacity=0][dash pattern={on 1.5pt off 1.5pt on 1.5pt off 1.5pt}][line width=1.5]  (267.55,88.62) .. controls (278.11,90.07) and (286.5,113.14) .. (286.5,141.38) .. controls (286.5,170.56) and (277.55,194.22) .. (266.5,194.22) .. controls (266.39,194.22) and (266.28,194.22) .. (266.18,194.21) -- (266.5,141.38) -- cycle ; \draw  [color={rgb, 255:red, 208; green, 2; blue, 27 }  ,draw opacity=1 ][dash pattern={on 1.5pt off 1.5pt on 1.5pt off 1.5pt}][line width=1.5]  (267.55,88.62) .. controls (278.11,90.07) and (286.5,113.14) .. (286.5,141.38) .. controls (286.5,170.56) and (277.55,194.22) .. (266.5,194.22) .. controls (266.39,194.22) and (266.28,194.22) .. (266.18,194.21) ;  
\draw   (200.57,106.62) .. controls (195.1,101.15) and (197.05,90.33) .. (204.93,82.46) .. controls (212.8,74.58) and (223.62,72.63) .. (229.09,78.1) .. controls (234.55,83.56) and (232.6,94.38) .. (224.73,102.26) .. controls (216.85,110.13) and (206.03,112.08) .. (200.57,106.62) -- cycle ;
\draw   (302.05,78.3) .. controls (307.51,72.83) and (318.33,74.78) .. (326.21,82.66) .. controls (334.08,90.53) and (336.03,101.35) .. (330.57,106.82) .. controls (325.1,112.29) and (314.28,110.33) .. (306.41,102.46) .. controls (298.53,94.58) and (296.58,83.77) .. (302.05,78.3) -- cycle ;
\draw   (229.37,204.77) .. controls (223.9,210.23) and (213.09,208.28) .. (205.21,200.41) .. controls (197.34,192.53) and (195.38,181.71) .. (200.85,176.25) .. controls (206.32,170.78) and (217.14,172.73) .. (225.01,180.61) .. controls (232.89,188.48) and (234.84,199.3) .. (229.37,204.77) -- cycle ;
\draw   (330.85,176.05) .. controls (336.32,181.51) and (334.37,192.33) .. (326.49,200.21) .. controls (318.62,208.08) and (307.8,210.03) .. (302.33,204.57) .. controls (296.86,199.1) and (298.82,188.28) .. (306.69,180.41) .. controls (314.57,172.53) and (325.38,170.58) .. (330.85,176.05) -- cycle ;
\draw    (229.09,78.1) .. controls (247.67,90.97) and (283.67,91.97) .. (302.05,78.3) ;
\draw    (332.31,104.36) .. controls (319.44,122.94) and (318.28,159.12) .. (331.95,177.5) ;
\draw    (302.33,204.57) .. controls (283.75,191.7) and (247.75,191.1) .. (229.37,204.77) ;
\draw    (199.25,178.18) .. controls (212.12,159.6) and (213.12,123.6) .. (199.45,105.22) ;
\draw [line width=1.5]    (317,90.54) -- (298.31,141.72) ;
\draw [line width=1.5]    (230.79,141.05) -- (298.31,141.72) ;
\draw [line width=1.5]    (298.31,141.72) -- (316.79,192.02) ;
\draw  [line width=0.75]  (298.31,136.35) .. controls (301.4,136.35) and (303.91,138.75) .. (303.91,141.72) .. controls (303.91,144.68) and (301.4,147.08) .. (298.31,147.08) .. controls (295.21,147.08) and (292.7,144.68) .. (292.7,141.72) .. controls (292.7,138.75) and (295.21,136.35) .. (298.31,136.35) -- cycle ;
\draw [line width=1.5]    (214.47,90.83) -- (230.79,141.05) ;
\draw [line width=1.5]    (230.79,141.05) -- (214.68,192.31) ;
\draw  [line width=0.75]  (230.79,135.68) .. controls (233.88,135.68) and (236.39,138.09) .. (236.39,141.05) .. controls (236.39,144.01) and (233.88,146.41) .. (230.79,146.41) .. controls (227.69,146.41) and (225.18,144.01) .. (225.18,141.05) .. controls (225.18,138.09) and (227.69,135.68) .. (230.79,135.68) -- cycle ;
\draw (254,145.18) node [anchor=north west][inner sep=0.75pt]    {$P_{0}$};
\draw (169.33,80.45) node [anchor=north west][inner sep=0.75pt]    {$P_{1}$};
\draw (338.33,183.78) node [anchor=north west][inner sep=0.75pt]    {$P_{3}$};
\draw (338.33,82.12) node [anchor=north west][inner sep=0.75pt]    {$P_{3}$};
\draw (169.67,181.12) node [anchor=north west][inner sep=0.75pt]    {$P_{1}$};
\end{tikzpicture}  = \int_0^\infty \dd \mu(P_2) \begin{Bmatrix}
  P_3 & P_3 & P_0 \\
  P_1 & P_1 & P_2 
 \end{Bmatrix}_{6j}\begin{tikzpicture}[x=0.75pt,y=0.75pt,yscale=-0.7,xscale=0.7,baseline={([yshift=-.5ex]current bounding box.center)}]
\draw  [draw opacity=0][dash pattern={on 1.5pt off 1.5pt on 1.5pt off 1.5pt}][line width=1.5]  (169.42,203.56) .. controls (171.02,193.03) and (195.37,184.67) .. (225.17,184.67) .. controls (256,184.67) and (281,193.62) .. (281,204.67) .. controls (281,204.78) and (281,204.89) .. (280.99,205.01) -- (225.17,204.67) -- cycle ; \draw  [color={rgb, 255:red, 208; green, 2; blue, 27 }  ,draw opacity=1 ][dash pattern={on 1.5pt off 1.5pt on 1.5pt off 1.5pt}][line width=1.5]  (169.42,203.56) .. controls (171.02,193.03) and (195.37,184.67) .. (225.17,184.67) .. controls (256,184.67) and (281,193.62) .. (281,204.67) .. controls (281,204.78) and (281,204.89) .. (280.99,205.01) ;  
\draw   (159.9,169.9) .. controls (154.43,164.43) and (156.38,153.62) .. (164.26,145.74) .. controls (172.14,137.86) and (182.95,135.91) .. (188.42,141.38) .. controls (193.89,146.85) and (191.93,157.66) .. (184.06,165.54) .. controls (176.18,173.41) and (165.37,175.37) .. (159.9,169.9) -- cycle ;
\draw   (261.38,141.58) .. controls (266.85,136.11) and (277.66,138.07) .. (285.54,145.94) .. controls (293.41,153.82) and (295.37,164.63) .. (289.9,170.1) .. controls (284.43,175.57) and (273.62,173.62) .. (265.74,165.74) .. controls (257.86,157.86) and (255.91,147.05) .. (261.38,141.58) -- cycle ;
\draw   (188.71,268.05) .. controls (183.24,273.52) and (172.42,271.56) .. (164.55,263.69) .. controls (156.67,255.81) and (154.72,245) .. (160.19,239.53) .. controls (165.65,234.06) and (176.47,236.01) .. (184.34,243.89) .. controls (192.22,251.77) and (194.17,262.58) .. (188.71,268.05) -- cycle ;
\draw   (290.19,239.33) .. controls (295.65,244.8) and (293.7,255.61) .. (285.82,263.49) .. controls (277.95,271.36) and (267.13,273.32) .. (261.67,267.85) .. controls (256.2,262.38) and (258.15,251.56) .. (266.03,243.69) .. controls (273.9,235.81) and (284.72,233.86) .. (290.19,239.33) -- cycle ;
\draw    (188.42,141.38) .. controls (207,154.25) and (243,155.25) .. (261.38,141.58) ;
\draw    (291.64,167.64) .. controls (278.77,186.22) and (277.62,222.41) .. (291.29,240.79) ;
\draw    (261.67,267.85) .. controls (243.08,254.98) and (207.08,254.38) .. (188.71,268.05) ;
\draw    (158.58,241.46) .. controls (171.46,222.88) and (172.46,186.88) .. (158.79,168.5) ;
\draw [line width=1.5]    (174.16,155.64) -- (225.33,173.53) ;
\draw [line width=1.5]    (224.67,238.17) -- (225.33,173.53) ;
\draw [line width=1.5]    (225.33,173.53) -- (275.64,155.84) ;
\draw  [line width=0.75]  (219.97,173.53) .. controls (219.97,170.57) and (222.37,168.17) .. (225.33,168.17) .. controls (228.3,168.17) and (230.7,170.57) .. (230.7,173.53) .. controls (230.7,176.5) and (228.3,178.9) .. (225.33,178.9) .. controls (222.37,178.9) and (219.97,176.5) .. (219.97,173.53) -- cycle ;
\draw [line width=1.5]    (174.45,253.79) -- (224.67,238.17) ;
\draw [line width=1.5]    (224.67,238.17) -- (275.93,253.59) ;
\draw  [draw opacity=0][line width=1.5]  (281,204.85) .. controls (280.72,215.81) and (255.83,224.67) .. (225.17,224.67) .. controls (194.51,224.67) and (169.63,215.82) .. (169.34,204.86) -- (225.17,204.67) -- cycle ; \draw  [color={rgb, 255:red, 208; green, 2; blue, 27 }  ,draw opacity=1 ][line width=1.5]  (281,204.85) .. controls (280.72,215.81) and (255.83,224.67) .. (225.17,224.67) .. controls (194.51,224.67) and (169.63,215.82) .. (169.34,204.86) ;  
\draw  [line width=0.75]  (219.3,238.17) .. controls (219.3,235.2) and (221.7,232.8) .. (224.67,232.8) .. controls (227.63,232.8) and (230.03,235.2) .. (230.03,238.17) .. controls (230.03,241.13) and (227.63,243.53) .. (224.67,243.53) .. controls (221.7,243.53) and (219.3,241.13) .. (219.3,238.17) -- cycle ;
\draw (228.17,195.23) node [anchor=north west][inner sep=0.75pt]    {$P_{2}$};
\draw (127.87,142.9) node [anchor=north west][inner sep=0.75pt]    {$P_{1}$};
\draw (297.17,248.57) node [anchor=north west][inner sep=0.75pt]    {$P_{3}$};
\draw (296.7,143.77) node [anchor=north west][inner sep=0.75pt]    {$P_{3}$};
\draw (131.73,248.23) node [anchor=north west][inner sep=0.75pt]    {$P_{1}$};
\end{tikzpicture}\,.
\end{equation}
The Virasoro $6j$ symbol coincides with the $6j$ symbol of the modular double of the non-compact quantum group $\mathcal{U}_q(\mathfrak{sl}(2,\R))$ \cite{Ponsot:2000mt}. In fact, the integration measure $\dd \mu(P) =  4\sqrt{2}\sinh(2\pi b P)\sinh(2\pi b^{-1}P)\dd P$ coincides with the Plancherel measure on the space of continuous representations of this quantum group. Moreover, the ratio of identity S-kernels that appears in \eqref{eq:VV_6j} can be viewed as the continuous analogue of the generalized `quantum dimension' of an anyonic excitation in the TQFT \cite{McGough:2013gka, Mertens:2022ujr}:
\begin{equation}
    \mathsf{d}^{\,P'}_P \coloneqq   \frac{\mathbb{S}_{P'P}[\bbi]}{\mathbb{S}_{\bbi P}[\bbi]} =\frac{\cos(4\pi PP')}{2\sinh(2\pi bP)\sinh(2\pi b^{-1}P)}.
\end{equation}
Although the quantum dimension $\mathsf{d}_{\bbi}^{P'}$ is ill-defined, its generalization to non-degenerate $P$ exists and is given by the above ratio of S-kernels. This ratio of S-kernels also appears in the FZZT boundary state of Liouville theory \cite{Fateev:2000ik,Teschner:2000md}. Moreover, the ratio is the eigenvalue of the Verlinde loop operator in quantum Teichm\"uller theory and Liouville CFT \cite{Verlinde:1989ua, Alday:2009fs, Drukker:2009id, Drukker:2010jp}. We return to both these facts in the discussion section.

 Given this discussion, it is natural to adopt the Racah-Wigner normalization and define the normalized fusion density:
\begin{equation}\label{eq:6jnormN}
    \widehat{\N}_{P_0}[P_1,P_2,P_3] \coloneqq \begin{Bmatrix}
  P_3 & P_3 & P_0 \\
  P_1 & P_1 & P_2 
 \end{Bmatrix}_{6j}. 
\end{equation}
Tracking the normalization factors, $\widehat{\N}$ is related to the old fusion density as: 
\begin{equation}\label{eq:relNNhat}
    \widehat{\N}_{P_0}[P_1,P_2,P_3]= \frac{1}{\rho_0(P_3)} \sqrt{\frac{C_0(P_0,P_1,P_1)}{C_0(P_0,P_3,P_3)}}\,\N_{P_0}[P_1,P_2,P_3].
\end{equation}
This normalization makes more symmetries of the fusion density manifest. Namely, the $6j$ symbol possesses a full \emph{tetrahedral symmetry}, generated by (1) swapping pairs of columns and (2) swapping the rows of pairs of entries in the same row:
\begin{equation}
    \begin{Bmatrix}
  P_3 & P_3 & P_0 \\
  P_1 & P_1 & P_2 
 \end{Bmatrix}_{6j} \stackrel{(1)}{=} 
 \begin{Bmatrix}
  P_3 & P_0 & P_3 \\
  P_1 & P_2 & P_1 
 \end{Bmatrix}_{6j} \stackrel{(2)}{=}\begin{Bmatrix}
  P_1 & P_1 & P_0 \\
  P_3 & P_3 & P_2 
 \end{Bmatrix}_{6j} \stackrel{(1)+(2)}{=}  \begin{Bmatrix}
  P_3 & P_1 & P_2 \\
  P_1 & P_3 & P_0 
 \end{Bmatrix}_{6j}\,.
\end{equation}

In fact, while $\widehat{\N}_{P_0}[P_1,P_2,P_3]$ has the symmetry group of a tetrahedron, the fusion density in the limit $P_0\to \bbi$ that the regulator is removed has a full permutation symmetry in $(P_1,P_2,P_3)$. Diagrammatically, we can see this by 
associating a tetrahedron to the fusion density, and noticing that the tetrahedron becomes a `sunset' diagram when we remove the $P_0$ edge:
\begin{equation}\label{eq:tetrahedron_diagram}
\begin{tikzpicture}[baseline={([yshift=-.5ex]current bounding box.center)}]
    \coordinate (A) at (0,0);
    \coordinate (B) at (3,0);
    \coordinate (C) at (1.5,{3*sqrt(3)/2});
    \coordinate (D) at (1.5,{sqrt(3)/2});
    \draw (A) -- (B) node[midway, below] {\small $2$};
    \draw (A) -- (C) node[midway, left] {\small $3$};
    \draw (A) -- (D) node[midway, above] {\small $ 1$};
    \draw (B) -- (C) node[midway, right] {\small $3$};
    \draw (B) -- (D) node[midway, above] {\small $1$};
    \draw (C) -- (D) node[midway, left] {\small $0$};   
    \fill (A) circle (2pt);
    \fill (B) circle (2pt);
    \fill (C) circle (2pt);
    \fill (D) circle (2pt);   
\end{tikzpicture} \quad \xrightarrow[]{P_0\to\bbi} \quad
\begin{tikzpicture}[baseline={([yshift=-.5ex]current bounding box.center)}]
    \draw[thick] (0,0) ellipse (1cm and 1cm);
    \coordinate (P1) at (-1, 0); 
    \coordinate (P2) at (1, 0);
    \draw[thick] (-1, 0) -- (1, 0);
    \fill[black] (P1) circle (2pt); 
    \fill[black] (P2) circle (2pt); 
    \node at (0, -1.3) {\small 2};  
    \node at (0, 0.3) {\small 1};    
    \node at (0, 1.3) {\small 3};   
\end{tikzpicture}\,.
\end{equation}
Besides making the tetrahedral symmetry manifest, the $6j$ normalization also makes the selection rules more transparent. For example, for non-degenerate representations $P_1,P_3$ we have 
\begin{equation}
    \begin{Bmatrix}
  P_3 & P_3 & P_0 \\
  P_1 & P_1 & \bbi 
 \end{Bmatrix}_{6j} = \frac{\delta(P_1-P_3)}{\rho_0(P_3)},
\end{equation}
which is equivalent to our earlier observation \eqref{eq:selection_rule}. Similar selection rules for degenerate representations involve the Wigner $3j$ symbol for the quantum group. Moreover, the Racah-Wigner normalization preserves the reality and positivity properties derived earlier, since both $\rho_0$ and $C_0$ are positive-definite functions. 

By analogy to the standard Verlinde formula, we can read our result \eqref{eq:VV_6j} as saying that \emph{the one-point S-kernel $\widehat{\mathbb{S}}_{PP'}[P_0]$ diagonalizes the Virasoro $6j$ symbol.} We will expand on this observation in the discussion section.

\paragraph{Commutativity of the fusion algebra.} 
The Racah-Wigner normalization makes it clear that $\widehat{\N}$ is invariant under exchanging $P_1\leftrightarrow P_3$, as it corresponds to one of the tetrahedral symmetries of the $6j$ symbol:
\begin{equation}
    \widehat{\N}_{P_0}[P_1,P_2,P_3] = \widehat{\N}_{P_0}[P_3,P_2,P_1].
\end{equation}
This is a desirable feature of the fusion density, since it reflects the fact that the fusion algebra is commutative, 
\begin{equation}
    N_{ij}\text{}^k = N_{ji}\text{}^k.
\end{equation}
In our previous normalization, the fusion density $\N_{P_0}$ is covariant under $P_1\leftrightarrow P_3$ with a specific proportionality constant. But as remarked above, the full permutation symmetry is restored when the regulator is removed, for both choices of normalization.
\paragraph{Associativity of the fusion algebra.}
Next to commutativity, the fusion algebra in any CFT$_2$ should also be associative. Associativity of the OPE of primary fields implies that the fusion coefficients should satisfy
\begin{equation}
    \sum_h N_{kj}\text{}^{h} N_{ih}\text{}^m = \sum_h N_{ij}\text{}^hN_{hk}\text{}^m.
\end{equation}
The analogous identity satisfied by the fusion density $\N_{P_0}$ can be shown to be:
\begin{multline}\label{eq:associativity2}
    \int_0^\infty \!\!\dd P_2\,\N_{P_0}[P_1,P_2,P_3]\,\N_{P_0}[P_4,P_2,P_5] = \int_0^\infty \!\!\dd P_2\,\N_{P_0}[P_4,P_2,P_3]\, \N_{P_0}[P_1,P_2,P_5].
\end{multline}
This is easily seen by substituting the Virasoro-Verlinde formula \eqref{eq:main_formula2} into \eqref{eq:associativity2} and using the fact that $\int_0^\infty \dd P_2\,\sker{P}{P_2}{\bbi}\sker{P_2}{P'}{\bbi} = \delta(P-P')$. Note that the associativity-like equation \eqref{eq:associativity2} holds for any value of the regulator $P_0$. 

Moreover, the Virasoro-Verlinde formula in the Racah-Wigner normalization \eqref{eq:VV_6j} allows us to derive the following nontrivial integral identities for the $6j$ symbol (we dropped the subscript $_{6j}$ for clarity):
\begin{align}
    \int_0^\infty \dd P_2 \begin{Bmatrix}
  P_3 & P_3 & P_0 \\
  P_1 & P_1 & P_2 
 \end{Bmatrix} \begin{Bmatrix}
  P_5 & P_5 & P_0 \\
  P_4 & P_4 & P_2 
 \end{Bmatrix} &= \int_0^\infty \dd P_2 \begin{Bmatrix}
  P_3 & P_3 & P_0 \\
  P_4 & P_4 & P_2 
 \end{Bmatrix} \begin{Bmatrix}
  P_5 & P_5 & P_0 \\
  P_1 & P_1 & P_2 
 \end{Bmatrix}\,, \\[1em]
 \int_0^\infty \dd \mu(P_3) \begin{Bmatrix}
  P_3 & P_3 & P_0 \\
  P_1 & P_1 & P_2 
 \end{Bmatrix} \begin{Bmatrix}
  P_3 & P_3 & P_0 \\
  P_4 & P_4 & P_5 
 \end{Bmatrix} &=\int_0^\infty \dd \mu(P_3) \begin{Bmatrix}
  P_3 & P_3 & P_0 \\
  P_1 & P_1 & P_5 
 \end{Bmatrix} \begin{Bmatrix}
  P_3 & P_3 & P_0 \\
  P_4 & P_4 & P_2 
 \end{Bmatrix}.
\end{align}
These equalities follow from the idempotency of the modular and one-point S-kernels. Many more new identities can be derived by translating the known $\mathrm{PSL}(2,\Z)$ identities satisfied by the S-kernel to integral identities satisfied by the $6j$ symbol.

\paragraph{Boundedness of the fusion density.} We can use the relation between the $6j$ symbol and the Virasoro fusion kernel to express the fusion density in the Racah-Wigner normalization as:
\begin{equation}\label{eq:boundedness}
    \widehat{\N}_{P_0}[P_1,P_2,P_3] = \sqrt{C_0(P_0,P_1,P_1)C_0(P_0,P_3,P_3)}\,\frac{\fker{P_0}{P_2}{P_3}{P_1}{P_1}{P_3}}{\fker{\bbi}{P_2}{P_3}{P_1}{P_1}{P_3}} \,.
\end{equation}
This allows us to analyze the asymptotic behavior of the fusion density when we take the operators to be very heavy,  $\Delta_{1,2,3}\to \infty$. The asymptotics of the $C_0$ function for $\O_0$ light and $\O_{1,2,3}$ heavy were given in equation (4.12) of \cite{Collier_2020}. From these asymptotics, one learns that the square root prefactor in \eqref{eq:boundedness} is exponentially suppressed at large $P_{1,3}$. Moreover, the ratio of fusion kernels has the asymptotic behavior \cite{Collier_2020}:
\begin{equation}
    \frac{\fker{P_0}{P_2}{P_3}{P_1}{P_1}{P_3}}{\fker{\bbi}{P_2}{P_3}{P_1}{P_1}{P_3}} \sim (\text{order-one})P^{-2\Delta_0}
\end{equation}
in the heavy-heavy-heavy limit $P_{1,2,3} \coloneqq P\to \infty$, with $0\leq \Delta_0 \leq \frac{c-1}{24}$ fixed. This shows that the fusion density remains \emph{bounded} as we take the conformal weights of $\O_{1,2,3}$ to infinity. So the fusion density is a normalizable density on $\mathbb{R}^3$. 

\paragraph{Analytic check at $c=25$.}
Lastly, let us discuss the Virasoro-Verlinde formula \eqref{eq:main_formula} at the special value of the central charge $c=25$, corresponding to $b=1$. A new closed form expression for the fusion kernel at $b=1$ was recently derived in \cite{Ribault:2023vqs}. This allows us to perform an analytic check of the formula \eqref{eq:main_formula} in the special case of $\Delta_0 = 1$ (which, at $c=25$, corresponds to $P_0 = 0$). In this special case, the one-point S-kernel significantly simplifies. We present the details of this analytic check in appendix \ref{sec:CFTproof}.

\section{Applications}\label{sec:III}  

\subsection{Open-closed duality in boundary Liouville CFT}\label{sec:open_closed}

Soon after Verlinde published his formula, it was realized by Cardy that the Verlinde formula is equivalent to the statement of open-closed duality for the annulus partition function in rational BCFT \cite{Cardy:1989ir}. In this subsection, we will show that the Virasoro-Verlinde formula has a similar interpretation as an open-closed duality for the boundary one-point function on the annulus, in $c>1$ Liouville BCFT. This duality can be summarized by the Weyl equivalence of the boundary-punctured annulus and cylinder:
\begin{equation}
   \begin{tikzpicture}[x=0.75pt,y=0.75pt,yscale=-1,xscale=1,baseline={([yshift=-.5ex]current bounding box.center)}]
\draw   (80,135) .. controls (80,110.15) and (100.15,90) .. (125,90) .. controls (149.85,90) and (170,110.15) .. (170,135) .. controls (170,159.85) and (149.85,180) .. (125,180) .. controls (100.15,180) and (80,159.85) .. (80,135) -- cycle ;
\draw   (105,135) .. controls (105,123.95) and (113.95,115) .. (125,115) .. controls (136.05,115) and (145,123.95) .. (145,135) .. controls (145,146.05) and (136.05,155) .. (125,155) .. controls (113.95,155) and (105,146.05) .. (105,135) -- cycle ;
\draw   (76.3,138.92) -- (83.79,131.52)(76.32,131.45) -- (83.77,138.99) ;
\draw   (240,135) .. controls (240,115.67) and (244.48,100) .. (250,100) .. controls (255.52,100) and (260,115.67) .. (260,135) .. controls (260,154.33) and (255.52,170) .. (250,170) .. controls (244.48,170) and (240,154.33) .. (240,135) -- cycle ;
\draw    (250,100) -- (340,100) ;
\draw  [draw opacity=0] (339.43,100.03) .. controls (339.56,100.01) and (339.69,100.01) .. (339.82,100.01) .. controls (345.35,100.01) and (349.82,115.68) .. (349.82,135.01) .. controls (349.82,154.13) and (345.44,169.67) .. (340,170) -- (339.82,135.01) -- cycle ; \draw   (339.43,100.03) .. controls (339.56,100.01) and (339.69,100.01) .. (339.82,100.01) .. controls (345.35,100.01) and (349.82,115.68) .. (349.82,135.01) .. controls (349.82,154.13) and (345.44,169.67) .. (340,170) ;  
\draw    (250,170) -- (340,170) ;
\draw  [draw opacity=0][dash pattern={on 1.5pt off 1.5pt on 1.5pt off 1.5pt}] (338.13,169.51) .. controls (333.41,166.69) and (329.82,152.32) .. (329.82,135.01) .. controls (329.82,115.68) and (334.3,100.01) .. (339.82,100.01) .. controls (340.33,100.01) and (340.82,100.14) .. (341.3,100.39) -- (339.82,135.01) -- cycle ; \draw  [dash pattern={on 1.5pt off 1.5pt on 1.5pt off 1.5pt}] (338.13,169.51) .. controls (333.41,166.69) and (329.82,152.32) .. (329.82,135.01) .. controls (329.82,115.68) and (334.3,100.01) .. (339.82,100.01) .. controls (340.33,100.01) and (340.82,100.14) .. (341.3,100.39) ;  
\draw   (256.24,139.17) -- (263.73,131.77)(256.25,131.7) -- (263.71,139.24) ;
\draw (193.77,130.26) node [anchor=north west][inner sep=0.75pt]    {$=$};
\end{tikzpicture}
\end{equation}
The LHS is the `open channel', and the RHS the `closed channel' of the boundary one-point function $\expval{\O_0}^{(\mathsf{ab})}_\tau$. Here $\mathsf{a},\mathsf{b}$ label conformal boundary conditions on the two circular boundaries, and $\tau = \frac{i\beta}{2\pi}$ with $\beta$ the length of the cylinder. The one-point function does not depend on the position of the boundary insertion $\O_0$. 

A concise review of BCFT$_2$ can be found in section 2 of \cite{Numasawa:2022cni}. In a general BCFT, the statement of open-closed duality of the cylinder one-point function is: 
\begin{equation}\label{eq:open-closed}
    \expval{\O_0}^{(\mathsf{ab})}_{-\frac{1}{\tau},\text{closed}} = (-i\tau)^{\Delta_0}\expval{\O_0}^{(\mathsf{ab})}_{\tau,\text{open}}\,.
\end{equation}
The right-hand side is expanded in conformal blocks of the BCFT in the open channel:
\begin{equation}\label{eq:BCFT_open}
  \expval{\O_0}^{(\mathsf{ab})}_{\tau,\text{open}} = \sum_{i\in \mathcal{H}_{\text{open}}}  \frac{1}{\mathfrak{g}_{ii}^{(\mathsf{ab})}} \,C_{i0i}^{(\mathsf{abb})} \mathcal{F}_{\text{open}}\!\left(\begin{tikzpicture}[x=0.75pt,y=0.75pt,yscale=-0.8,xscale=0.8,baseline={([yshift=-.5ex]current bounding box.center)}]
\draw [color={rgb, 255:red, 208; green, 2; blue, 27 }  ,draw opacity=1 ]   (184.38,180.14) -- (213.14,180) ;
\draw  [draw opacity=0] (139.66,145.29) .. controls (147.41,138.95) and (157.33,135.14) .. (168.13,135.14) .. controls (192.98,135.14) and (213.13,155.29) .. (213.13,180.14) .. controls (213.13,205) and (192.98,225.14) .. (168.13,225.14) .. controls (157.59,225.14) and (147.91,221.52) .. (140.24,215.46) -- (168.13,180.14) -- cycle ; \draw   (139.66,145.29) .. controls (147.41,138.95) and (157.33,135.14) .. (168.13,135.14) .. controls (192.98,135.14) and (213.13,155.29) .. (213.13,180.14) .. controls (213.13,205) and (192.98,225.14) .. (168.13,225.14) .. controls (157.59,225.14) and (147.91,221.52) .. (140.24,215.46) ;  
\draw   (151.88,180.14) .. controls (151.88,171.17) and (159.15,163.89) .. (168.13,163.89) .. controls (177.1,163.89) and (184.38,171.17) .. (184.38,180.14) .. controls (184.38,189.12) and (177.1,196.39) .. (168.13,196.39) .. controls (159.15,196.39) and (151.88,189.12) .. (151.88,180.14) -- cycle ;
\draw  [dash pattern={on 1.5pt off 1.5pt on 1.5pt off 1.5pt}]  (113.58,164.93) -- (113.67,195.93) ;
\draw    (113.78,165.4) .. controls (125.43,165.24) and (132.23,151.24) .. (139.66,145.29) ;
\draw    (113.63,195.36) .. controls (125.6,195.53) and (132.6,209.52) .. (140.24,215.46) ;
\draw (85.25,169.79) node [anchor=north west][inner sep=0.75pt]    {$\O_{0}$};
\draw (137,174.94) node [anchor=north west][inner sep=0.75pt]  [font=\small]  {$\mathsf{a}$};
\draw (160,135.54) node [anchor=north west][inner sep=0.75pt]  [font=\small]  {$\mathsf{b}$};
\draw (194,163.29) node [anchor=north west][inner sep=0.75pt]  [font=\small]  {$i$};
\end{tikzpicture}\right)\,.
\end{equation}
Here we represented the insertion of $\O_0$ by a dotted edge, and labeled the boundary states by $\mathsf{a},\mathsf{b}$ and the intermediate bulk state by the index $i$. The structure constants that appear are the boundary 2-point normalization coefficients $\mathfrak{g}_{ij}^{(\mathsf{ab})}$ (which serve to raise and lower bulk indices) and the boundary OPE coefficient $C^{(\mathsf{abc})}_{ijk}$ (which describes the boundary 3-point function on the disk). 

The left-hand side is expanded in closed channel conformal blocks:
\begin{equation}
    \expval{\O_0}^{(\mathsf{ab})}_{-\frac{1}{\tau},\text{closed}} = \sum_{j\in \mathcal{H}_{\text{closed}}} \,C_{\bbi j}^{(\mathsf{a})} C_{j 0}^{(\mathsf{b})}\, \mathcal{F}_{\text{closed}}\!\left(\begin{tikzpicture}[x=0.75pt,y=0.75pt,yscale=-0.8,xscale=0.8,baseline={([yshift=-.5ex]current bounding box.center)}]
\draw  [draw opacity=0] (116.53,120.15) .. controls (124.29,113.8) and (134.2,110) .. (145,110) .. controls (169.85,110) and (190,130.15) .. (190,155) .. controls (190,179.85) and (169.85,200) .. (145,200) .. controls (134.47,200) and (124.78,196.38) .. (117.11,190.32) -- (145,155) -- cycle ; \draw   (116.53,120.15) .. controls (124.29,113.8) and (134.2,110) .. (145,110) .. controls (169.85,110) and (190,130.15) .. (190,155) .. controls (190,179.85) and (169.85,200) .. (145,200) .. controls (134.47,200) and (124.78,196.38) .. (117.11,190.32) ;  
\draw   (128.75,155) .. controls (128.75,146.03) and (136.03,138.75) .. (145,138.75) .. controls (153.97,138.75) and (161.25,146.03) .. (161.25,155) .. controls (161.25,163.97) and (153.97,171.25) .. (145,171.25) .. controls (136.03,171.25) and (128.75,163.97) .. (128.75,155) -- cycle ;
\draw  [dash pattern={on 1.5pt off 1.5pt on 1.5pt off 1.5pt}]  (90.46,139.79) -- (90.54,170.79) ;
\draw    (90.65,140.26) .. controls (102.3,140.1) and (109.1,126.1) .. (116.53,120.15) ;
\draw    (90.5,170.22) .. controls (102.48,170.38) and (109.47,184.37) .. (117.11,190.32) ;
\draw  [color={rgb, 255:red, 208; green, 2; blue, 27 }  ,draw opacity=1 ] (114.36,155) .. controls (114.36,138.08) and (128.08,124.36) .. (145,124.36) .. controls (161.92,124.36) and (175.64,138.08) .. (175.64,155) .. controls (175.64,171.92) and (161.92,185.64) .. (145,185.64) .. controls (128.08,185.64) and (114.36,171.92) .. (114.36,155) -- cycle ;
\draw (62.13,144.65) node [anchor=north west][inner sep=0.75pt]    {$\O_{0}$};
\draw (147.38,150.4) node [anchor=north west][inner sep=0.75pt]  [font=\small]  {$\mathsf{a}$};
\draw (192.88,146.4) node [anchor=north west][inner sep=0.75pt]  [font=\small]  {$\mathsf{b}$};
\draw (101.63,146.4) node [anchor=north west][inner sep=0.75pt]  [font=\small]  {$j$};
\end{tikzpicture}\right).
\end{equation}
Here $C_{\bbi j}^{(\mathsf{a})} = \mathcal{B}_{j}^{\,\mathsf{a}} $ is the structure constant of the bulk one-point function on the disk  and $C_{j0}^{(\mathsf{b})}$ is the structure constant of the bulk-boundary two-point function. 

The open and closed channel conformal blocks of the punctured cylinder are related to each other by a crossing transformation. Specializing to $c>1$ non-rational BCFT's with only Virasoro symmetry, the relevant crossing kernel is one (holomorphic) copy of the Ponsot-Teschner one-point S-kernel:
\begin{equation}\label{eq:BCFT_crossing}
    (-i\tau)^{\Delta_0}\mathcal{F}_{\text{open}}\!\left(\begin{tikzpicture}[x=0.75pt,y=0.75pt,yscale=-0.7,xscale=0.7,baseline={([yshift=-.5ex]current bounding box.center)}]
\draw [color={rgb, 255:red, 208; green, 2; blue, 27 }  ,draw opacity=1 ]   (184.38,180.14) -- (213.14,180) ;
\draw  [draw opacity=0] (139.66,145.29) .. controls (147.41,138.95) and (157.33,135.14) .. (168.13,135.14) .. controls (192.98,135.14) and (213.13,155.29) .. (213.13,180.14) .. controls (213.13,205) and (192.98,225.14) .. (168.13,225.14) .. controls (157.59,225.14) and (147.91,221.52) .. (140.24,215.46) -- (168.13,180.14) -- cycle ; \draw   (139.66,145.29) .. controls (147.41,138.95) and (157.33,135.14) .. (168.13,135.14) .. controls (192.98,135.14) and (213.13,155.29) .. (213.13,180.14) .. controls (213.13,205) and (192.98,225.14) .. (168.13,225.14) .. controls (157.59,225.14) and (147.91,221.52) .. (140.24,215.46) ;  
\draw   (151.88,180.14) .. controls (151.88,171.17) and (159.15,163.89) .. (168.13,163.89) .. controls (177.1,163.89) and (184.38,171.17) .. (184.38,180.14) .. controls (184.38,189.12) and (177.1,196.39) .. (168.13,196.39) .. controls (159.15,196.39) and (151.88,189.12) .. (151.88,180.14) -- cycle ;
\draw  [dash pattern={on 1.5pt off 1.5pt on 1.5pt off 1.5pt}]  (113.58,164.93) -- (113.67,195.93) ;
\draw    (113.78,165.4) .. controls (125.43,165.24) and (132.23,151.24) .. (139.66,145.29) ;
\draw    (113.63,195.36) .. controls (125.6,195.53) and (132.6,209.52) .. (140.24,215.46) ;
\draw (85.25,169.79) node [anchor=north west][inner sep=0.75pt] [font=\footnotesize]   {$\O_{0}$};
\draw (137,174.94) node [anchor=north west][inner sep=0.75pt]  [font=\footnotesize]  {$\mathsf{a}$};
\draw (160,135.54) node [anchor=north west][inner sep=0.75pt]  [font=\footnotesize]  {$\mathsf{b}$};
\draw (190,163.29) node [anchor=north west][inner sep=0.75pt]  [font=\footnotesize]  {$P$};
\end{tikzpicture}\right) = \int_0^\infty \dd P' \,\mathbb{S}_{PP'}[P_0]\, \mathcal{F}_{\text{closed}}\!\left(\begin{tikzpicture}[x=0.75pt,y=0.75pt,yscale=-0.7,xscale=0.7,baseline={([yshift=-.5ex]current bounding box.center)}]
\draw  [draw opacity=0] (116.53,120.15) .. controls (124.29,113.8) and (134.2,110) .. (145,110) .. controls (169.85,110) and (190,130.15) .. (190,155) .. controls (190,179.85) and (169.85,200) .. (145,200) .. controls (134.47,200) and (124.78,196.38) .. (117.11,190.32) -- (145,155) -- cycle ; \draw   (116.53,120.15) .. controls (124.29,113.8) and (134.2,110) .. (145,110) .. controls (169.85,110) and (190,130.15) .. (190,155) .. controls (190,179.85) and (169.85,200) .. (145,200) .. controls (134.47,200) and (124.78,196.38) .. (117.11,190.32) ;  
\draw   (128.75,155) .. controls (128.75,146.03) and (136.03,138.75) .. (145,138.75) .. controls (153.97,138.75) and (161.25,146.03) .. (161.25,155) .. controls (161.25,163.97) and (153.97,171.25) .. (145,171.25) .. controls (136.03,171.25) and (128.75,163.97) .. (128.75,155) -- cycle ;
\draw  [dash pattern={on 1.5pt off 1.5pt on 1.5pt off 1.5pt}]  (90.46,139.79) -- (90.54,170.79) ;
\draw    (90.65,140.26) .. controls (102.3,140.1) and (109.1,126.1) .. (116.53,120.15) ;
\draw    (90.5,170.22) .. controls (102.48,170.38) and (109.47,184.37) .. (117.11,190.32) ;
\draw  [color={rgb, 255:red, 208; green, 2; blue, 27 }  ,draw opacity=1 ] (114.36,155) .. controls (114.36,138.08) and (128.08,124.36) .. (145,124.36) .. controls (161.92,124.36) and (175.64,138.08) .. (175.64,155) .. controls (175.64,171.92) and (161.92,185.64) .. (145,185.64) .. controls (128.08,185.64) and (114.36,171.92) .. (114.36,155) -- cycle ;
\draw (62.13,144.65) node [anchor=north west][inner sep=0.75pt]  [font=\footnotesize]  {$\O_{0}$};
\draw (145.38,150.4) node [anchor=north west][inner sep=0.75pt]  [font=\footnotesize]  {$\mathsf{a}$};
\draw (192.88,146.4) node [anchor=north west][inner sep=0.75pt]  [font=\footnotesize]  {$\mathsf{b}$};
\draw (91.63,144.4) node [anchor=north west][inner sep=0.75pt]  [font=\footnotesize]  {$P'$};
\end{tikzpicture}\right).
\end{equation}
This can be shown by using the doubling trick to map the punctured annulus to the punctured torus, and then applying the S-transform on the torus \cite{Numasawa:2022cni}. 

\paragraph{Boundary Liouville.}
A prominent example of a BCFT with only Virasoro symmetry is boundary Liouville theory, which was studied, among others, in \cite{Ponsot:1999uf,Fateev:2000ik, Teschner:2000md, Ponsot:2001ng, Hosomichi:2001xc, Numasawa:2022cni, Collier:2021ngi, Kusuki:2021gpt}. For $c>1$ the spectrum is continuous and coincides with the non-degenerate representations $P\in \mathbb{R}$ of the Virasoro algebra, and its structure constants will also depend smoothly on the Liouville momenta. Let us explore the consequence of the open-closed duality \eqref{eq:open-closed} for Liouville BCFT with FZZT boundary conditions.

Plugging the S-transformation \eqref{eq:BCFT_crossing} into equation \eqref{eq:BCFT_open} and imposing the open-closed duality relation \eqref{eq:open-closed} results in the following condition that the boundary Liouville structure constants have to satisfy:
\begin{equation}\label{eq:BCFT_consistency}
    \int_0^\infty \dd P \,C_{PP_0P}^{(\mathsf{abb})} \,\mathbb{S}_{PP'}[P_0] = C_{\bbi P'}^{(\mathsf{a})} C_{P' P_0}^{(\mathsf{b})}.
\end{equation}
Here we used that the integration measure over the Liouville spectrum is flat in $P$, if we normalize  the two-point function to unity (i.e.\! $\mathfrak{g}_{PP}^{(\mathsf{ab})} = 1$). 
The structure constants, which now depend on continuous Liouville momenta, are known:
\begin{itemize}
\item Boundary three-point function on the disk \cite{Ponsot:2001ng}:
\begin{equation}\label{eq:bdy_3pt}
    C_{P_1P_2P_3}^{(\mathsf{abc})} = \frac{g^{(\mathsf{ba})}_{P_1}}{g_{P_2}^{(\mathsf{bc})}g_{P_3}^{(\mathsf{ca})}}\fker{P_{c}}{P_1}{P_3}{P_2}{P_b}{P_a}.
\end{equation}
Here $\mathbb{F}$ is the Virasoro fusion kernel, and the coefficients $g_P^{(\mathsf{ab})}$ are given by a ratio of double gamma functions (see e.g.\! equation (26) in \cite{Ponsot:2001ng} with $\beta \coloneqq \frac{Q}{2}+iP$). 

\item Bulk one-point function on the disk \cite{Teschner:2000md}:
\begin{equation}\label{eq:bulk_A}
    C_{\bbi P}^{(\mathsf{a})} =  A(P | P_a) = 2^{1/4}\,\e^{i\delta(P)} \frac{\mathbb{S}_{P_a P}[\bbi]}{\sqrt{\sker{\bbi}{P}{\bbi}}}.
\end{equation}
Here $\sker{P_a}{P}{\bbi}$ is the modular S-kernel \eqref{eq:torus_mod_S} (for FZZT boundary conditions) and the phase factor is the square root of the bulk Liouville reflection amplitude \cite{Zamolodchikov:1995aa}:
\begin{equation}
    \e^{2i \delta(P)} = R(P),\quad \Psi_{ZZ}^{(1,1)}(P) = R(P) \,\Psi_{ZZ}^{(1,1)}(-P),
\end{equation}
where $\Psi_{ZZ}^{(1,1)}(P)$ is the identity ZZ-brane wavefunction:
\begin{equation}
    \Psi_{ZZ}^{(1,1)}(P) = (\pi \mu \gamma(b^2))^{-iP/b} \frac{2^{3/4}\,2\pi i P}{\Gamma(1-2ib P)\Gamma(1-2i b^{-1}P)}.
\end{equation}
\item Bulk-boundary two-point function on the disk \cite{Hosomichi:2001xc}:
\begin{equation}\label{eq:bb_2pt}
    C_{PP_0}^{(\mathsf{a})} = A(P;P_0 | P_a).
\end{equation}
The precise expression for $A(P;P_0 | P_a)$ can be found in appendix D of \cite{Numasawa:2022cni}. There it was shown by an explicit computation that the bulk-boundary two-point structure constant \eqref{eq:bb_2pt} is proportional to the one-point S-kernel\footnote{
The internal indices of $\mathbb{S}$ are swapped compared to \cite{Numasawa:2022cni}. Explicitly, $\mathbb{S}_{P'P}[P_0]^{\text{here}} = \mathbb{S}_{PP'}[P_0]^{\text{there}}$.
}:
\begin{equation}\label{eq:S_vs_A}
    \sker{P_b}{P}{P_0} = \frac{1}{2^{1/4}}\Psi_{ZZ}^{(1,1)}(P)\, g_{P_0}^{(\mathsf{bb})}A(P;P_0|P_b).
\end{equation}
\end{itemize}
Having collected the ingredients that go into the open-closed duality condition \eqref{eq:BCFT_consistency}, we make the following observation: the boundary three-point function that appears is proportional to the fusion density: 
\begin{equation}\label{eq:propto}
    C^{(\mathsf{abb})}_{PP_0P} = \frac{1}{g_{P_0}^{(\mathsf{bb})}} \,\fker{P_b}{P}{P}{P_0}{P_b}{P_a} = \frac{1}{g_{P_0}^{(\mathsf{bb})}}\, \N_{P_0}[P_b,P_a,P].
\end{equation}
The proportionality constant is independent of $P$, so it can be taken outside the $P$-integral. Substituting \eqref{eq:bulk_A}, \eqref{eq:S_vs_A} and \eqref{eq:propto} into equation \eqref{eq:BCFT_consistency}, and using the relation between the ZZ brane wavefunction and $\rho_0$,
\begin{equation}
    \big|\Psi_{ZZ}^{(1,1)}(P)\big|^2 =\Psi_{ZZ}^{(1,1)}(P) \Psi_{ZZ}^{(1,1)}(-P)=  2^{-1/2}\rho_0(P),
\end{equation}
the open-closed duality condition can be written as:
\begin{equation}
    \int_0^\infty \dd P\, \N_{P_0}[P_b,P_a,P] \,\sker{P}{P'}{P_0} = \sker{P_b}{P'}{P_0} \frac{ \sker{P_a}{P'}{\bbi} }{\rho_0(P')}.
\end{equation}
This is precisely the Virasoro-Verlinde formula! So our formula demonstrates that boundary Liouville CFT indeed obeys the open-closed duality of the annulus one-point function, at a fully non-perturbative level.\footnote{One might wonder whether other interesting relations between $\mathbb{S},\mathbb{F}$ could emerge from additional consistency constraints in boundary Liouville theory. Cardy and Lewellen \cite{Cardy:1991tv, Lewellen:1991tb} showed there are \textit{four} such independent constraints: three coming from the disk and one from the annulus (the latter essentially yielding our formula). Investigating the bulk-to-boundary three-point function constraint on the disk would be particularly interesting. We thank S. Ribault for pointing this out to us. } Our analysis can be straightforwardly modified to ZZ or mixed boundary conditions, in which case we land on the Virasoro-Verlinde formula \eqref{eq:degenerate_Verlinde_formula} with degenerate external momenta.
\enlargethispage{1\baselineskip}
\enlargethispage{1\baselineskip}

\subsection{Knot and link partition functions}\label{sec:whitehead}

For the next application, we shift gears and turn to knots and links in 3d gravity. As explained in \cite{Collier_2023}, the Virasoro TQFT is a building block for evaluating the exact Euclidean AdS$_3$ gravitational path integral on a fixed hyperbolic 3-manifold $M$:
\begin{equation}\label{eq:grav_part}
    Z_\text{grav}(M) = \frac{1}{|\text{Map}(M)|} Z_\text{Vir}(M)Z^*_{\overline{\text{Vir}}}(M). 
\end{equation}
Here $\text{Map}(M)$ is the mapping class group of the 3-manifold, which, for hyperbolic manifolds, is isomorphic to the isometry group of $M$. In particular, this group is finite, so it just gives an overall symmetry factor. If $M$ has a conformal boundary, one should also sum over the \emph{boundary} mapping class group (the `modular images'). 

\paragraph{Figure-8 knot.} In \cite{Collier:2024mgv}, this recipe was used to compute the exact partition function of the figure-8 knot complement $\mathbf{4}_1$, which is a prominent example of a hyperbolic 3-manifold.\footnote{Here we used the standard nomenclature $\mathbf{4}_1$ that counts the number of crossings of the knot.} Two expressions were found for the partition function, using two different techniques. On the one hand, the \emph{Heegaard splitting} of the knot gives the following answer:\footnote{To compare to equation (4.5) in \cite{Collier:2024mgv}, use the fact that $\mathbb{S}_{P_0P_0}^*[P] = e^{-\pi i (\frac{c-1}{24}+P^2)}\mathbb{S}_{P_0P_0}[P]$.}
\begin{equation}\label{eq:fig8_1}
    Z_\text{Vir}\!\left(\begin{tikzpicture}[x=0.75pt,y=0.75pt,yscale=-0.6,xscale=0.6,baseline={([yshift=-0.5ex]current bounding box.center)}]
\draw [line width=1.5]    (213.91,112.69) .. controls (216.8,80.74) and (265.95,96.99) .. (232.49,142.89) ;
\draw [line width=1.5]    (213.91,123.16) .. controls (214.73,140.47) and (222.84,144.7) .. (234.97,155.77) .. controls (247.11,166.84) and (249.53,181.78) .. (232.6,202.11) ;
\draw [line width=1.5]    (249.02,119.13) .. controls (287.02,136.04) and (268.23,200.55) .. (248.71,207.85) .. controls (229.19,215.14) and (199.03,191.19) .. (222.58,155.36) ;
\draw [line width=1.5]    (223.35,207.96) .. controls (204.76,216.11) and (186.64,184.09) .. (186.77,158.56) .. controls (186.9,133.03) and (200.79,112.01) .. (238.07,116.96) ;
\draw (174,105.4) node [anchor=west][inner sep=0.75pt]  [font=\normalsize]  {$P_0$};
\end{tikzpicture}\right) = \frac{1}{\rho_0(P_0)} \int_0^\infty \dd P\,\rho_0(P) \mathbb{S}_{P_0P_0}^*[P]\,\e^{-2\pi i (\Delta_P - 2\Delta_0) }\,.
\end{equation}
On the other hand, the figure-8 knot complement is known to be a surface bundle over the circle $\Sigma_{1,1}\times_\varphi S^1$, where the monodromy $\varphi = ST^3$ is an element of the mapping class group of the once-punctured torus, $\mathrm{PSL}(2,\mathbb{Z})$. Therefore, using the standard rules of TQFT, the partition function should be given by the following answer:
\begin{equation}\label{eq:fig8_2}
Z'_\text{Vir}\!\left(\Sigma_{1,1}\times_{\varphi}S^1;P_0'\right) = \Tr_{\mathcal{H}_{1,1}}(\mathbb{S}\mathbb{T}^3) = \int_0^\infty \dd P \,\sker{P}{P}{P_0'}\,\big(\e^{-2\pi i (P^2-\frac{1}{24})}\big)^3\,. 
\end{equation}
However, it was observed in \cite{Collier:2024mgv} that the momentum $P_0$ in equation \eqref{eq:fig8_1} runs along the longitudinal cycle of a torus neighborhood of the knot, whereas $P_0'$ in equation \eqref{eq:fig8_2} runs along a meridian cycle of the boundary torus. It was conjectured that exchanging the two cycles of the boundary torus, using the modular S-transform, relates the partition functions $Z_\text{Vir}$ and $Z_\text{Vir}'$:
\begin{equation}\label{eq:conjecture}
    Z_\text{Vir}\left(\mathbf{4}_1;P_0\right) \stackrel{?}{=}  \int_0^\infty \dd P_0' \,\mathbb{S}_{P_0P_0'}[\bbi] \,Z'_\text{Vir}\left(\mathbf{4}_1;P_0'\right).
\end{equation}
This was checked to hold true in a large-$c$ saddle point approximation in \cite{Collier:2024mgv}. However, we can use the Virasoro-Verlinde formula to show that \eqref{eq:conjecture} holds exactly. 

To do so, let us first use the standard identity that expresses the one-point S-kernel as an integral over the fusion kernel:\footnote{To arrive at this form, we took the complex conjugate of the equivalent identity (2.33) in \cite{Collier:2024mgv}, and simplified the integrand using the tetrahedral symmetry of the fusion kernel.}
\begin{equation}\label{eq:S_identity}
   \sker{P}{P}{P_0'} = \int_0^\infty \dd P_s\, \rho_0(P_s)\,\e^{-2\pi i(\Delta_s-2\Delta_P)}\,\fker{P}{P}{P}{P_s}{P}{P_0'}.
\end{equation}
We now recognize the fusion density $\N_{P_s}[P,P_0',P]$ inside the integrand. Plugging \eqref{eq:S_identity} into the expression for $Z_\text{Vir}'$ in \eqref{eq:fig8_2}, we obtain the intermediate result:
\begin{equation}
    Z'_\text{Vir}(\mathbf{4}_1;P_0') = \e^{6\pi i (\frac{c}{24})}\int_0^\infty \dd P \dd P_s\,\rho_0(P_s) \, \e^{-2\pi i (\Delta_s+\Delta_P)}\N_{P_s}[P,P_0',P].
\end{equation}
Next, we use the Virasoro-Verlinde formula \eqref{eq:main_formula} for $\N_{P_s}$, which allows us to compute the modular S-transform in $P_0'$:
\begin{equation}\label{eq:VV_S}
    \int_0^\infty \dd P_0'\,\sker{P_0}{P_0'}{\bbi}\,\N_{P_s}[P,P_0',P] = \frac{\sker{P}{P_0}{P_s}\mathbb{S}^*_{P_0P}[P_s]}{\sker{\bbi}{P_0}{\bbi}}\,.
\end{equation}
Putting everything together, we have shown that:
\begin{multline}
    \int_0^\infty \dd P_0' \,\mathbb{S}_{P_0P_0'}[\bbi] \,Z'_\text{Vir}\left(\mathbf{4}_1;P_0'\right) = \frac{\e^{6\pi i (\frac{c}{24})}}{\rho_0(P_0)}\int_0^\infty \dd P \dd P_s\,\rho_0(P_s) \mathbb{T}_s \sker{P}{P_0}{P_s}\mathbb{T}_P \mathbb{S}^*_{P_0P}[P_s]
\end{multline}
where $\mathbb{T}_P = e^{-2\pi i\Delta_P}$. The conjectural equality \eqref{eq:conjecture} now follows from the fundamental $\mathrm{PSL}(2,\Z)$ identity $\mathbb{S}\mathbb{T}\mathbb{S}^* = \mathbb{T}^*\mathbb{S}^*\mathbb{T}^*$ (see eq.\! (2.49) in \cite{Eberhardt:2023mrq}), which allows us to perform the $P$-integral. Indeed, all the factors (including the phases) assemble nicely into \eqref{eq:fig8_1}.

\paragraph{Whitehead link.} As a next application, let us consider a slightly more complicated hyperbolic 3-manifold. The Whitehead link complement, which is denoted by $\mathbf{5}^2_1$, is one of the two 3-manifolds with the smallest hyperbolic volume among all hyperbolic 3-manifolds with two torus boundaries \cite{Agol_2010}. 

In a forthcoming work \cite{toappear}, it will be shown that the Virasoro TQFT partition function on the Whitehead link can also be computed in two ways. First, a genus-two Heegaard splitting of the manifold gives the following exact result:
\begin{equation}\label{eq:wh_1}
   Z_\text{Vir}\!\left(\hspace{-1mm}\begin{tikzpicture}[x=0.75pt,y=0.75pt,yscale=-0.6,xscale=0.6,baseline={([yshift=-0.5ex]current bounding box.center)}]
\draw  [draw opacity=0][line width=1.5]  (358.64,149.55) .. controls (353.36,168.8) and (340.88,182.33) .. (326.33,182.33) .. controls (307,182.33) and (291.33,158.46) .. (291.33,129) .. controls (291.33,125.34) and (291.58,121.76) .. (292.04,118.31) -- (326.33,129) -- cycle ; \draw  [line width=1.5]  (358.64,149.55) .. controls (353.36,168.8) and (340.88,182.33) .. (326.33,182.33) .. controls (307,182.33) and (291.33,158.46) .. (291.33,129) .. controls (291.33,125.34) and (291.58,121.76) .. (292.04,118.31) ;  
\draw  [draw opacity=0][line width=1.5]  (293.95,108.75) .. controls (299.18,89.34) and (311.71,75.67) .. (326.33,75.67) .. controls (345.66,75.67) and (361.33,99.54) .. (361.33,129) .. controls (361.33,132.85) and (361.07,136.61) .. (360.56,140.23) -- (326.33,129) -- cycle ; \draw  [line width=1.5]  (293.95,108.75) .. controls (299.18,89.34) and (311.71,75.67) .. (326.33,75.67) .. controls (345.66,75.67) and (361.33,99.54) .. (361.33,129) .. controls (361.33,132.85) and (361.07,136.61) .. (360.56,140.23) ;  
\draw [line width=1.5]    (287.29,144.43) .. controls (245.29,146.71) and (251.75,84.38) .. (322.14,126.14) ;
\draw [line width=1.5]    (297.71,143.14) .. controls (315.86,139.86) and (335.94,120.9) .. (352.8,114.9) ;
\draw [line width=1.5]    (331.71,132.71) .. controls (395.57,172.43) and (414.77,106.7) .. (366.2,112.7) ;
\draw (317.8,80) node [anchor=north west][inner sep=0.75pt]    {\small $P_{1}$};
\draw (396.8,118.6) node [anchor=north west][inner sep=0.75pt]    {\small $P_{2}$};
\end{tikzpicture}\right) = \frac{\e^{-6\pi i (\frac{c}{24})}}{\rho_0(P_1)\rho_0(P_2)}\int_0^\infty \dd P\,\rho_0(P) \sker{P_1}{P_2}{P}\mathbb{S}^*_{P_2P_1}[P]\,\e^{2\pi i \Delta_P}.
\end{equation}
On the other hand, the Whitehead link complement also admits a description as a fibration of the twice-punctured torus over the circle $\mathbf{5}_1^2 \cong \Sigma_{1,2}\times_\varphi S^1$, where $\varphi$ is a composition of three Dehn twists along the standard cycles on $\Sigma_{1,2}$. Leaving the details to \cite{toappear}, the trace of this mapping class evaluates to the following simple expression:
\begin{equation}\label{eq:wh_2}
Z'_\text{Vir}\!\left(\Sigma_{1,2}\times_{\varphi}S^1;P_1',P_2'\right) = \Tr_{\mathcal{H}_{1,2}}(\varphi) = \int_0^\infty \dd P \,\fker{P}{P}{P}{P_2'}{P}{P_1'}\,\e^{-2\pi i \Delta_P}\,.
\end{equation}

As in the previous example, the partition functions $Z_\text{Vir}$ and $Z_\text{Vir}'$ are not equal: the momenta $P_1,P_2$ going along the link components in \eqref{eq:wh_1} describe the S-dual cycles of the momenta $P_1',P_2'$ that run along the boundary circles of the Seifert surface. Hence, a natural guess is that $Z_\text{Vir}$ and $Z_\text{Vir}'$ are related by a double modular S-transform:
\begin{equation}\label{eq:wh_conjecture}
Z_\text{Vir}\left(\mathbf{5}^2_1;P_1,P_2\right) \stackrel{?}{=}  \int_0^\infty \dd P_1' \dd P_2'\,\mathbb{S}_{P_1P_1'}[\bbi] \mathbb{S}_{P_2P_2'}[\bbi]\,Z'_\text{Vir}\left(\mathbf{5}^2_1;P_1',P_2'\right).
\end{equation}
Again, the key to showing that \eqref{eq:wh_conjecture} is indeed true, is the Virasoro-Verlinde formula. Namely, notice that the integrand of eq.\! \eqref{eq:wh_2} contains the fusion density $\N_{P_2'}[P,P_1',P]$, so we can use \eqref{eq:VV_S} to evaluate the first modular S-transform:
\begin{equation}
    \int_0^\infty \dd P_1'\,\mathbb{S}_{P_1P_1'}[\bbi] Z'_\text{Vir}\left(\mathbf{5}^2_1;P_1',P_2'\right) = \int_0^\infty \dd P \,\frac{\sker{P}{P_1}{P_2'}\,\mathbb{S}^*_{P_1P}[P_2']}{\sker{\bbi}{P_1}{\bbi}} \,\e^{-2\pi i\Delta_P}.
\end{equation}
Next, we evaluate the $P$-integral using the $\mathrm{PSL}(2,\mathbb{Z})$ identity $\mathbb{S}\mathbb{T}\mathbb{S}^* = \mathbb{T}^*\mathbb{S}^*\mathbb{T}^*$, which gives a single one-point S-kernel $\mathbb{S}_{P_1P_1}^*[P_2']$ times a phase. As before, the one-point S-kernel can be represented as an integral over a fusion kernel via \eqref{eq:S_identity}, which allows us to also compute the second modular S-transform:
\begin{multline}
     \int_0^\infty \dd P_1'\dd P_2'\,\mathbb{S}_{P_1P_1'}[\bbi]   \,\mathbb{S}_{P_2P_2'}[\bbi]\,Z'_\text{Vir}\left(\mathbf{5}^2_1;P_1',P_2'\right) \\[1em] = \frac{\e^{-6\pi i (\frac{c}{24})}\,\e^{4\pi i \Delta_1}}{\rho_0(P_1)} \int_0^\infty \dd P_s \,\rho_0(P_s)\,\e^{2\pi i (\Delta_s - 2 \Delta_1)}\int_0^\infty \dd P_2'\,\mathbb{S}_{P_2P_2'}[\bbi]\fker{P_1}{P_1}{P_1}{P_s}{P_1}{P_2'}.
\end{multline}
Applying the Virasoro-Verlinde formula once more, the integral over $P_2'$ gives a ratio of S-kernels as in \eqref{eq:VV_S}, and we see that we precisely land on the expression \eqref{eq:wh_1} for $Z_\text{Vir}(\mathbf{5}^2_1)$. This shows that the conjecture \eqref{eq:wh_conjecture} is true, meaning that $P_{1,2}$ are indeed the S-dual holonomies of $P_{1,2}'$. 

The above two examples show that the Virasoro-Verlinde formula is a useful identity in Virasoro TQFT when computing the Euclidean gravity path integral on hyperbolic 3-manifolds with finite volume. This is an interesting application, since the gravitational partition function is a topological invariant of the 3-manifold. In the next section, we will discuss an application where the 3-manifold has asymptotic (conformal) boundaries, which is more natural in the context of AdS$_3$/CFT$_2$.

\subsection{Three-boundary wormhole in 3d gravity}\label{sec:3boundary}

As a third application, we will show that the Virasoro-Verlinde formula has a nice interpretation in 3d gravity: it computes the partition function of a Euclidean wormhole with three asymptotic torus boundaries. We will explain that the regulator $\O_0$ can be seen as a conical defect going through the bulk that makes the wormhole partition function finite. We then discuss the limit in which the regulator is removed and obtain a finite answer that is compatible with the intuition from random matrix theory. 

To set up the calculation, consider the following three-boundary wormhole:
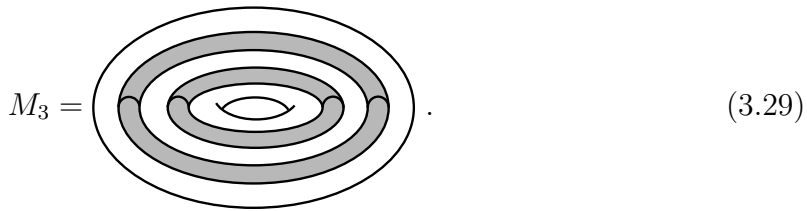
\begin{equation}\label{eq:M3}
   M_3 =  \begin{tikzpicture}[x=0.75pt,y=0.75pt,yscale=-0.61,xscale=0.61,baseline={([yshift=-.5ex]current bounding box.center)}]
\draw  [fill={rgb, 255:red, 255; green, 255; blue, 255 }  ,fill opacity=1 ] (84.83,185.82) .. controls (84.83,140.08) and (144.16,102.99) .. (217.33,102.99) .. controls (290.51,102.99) and (349.83,140.08) .. (349.83,185.82) .. controls (349.83,231.57) and (290.51,268.66) .. (217.33,268.66) .. controls (144.16,268.66) and (84.83,231.57) .. (84.83,185.82) -- cycle ;
\draw  [fill={rgb, 255:red, 155; green, 155; blue, 155 }  ,fill opacity=0.71 ] (105.67,185.82) .. controls (105.67,151.21) and (155.66,123.16) .. (217.33,123.16) .. controls (279.01,123.16) and (329,151.21) .. (329,185.82) .. controls (329,220.43) and (279.01,248.49) .. (217.33,248.49) .. controls (155.66,248.49) and (105.67,220.43) .. (105.67,185.82) -- cycle ;
\draw  [fill={rgb, 255:red, 255; green, 255; blue, 255 }  ,fill opacity=1 ] (123,185.82) .. controls (123,159.5) and (165.24,138.16) .. (217.33,138.16) .. controls (269.43,138.16) and (311.67,159.5) .. (311.67,185.82) .. controls (311.67,212.15) and (269.43,233.49) .. (217.33,233.49) .. controls (165.24,233.49) and (123,212.15) .. (123,185.82) -- cycle ;
\draw  [fill={rgb, 255:red, 155; green, 155; blue, 155 }  ,fill opacity=0.69 ] (146.34,185.82) .. controls (146.34,167.49) and (178.87,152.62) .. (219,152.62) .. controls (259.13,152.62) and (291.67,167.49) .. (291.67,185.82) .. controls (291.67,204.16) and (259.13,219.02) .. (219,219.02) .. controls (178.87,219.02) and (146.34,204.16) .. (146.34,185.82) -- cycle ;
\draw  [fill={rgb, 255:red, 255; green, 255; blue, 255 }  ,fill opacity=1 ] (163.67,185.82) .. controls (163.67,174.78) and (188.44,165.82) .. (219,165.82) .. controls (249.56,165.82) and (274.34,174.78) .. (274.34,185.82) .. controls (274.34,196.87) and (249.56,205.82) .. (219,205.82) .. controls (188.44,205.82) and (163.67,196.87) .. (163.67,185.82) -- cycle ;
\draw  [draw opacity=0] (251.2,183.99) .. controls (245.61,190.79) and (233.52,195.5) .. (219.5,195.5) .. controls (204.15,195.5) and (191.1,189.85) .. (186.38,181.99) -- (219.5,175.5) -- cycle ; \draw   (251.2,183.99) .. controls (245.61,190.79) and (233.52,195.5) .. (219.5,195.5) .. controls (204.15,195.5) and (191.1,189.85) .. (186.38,181.99) ;  
\draw  [draw opacity=0] (191.37,186.95) .. controls (194.98,181.38) and (205.76,177.33) .. (218.5,177.33) .. controls (231.75,177.33) and (242.88,181.71) .. (246.03,187.62) -- (218.5,191.04) -- cycle ; \draw   (191.37,186.95) .. controls (194.98,181.38) and (205.76,177.33) .. (218.5,177.33) .. controls (231.75,177.33) and (242.88,181.71) .. (246.03,187.62) ;  
\draw  [draw opacity=0] (105.66,185.29) .. controls (105.73,180.64) and (109.58,176.89) .. (114.33,176.89) .. controls (118.8,176.89) and (122.48,180.22) .. (122.94,184.5) -- (114.33,185.42) -- cycle ; \draw   (105.66,185.29) .. controls (105.73,180.64) and (109.58,176.89) .. (114.33,176.89) .. controls (118.8,176.89) and (122.48,180.22) .. (122.94,184.5) ;  
\draw  [draw opacity=0] (146.34,185.57) .. controls (146.47,180.63) and (150.3,176.67) .. (155,176.67) .. controls (159.69,176.67) and (163.51,180.6) .. (163.66,185.51) -- (155,185.82) -- cycle ; \draw   (146.34,185.57) .. controls (146.47,180.63) and (150.3,176.67) .. (155,176.67) .. controls (159.69,176.67) and (163.51,180.6) .. (163.66,185.51) ;  
\draw  [draw opacity=0] (274.34,185.54) .. controls (274.48,180.62) and (278.3,176.67) .. (283,176.67) .. controls (287.79,176.67) and (291.67,180.77) .. (291.67,185.82) .. controls (291.67,185.94) and (291.66,186.06) .. (291.66,186.18) -- (283,185.82) -- cycle ; \draw   (274.34,185.54) .. controls (274.48,180.62) and (278.3,176.67) .. (283,176.67) .. controls (287.79,176.67) and (291.67,180.77) .. (291.67,185.82) .. controls (291.67,185.94) and (291.66,186.06) .. (291.66,186.18) ;  
\draw  [draw opacity=0] (311.67,185.86) .. controls (311.67,185.85) and (311.67,185.84) .. (311.67,185.82) .. controls (311.67,180.77) and (315.55,176.67) .. (320.33,176.67) .. controls (325.05,176.67) and (328.89,180.65) .. (329,185.6) -- (320.33,185.82) -- cycle ; \draw   (311.67,185.86) .. controls (311.67,185.85) and (311.67,185.84) .. (311.67,185.82) .. controls (311.67,180.77) and (315.55,176.67) .. (320.33,176.67) .. controls (325.05,176.67) and (328.89,180.65) .. (329,185.6) ;  
\end{tikzpicture}\,\,.
\end{equation}
Topologically, $M_3$ is a three-holed sphere times a circle $\Sigma_{0,3}\times S^1$. Its boundary consists of 3 tori (one has the opposite orientation to the other two), which we label by complex structure moduli $\tau_1,\tau_2$ and $\tau_3$. We can alternatively think of $M_3$ as a solid torus with two solid tori removed from the interior, as illustrated in \eqref{eq:M3}. 

\paragraph{Warm-up: Chern-Simons theory.} Before computing the 3d gravity path integral on this topology, let us study the simpler problem of compact Chern-Simons theory.\footnote{We would like to thank Jan de Boer for discussions on this topic.} In that case, the space of conformal blocks on the three-holed sphere has a dimension given by the fusion coefficient, so \cite{Witten:1988hf}
\begin{equation}\label{eq:CS_simple}
    Z_{\mathrm{CS}}(\Sigma_{0,3}\times S^1) = \Tr_{\mathcal{H}_{0,3}} (\bbi) = \mathrm{dim} \mathcal{H}_{0,3} = N_{ij}\text{}^k.
\end{equation}
This is the `microcanonical' partition function, where we fix boundary holonomies in irreducible representations $i,j,k$ of the gauge group. To obtain the `canonical' partition function, we multiply by the characters of the chiral algebra and sum over the representations:
\begin{equation}\label{eq:CS_comp}
    \sum_{i,j,k} N_{ij}\text{}^k \,\chi_i(\tau_1) \chi_j(\tau_2)\chi^*_k(\tau_3).
\end{equation}
Plugging in the Verlinde formula \eqref{eq:original_formula}, one obtains a sum of products of characters in the S-dual channel:
\begin{equation}\label{eq:CS_result}
    \sum_h (S_{\bbi h})^{-1} \chi_h(-\tfrac{1}{\tau_1})\chi_h(-\tfrac{1}{\tau_2})\chi^*_h(-\tfrac{1}{\tau_3}).
\end{equation}
Notice that the three characters are all paired in the primary conformal weight $h$, and weighted by the inverse modular S-matrix. Also note that \eqref{eq:CS_result} need not be modular invariant. One way to construct a modular invariant object is to simply sum over the modular images of $\tau_{1,2,3}$. However, from the Chern-Simons point of view this is not required. For further connections between torus wormholes, the Verlinde formula and topological entanglement in Chern-Simons theory, see \cite{Salton:2016qpp,Balasubramanian:2016sro}. 

\paragraph{The 3d gravity computation: adding a regulator.} 
We would like to mimic the above Chern-Simons calculation in the non-compact case of pure 3d gravity. However, a naive application of the logic does not hold: as we saw in the previous section, the integral 
\begin{equation}
    \int_0^\infty \dd P\,\frac{\sker{P_1}{P}{\bbi}\sker{P_2}{P}{\bbi}\mathbb{S}_{PP_3}^*[\bbi]}{\sker{\bbi}{P}{\bbi}}
\end{equation}
diverges for all real values of $P_{1,2,3}$. Since the gravitational partition function can be written as an integral over Virasoro characters in the continuous black hole band, the naive generalization of the formula \eqref{eq:CS_comp} for the three-boundary partition function is clearly ill-defined in pure 3d gravity.

The way forward is to couple AdS$_3$ gravity to a matter field dual to a CFT operator $\O_0$. We will think of $\O_0$ as a probe of the wormhole geometry, which in the bulk extends to a Wilson line $\mathcal{W}_0$ that ends on two of the three asymptotic boundaries:
\begin{equation}\label{eq:regulated_manifold}
    M_3^{\O_0} = \begin{tikzpicture}[x=0.75pt,y=0.75pt,yscale=-0.65,xscale=0.65,baseline={([yshift=-.5ex]current bounding box.center)}]
\draw  [fill={rgb, 255:red, 255; green, 255; blue, 255 }  ,fill opacity=1 ] (84.83,185.82) .. controls (84.83,140.08) and (144.16,102.99) .. (217.33,102.99) .. controls (290.51,102.99) and (349.83,140.08) .. (349.83,185.82) .. controls (349.83,231.57) and (290.51,268.66) .. (217.33,268.66) .. controls (144.16,268.66) and (84.83,231.57) .. (84.83,185.82) -- cycle ;
\draw  [fill={rgb, 255:red, 128; green, 128; blue, 128 }  ,fill opacity=0.3 ] (105.67,185.82) .. controls (105.67,151.21) and (155.66,123.16) .. (217.33,123.16) .. controls (279.01,123.16) and (329,151.21) .. (329,185.82) .. controls (329,220.43) and (279.01,248.49) .. (217.33,248.49) .. controls (155.66,248.49) and (105.67,220.43) .. (105.67,185.82) -- cycle ;
\draw  [fill={rgb, 255:red, 255; green, 255; blue, 255 }  ,fill opacity=1 ] (123,185.82) .. controls (123,159.5) and (165.24,138.16) .. (217.33,138.16) .. controls (269.43,138.16) and (311.67,159.5) .. (311.67,185.82) .. controls (311.67,212.15) and (269.43,233.49) .. (217.33,233.49) .. controls (165.24,233.49) and (123,212.15) .. (123,185.82) -- cycle ;
\draw  [fill={rgb, 255:red, 155; green, 155; blue, 155 }  ,fill opacity=0.3 ] (146.34,185.82) .. controls (146.34,167.49) and (178.87,152.62) .. (219,152.62) .. controls (259.13,152.62) and (291.67,167.49) .. (291.67,185.82) .. controls (291.67,204.16) and (259.13,219.02) .. (219,219.02) .. controls (178.87,219.02) and (146.34,204.16) .. (146.34,185.82) -- cycle ;
\draw  [fill={rgb, 255:red, 255; green, 255; blue, 255 }  ,fill opacity=1 ] (163.67,185.82) .. controls (163.67,174.78) and (188.44,165.82) .. (219,165.82) .. controls (249.56,165.82) and (274.34,174.78) .. (274.34,185.82) .. controls (274.34,196.87) and (249.56,205.82) .. (219,205.82) .. controls (188.44,205.82) and (163.67,196.87) .. (163.67,185.82) -- cycle ;
\draw  [draw opacity=0] (251.2,183.99) .. controls (245.61,190.79) and (233.52,195.5) .. (219.5,195.5) .. controls (204.15,195.5) and (191.1,189.85) .. (186.38,181.99) -- (219.5,175.5) -- cycle ; \draw   (251.2,183.99) .. controls (245.61,190.79) and (233.52,195.5) .. (219.5,195.5) .. controls (204.15,195.5) and (191.1,189.85) .. (186.38,181.99) ;  
\draw  [draw opacity=0] (191.37,186.95) .. controls (194.98,181.38) and (205.76,177.33) .. (218.5,177.33) .. controls (231.75,177.33) and (242.88,181.71) .. (246.03,187.62) -- (218.5,191.04) -- cycle ; \draw   (191.37,186.95) .. controls (194.98,181.38) and (205.76,177.33) .. (218.5,177.33) .. controls (231.75,177.33) and (242.88,181.71) .. (246.03,187.62) ;  
\draw  [draw opacity=0] (105.66,185.29) .. controls (105.73,180.64) and (109.58,176.89) .. (114.33,176.89) .. controls (118.8,176.89) and (122.48,180.22) .. (122.94,184.5) -- (114.33,185.42) -- cycle ; \draw   (105.66,185.29) .. controls (105.73,180.64) and (109.58,176.89) .. (114.33,176.89) .. controls (118.8,176.89) and (122.48,180.22) .. (122.94,184.5) ;  
\draw  [draw opacity=0] (146.34,185.57) .. controls (146.47,180.63) and (150.3,176.67) .. (155,176.67) .. controls (159.69,176.67) and (163.51,180.6) .. (163.66,185.51) -- (155,185.82) -- cycle ; \draw   (146.34,185.57) .. controls (146.47,180.63) and (150.3,176.67) .. (155,176.67) .. controls (159.69,176.67) and (163.51,180.6) .. (163.66,185.51) ;  
\draw  [draw opacity=0] (274.34,185.54) .. controls (274.48,180.62) and (278.3,176.67) .. (283,176.67) .. controls (287.79,176.67) and (291.67,180.77) .. (291.67,185.82) .. controls (291.67,185.94) and (291.66,186.06) .. (291.66,186.18) -- (283,185.82) -- cycle ; \draw   (274.34,185.54) .. controls (274.48,180.62) and (278.3,176.67) .. (283,176.67) .. controls (287.79,176.67) and (291.67,180.77) .. (291.67,185.82) .. controls (291.67,185.94) and (291.66,186.06) .. (291.66,186.18) ;  
\draw  [draw opacity=0] (311.67,185.86) .. controls (311.67,185.85) and (311.67,185.84) .. (311.67,185.82) .. controls (311.67,180.77) and (315.55,176.67) .. (320.33,176.67) .. controls (325.05,176.67) and (328.89,180.65) .. (329,185.6) -- (320.33,185.82) -- cycle ; \draw   (311.67,185.86) .. controls (311.67,185.85) and (311.67,185.84) .. (311.67,185.82) .. controls (311.67,180.77) and (315.55,176.67) .. (320.33,176.67) .. controls (325.05,176.67) and (328.89,180.65) .. (329,185.6) ;  
\draw [color={rgb, 255:red, 208; green, 2; blue, 27 }  ,draw opacity=1 ][line width=1.5]    (228.83,130.33) .. controls (238.5,138.17) and (221.83,150.5) .. (229.5,159.17) ;
\draw [shift={(229.5,159.17)}, rotate = 48.5] [color={rgb, 255:red, 208; green, 2; blue, 27 }  ,draw opacity=1 ][fill={rgb, 255:red, 208; green, 2; blue, 27 }  ,fill opacity=1 ][line width=1.5]      (0, 0) circle [x radius= 1.74, y radius= 1.74]   ;
\draw [shift={(228.83,130.33)}, rotate = 39.02] [color={rgb, 255:red, 208; green, 2; blue, 27 }  ,draw opacity=1 ][fill={rgb, 255:red, 208; green, 2; blue, 27 }  ,fill opacity=1 ][line width=1.5]      (0, 0) circle [x radius= 1.74, y radius= 1.74]   ;
\draw [color={rgb, 255:red, 208; green, 2; blue, 27 }  ,draw opacity=1 ]   (238.95,147) .. controls (313.89,125.73) and (341.67,161.43) .. (374.67,137.67) ;
\draw [shift={(236.67,147.67)}, rotate = 343.53] [color={rgb, 255:red, 208; green, 2; blue, 27 }  ,draw opacity=1 ][line width=0.75]    (10.93,-3.29) .. controls (6.95,-1.4) and (3.31,-0.3) .. (0,0) .. controls (3.31,0.3) and (6.95,1.4) .. (10.93,3.29)   ;
\draw (376.5,127.32) node [anchor=north west][inner sep=0.75pt]  [font=\large,color={rgb, 255:red, 208; green, 2; blue, 27 }  ,opacity=1 ]  {$\mathcal{W}_0$};
\end{tikzpicture} \,\,.
\end{equation}

In the same way that the addition of a matter field makes the \emph{two}-boundary wormhole in 3d gravity on-shell \cite{Chandra:2022bqq}, the three-boundary wormhole $M_3^{\O_0}$ can be stabilized for heavy enough operators $\O_0$. In particular, a semiclassical description is valid when $\O_0$ is in the conical defect regime, meaning that $\Delta_0$ and $\bar \Delta_0$ scale with $c$ while remaining below the black hole threshold $\frac{c-1}{24}$ (so $P_0,\bar P_0$ are purely imaginary) \cite{Benjamin:2020mfz}. 

The semiclassical geometry is difficult to describe with an explicit metric that includes the backreacted matter field. Fortunately, we can use the Virasoro TQFT to derive the exact partition function on $M_3^{\O_0}$ without having to construct a  metric. As a first step, we compute the `microcanonical' partition function:
\begin{equation}\label{eq:3bdy_step1}
    Z_\text{Vir}(M_3^{\O_0};P_{1,2,3}) = \int_0^\infty \dd P\,\sker{P_2}{P}{\bbi}\,  Z_\text{Vir}\!\left(\begin{tikzpicture}[x=0.75pt,y=0.75pt,yscale=-0.75,xscale=0.75,baseline={([yshift=-.5ex]current bounding box.center)}]
\draw [color={rgb, 255:red, 208; green, 2; blue, 27 }  ,draw opacity=1 ][line width=1.5]    (160,100) -- (220,100) ;
\draw  [draw opacity=0][line width=1.5]  (144.12,73.52) .. controls (153.57,78.57) and (160,88.54) .. (160,100) .. controls (160,116.57) and (146.57,130) .. (130,130) .. controls (113.43,130) and (100,116.57) .. (100,100) .. controls (100,83.57) and (113.2,70.23) .. (129.57,70) -- (130,100) -- cycle ; \draw  [line width=1.5]  (144.12,73.52) .. controls (153.57,78.57) and (160,88.54) .. (160,100) .. controls (160,116.57) and (146.57,130) .. (130,130) .. controls (113.43,130) and (100,116.57) .. (100,100) .. controls (100,83.57) and (113.2,70.23) .. (129.57,70) ;  
\draw  [draw opacity=0][line width=1.5]  (248.05,70.06) .. controls (248.69,70.02) and (249.34,70) .. (250,70) .. controls (266.57,70) and (280,83.43) .. (280,100) .. controls (280,116.57) and (266.57,130) .. (250,130) .. controls (233.43,130) and (220,116.57) .. (220,100) .. controls (220,88.32) and (226.68,78.2) .. (236.42,73.24) -- (250,100) -- cycle ; \draw  [line width=1.5]  (248.05,70.06) .. controls (248.69,70.02) and (249.34,70) .. (250,70) .. controls (266.57,70) and (280,83.43) .. (280,100) .. controls (280,116.57) and (266.57,130) .. (250,130) .. controls (233.43,130) and (220,116.57) .. (220,100) .. controls (220,88.32) and (226.68,78.2) .. (236.42,73.24) ;  
\draw  [draw opacity=0][line width=1.5]  (133.86,121.22) .. controls (131.37,114.63) and (130,107.47) .. (130,100) .. controls (130,66.86) and (156.86,40) .. (190,40) .. controls (223.14,40) and (250,66.86) .. (250,100) .. controls (250,107.59) and (248.59,114.86) .. (246.01,121.55) -- (190,100) -- cycle ; \draw  [line width=1.5]  (133.86,121.22) .. controls (131.37,114.63) and (130,107.47) .. (130,100) .. controls (130,66.86) and (156.86,40) .. (190,40) .. controls (223.14,40) and (250,66.86) .. (250,100) .. controls (250,107.59) and (248.59,114.86) .. (246.01,121.55) ;  
\draw  [draw opacity=0][line width=1.5]  (237.55,136.59) .. controls (226.58,150.83) and (209.36,160) .. (190,160) .. controls (170.42,160) and (153.04,150.62) .. (142.09,136.12) -- (190,100) -- cycle ; \draw  [line width=1.5]  (237.55,136.59) .. controls (226.58,150.83) and (209.36,160) .. (190,160) .. controls (170.42,160) and (153.04,150.62) .. (142.09,136.12) ;  
\draw (183.5,76.9) node [anchor=north west][inner sep=0.75pt]    {$P_{0}$};
\draw (104,89.4) node [anchor=north west][inner sep=0.75pt]    {$P_{1}$};
\draw (252.5,89.4) node [anchor=north west][inner sep=0.75pt]    {$P_{3}$};
\draw (184.5,44.4) node [anchor=north west][inner sep=0.75pt]    {$P$};
\end{tikzpicture}\right)\,.
\end{equation}
To arrive at the above expression, we first represented $M_3^{\O_0}$ as the complement of three solid tori inside $S^3$. The solid tori can be thought of as tubular neighborhoods of Wilson loops $\mathcal{L}_1,\mathcal{L}_2$ and $\mathcal{L}_3$. We specified the holonomies around the meridian (non-contractible) cycles of the boundary tori by $P_1, P_2$ and $P_3$ (in $\mathbb{R}_+$). We also attached a Wilson line $\mathcal{W}_0$ between $\mathcal{L}_1$ and $\mathcal{L}_3$. 

Then, we used the modular S-transform to exchange the meridian and longitudinal cycles of one of the boundary tori, corresponding to $\mathcal{L}_2$. Topologically, this step is tantamount to performing a Dehn surgery \cite{Collier:2024mgv}. The fact that toric Dehn surgery is implemented in the Virasoro TQFT by $\mathbb{S}_{PP_2}[\bbi]$ was already used in the previous section, recall equations \eqref{eq:conjecture} and \eqref{eq:wh_conjecture}.

After this Dehn surgery, we compute the Virasoro TQFT partition function on the right-hand side of \eqref{eq:3bdy_step1} by finding a so-called Heegaard splitting of the manifold inside the brackets. We do so by using the TQFT identities \eqref{eq:linked_lines} and \eqref{eq:unlinking2}:
\begin{align}
\begin{split}
    &Z_\text{Vir}\!\left(\begin{tikzpicture}[x=0.75pt,y=0.75pt,yscale=-0.6,xscale=0.6,baseline={([yshift=-.5ex]current bounding box.center)}]
\draw [color={rgb, 255:red, 208; green, 2; blue, 27 }  ,draw opacity=1 ][line width=1.5]    (160,100) -- (220,100) ;
\draw  [draw opacity=0][line width=1.5]  (144.12,73.52) .. controls (153.57,78.57) and (160,88.54) .. (160,100) .. controls (160,116.57) and (146.57,130) .. (130,130) .. controls (113.43,130) and (100,116.57) .. (100,100) .. controls (100,83.57) and (113.2,70.23) .. (129.57,70) -- (130,100) -- cycle ; \draw  [line width=1.5]  (144.12,73.52) .. controls (153.57,78.57) and (160,88.54) .. (160,100) .. controls (160,116.57) and (146.57,130) .. (130,130) .. controls (113.43,130) and (100,116.57) .. (100,100) .. controls (100,83.57) and (113.2,70.23) .. (129.57,70) ;  
\draw  [draw opacity=0][line width=1.5]  (248.05,70.06) .. controls (248.69,70.02) and (249.34,70) .. (250,70) .. controls (266.57,70) and (280,83.43) .. (280,100) .. controls (280,116.57) and (266.57,130) .. (250,130) .. controls (233.43,130) and (220,116.57) .. (220,100) .. controls (220,88.32) and (226.68,78.2) .. (236.42,73.24) -- (250,100) -- cycle ; \draw  [line width=1.5]  (248.05,70.06) .. controls (248.69,70.02) and (249.34,70) .. (250,70) .. controls (266.57,70) and (280,83.43) .. (280,100) .. controls (280,116.57) and (266.57,130) .. (250,130) .. controls (233.43,130) and (220,116.57) .. (220,100) .. controls (220,88.32) and (226.68,78.2) .. (236.42,73.24) ;  
\draw  [draw opacity=0][line width=1.5]  (133.86,121.22) .. controls (131.37,114.63) and (130,107.47) .. (130,100) .. controls (130,66.86) and (156.86,40) .. (190,40) .. controls (223.14,40) and (250,66.86) .. (250,100) .. controls (250,107.59) and (248.59,114.86) .. (246.01,121.55) -- (190,100) -- cycle ; \draw  [line width=1.5]  (133.86,121.22) .. controls (131.37,114.63) and (130,107.47) .. (130,100) .. controls (130,66.86) and (156.86,40) .. (190,40) .. controls (223.14,40) and (250,66.86) .. (250,100) .. controls (250,107.59) and (248.59,114.86) .. (246.01,121.55) ;  
\draw  [draw opacity=0][line width=1.5]  (237.55,136.59) .. controls (226.58,150.83) and (209.36,160) .. (190,160) .. controls (170.42,160) and (153.04,150.62) .. (142.09,136.12) -- (190,100) -- cycle ; \draw  [line width=1.5]  (237.55,136.59) .. controls (226.58,150.83) and (209.36,160) .. (190,160) .. controls (170.42,160) and (153.04,150.62) .. (142.09,136.12) ;  
\draw (183.5,76.9) node [anchor=north west][inner sep=0.75pt]    {\scriptsize $P_{0}$};
\draw (104,89.4) node [anchor=north west][inner sep=0.75pt]    {\scriptsize $P_{1}$};
\draw (252.5,89.4) node [anchor=north west][inner sep=0.75pt]    {\scriptsize $P_{3}$};
\draw (184.5,44.4) node [anchor=north west][inner sep=0.75pt]    {\scriptsize $P$};
\end{tikzpicture}\right) =\\& \int_0^\infty \dd P_4 \dd P_5\, \fker{\bbi}{P_4}{P_1}{P_1}{P_1}{P_1}\frac{\sker{P_1}{P}{P_4}}{\sker{\bbi}{P}{\bbi}} \fker{\bbi}{P_5}{P_3}{P_3}{P_3}{P_3}\frac{\mathbb{S}_{P_3P}^*[P_5]}{\sker{\bbi}{P}{\bbi}} \,Z_\text{Vir}\!\left(\begin{tikzpicture}[x=0.75pt,y=0.75pt,yscale=-0.65,xscale=0.65,baseline={([yshift=-.5ex]current bounding box.center)}]
\draw [color={rgb, 255:red, 208; green, 2; blue, 27 }  ,draw opacity=1 ][line width=1.5]    (227.5,126.67) -- (257,126.67) ;
\draw [color={rgb, 255:red, 208; green, 2; blue, 27 }  ,draw opacity=1 ][line width=1.5]    (80,126.67) -- (109.5,126.67) ;
\draw [color={rgb, 255:red, 208; green, 2; blue, 27 }  ,draw opacity=1 ][line width=1.5]    (153.75,126.67) -- (183.25,126.67) ;
\draw  [line width=1.5]  (109.5,126.67) .. controls (109.5,111.68) and (119.41,99.52) .. (131.63,99.52) .. controls (143.84,99.52) and (153.75,111.68) .. (153.75,126.67) .. controls (153.75,141.66) and (143.84,153.81) .. (131.63,153.81) .. controls (119.41,153.81) and (109.5,141.66) .. (109.5,126.67) -- cycle ;
\draw  [line width=1.5]  (183.25,126.67) .. controls (183.25,111.68) and (193.16,99.52) .. (205.38,99.52) .. controls (217.59,99.52) and (227.5,111.68) .. (227.5,126.67) .. controls (227.5,141.66) and (217.59,153.81) .. (205.38,153.81) .. controls (193.16,153.81) and (183.25,141.66) .. (183.25,126.67) -- cycle ;
\draw  [line width=1.5]  (80,126.67) .. controls (80,91.69) and (119.62,63.33) .. (168.5,63.33) .. controls (217.38,63.33) and (257,91.69) .. (257,126.67) .. controls (257,161.64) and (217.38,190) .. (168.5,190) .. controls (119.62,190) and (80,161.64) .. (80,126.67) -- cycle ;
\draw (157.37,105.15) node [anchor=north west][inner sep=0.75pt]    {\scriptsize $P_{0}$};
\draw (122.99,104.54) node [anchor=north west][inner sep=0.75pt]    {\scriptsize $P_{1}$};
\draw (198.95,107.56) node [anchor=north west][inner sep=0.75pt]    {\scriptsize $P_{3}$};
\draw (160.48,68.05) node [anchor=north west][inner sep=0.75pt]    {\scriptsize $P$};
\draw (85.59,106.35) node [anchor=north west][inner sep=0.75pt]    {\scriptsize $P_{4}$};
\draw (230.66,107.26) node [anchor=north west][inner sep=0.75pt]    {\scriptsize $P_{5}$};
\end{tikzpicture}\right)\,.\end{split} 
\end{align}
These identities were derived in appendix \ref{app:TQFT_identities} by inserting identity lines and doing fusion and braiding. Topologically, inserting the identity line corresponds to adding what is known in the knot theory literature as an `unknotting tunnel' \cite{rolfsen2003knots}. Next, we remove the `Wilson bubbles' using the identity \eqref{eq:WilsonBubble}, obtaining: 
\begin{equation}
    Z_\text{Vir}\!\left(\begin{tikzpicture}[x=0.75pt,y=0.75pt,yscale=-0.6,xscale=0.6,baseline={([yshift=-.5ex]current bounding box.center)}]
\draw [color={rgb, 255:red, 208; green, 2; blue, 27 }  ,draw opacity=1 ][line width=1.5]    (160,100) -- (220,100) ;
\draw  [draw opacity=0][line width=1.5]  (144.12,73.52) .. controls (153.57,78.57) and (160,88.54) .. (160,100) .. controls (160,116.57) and (146.57,130) .. (130,130) .. controls (113.43,130) and (100,116.57) .. (100,100) .. controls (100,83.57) and (113.2,70.23) .. (129.57,70) -- (130,100) -- cycle ; \draw  [line width=1.5]  (144.12,73.52) .. controls (153.57,78.57) and (160,88.54) .. (160,100) .. controls (160,116.57) and (146.57,130) .. (130,130) .. controls (113.43,130) and (100,116.57) .. (100,100) .. controls (100,83.57) and (113.2,70.23) .. (129.57,70) ;  
\draw  [draw opacity=0][line width=1.5]  (248.05,70.06) .. controls (248.69,70.02) and (249.34,70) .. (250,70) .. controls (266.57,70) and (280,83.43) .. (280,100) .. controls (280,116.57) and (266.57,130) .. (250,130) .. controls (233.43,130) and (220,116.57) .. (220,100) .. controls (220,88.32) and (226.68,78.2) .. (236.42,73.24) -- (250,100) -- cycle ; \draw  [line width=1.5]  (248.05,70.06) .. controls (248.69,70.02) and (249.34,70) .. (250,70) .. controls (266.57,70) and (280,83.43) .. (280,100) .. controls (280,116.57) and (266.57,130) .. (250,130) .. controls (233.43,130) and (220,116.57) .. (220,100) .. controls (220,88.32) and (226.68,78.2) .. (236.42,73.24) ;  
\draw  [draw opacity=0][line width=1.5]  (133.86,121.22) .. controls (131.37,114.63) and (130,107.47) .. (130,100) .. controls (130,66.86) and (156.86,40) .. (190,40) .. controls (223.14,40) and (250,66.86) .. (250,100) .. controls (250,107.59) and (248.59,114.86) .. (246.01,121.55) -- (190,100) -- cycle ; \draw  [line width=1.5]  (133.86,121.22) .. controls (131.37,114.63) and (130,107.47) .. (130,100) .. controls (130,66.86) and (156.86,40) .. (190,40) .. controls (223.14,40) and (250,66.86) .. (250,100) .. controls (250,107.59) and (248.59,114.86) .. (246.01,121.55) ;  
\draw  [draw opacity=0][line width=1.5]  (237.55,136.59) .. controls (226.58,150.83) and (209.36,160) .. (190,160) .. controls (170.42,160) and (153.04,150.62) .. (142.09,136.12) -- (190,100) -- cycle ; \draw  [line width=1.5]  (237.55,136.59) .. controls (226.58,150.83) and (209.36,160) .. (190,160) .. controls (170.42,160) and (153.04,150.62) .. (142.09,136.12) ;  
\draw (183.5,76.9) node [anchor=north west][inner sep=0.75pt]    {\scriptsize $P_{0}$};
\draw (104,89.4) node [anchor=north west][inner sep=0.75pt]    {\scriptsize $P_{1}$};
\draw (252.5,89.4) node [anchor=north west][inner sep=0.75pt]    {\scriptsize $P_{3}$};
\draw (184.5,44.4) node [anchor=north west][inner sep=0.75pt]    {\scriptsize $P$};
\end{tikzpicture}\right) = \frac{\sker{P_1}{P}{P_0}}{\sker{\bbi}{P}{\bbi}} \frac{\mathbb{S}_{P_3P}^*[P_0]}{\sker{\bbi}{P}{\bbi}} \,Z_\text{Vir}\!\left(\begin{tikzpicture}[x=0.75pt,y=0.75pt,yscale=-0.7,xscale=0.7,baseline={([yshift=-0.5ex]current bounding box.center)}]
\draw [color={rgb, 255:red, 208; green, 2; blue, 27 }  ,draw opacity=1 ][line width=1.5]    (321,187) -- (420.67,187.17) ;
\draw  [draw opacity=0][line width=1.5]  (417.99,206.3) .. controls (411.22,227.26) and (392.5,242.33) .. (370.47,242.33) .. controls (342.73,242.33) and (320.24,218.46) .. (320.24,189) .. controls (320.24,159.54) and (342.73,135.67) .. (370.47,135.67) .. controls (398.2,135.67) and (420.69,159.54) .. (420.69,189) .. controls (420.69,195.05) and (419.74,200.87) .. (417.99,206.29) -- (370.47,189) -- cycle ; \draw  [line width=1.5]  (417.99,206.3) .. controls (411.22,227.26) and (392.5,242.33) .. (370.47,242.33) .. controls (342.73,242.33) and (320.24,218.46) .. (320.24,189) .. controls (320.24,159.54) and (342.73,135.67) .. (370.47,135.67) .. controls (398.2,135.67) and (420.69,159.54) .. (420.69,189) .. controls (420.69,195.05) and (419.74,200.87) .. (417.99,206.29) ;  
\draw (363,139.2) node [anchor=north west][inner sep=0.75pt]  [font=\footnotesize]  {$P$};
\draw (363,189.73) node [anchor=north west][inner sep=0.75pt]  [font=\footnotesize]  {$P_0$};
\end{tikzpicture}
\right).
\end{equation}
 Finally, we use the result of \cite{Collier:2024mgv} for evaluating the Virasoro TQFT partition function on the `sunset' diagram:
 \begin{equation}
     Z_\text{Vir}\!\left(\begin{tikzpicture}[x=0.75pt,y=0.75pt,yscale=-0.5,xscale=0.5,baseline={([yshift=-0.5ex]current bounding box.center)}]
\draw [color={rgb, 255:red, 208; green, 2; blue, 27 }  ,draw opacity=1 ][line width=1.5]    (321,187) -- (420.67,187.17) ;
\draw  [draw opacity=0][line width=1.5]  (417.99,206.3) .. controls (411.22,227.26) and (392.5,242.33) .. (370.47,242.33) .. controls (342.73,242.33) and (320.24,218.46) .. (320.24,189) .. controls (320.24,159.54) and (342.73,135.67) .. (370.47,135.67) .. controls (398.2,135.67) and (420.69,159.54) .. (420.69,189) .. controls (420.69,195.05) and (419.74,200.87) .. (417.99,206.29) -- (370.47,189) -- cycle ; \draw  [line width=1.5]  (417.99,206.3) .. controls (411.22,227.26) and (392.5,242.33) .. (370.47,242.33) .. controls (342.73,242.33) and (320.24,218.46) .. (320.24,189) .. controls (320.24,159.54) and (342.73,135.67) .. (370.47,135.67) .. controls (398.2,135.67) and (420.69,159.54) .. (420.69,189) .. controls (420.69,195.05) and (419.74,200.87) .. (417.99,206.29) ;  
\draw (363,139.2) node [anchor=north west][inner sep=0.75pt]  [font=\footnotesize]  {$P$};
\draw (363,189.73) node [anchor=north west][inner sep=0.75pt]  [font=\footnotesize]  {$P_0$};
\end{tikzpicture}
\right) = \frac{1}{C_0(P,P,P_0)}.
 \end{equation}
We can put everything together to obtain the answer for $Z_\text{Vir}(M_3^{\O_0})$ in \eqref{eq:3bdy_step1}. The result can be further massaged by writing $\sker{\bbi}{P}{\bbi}= \rho_0(P)$ and using the formula for swapping the internal indices of the one-point S-kernel \cite{Eberhardt:2023mrq}: 
\begin{equation}\label{eq:S_swap}
    \frac{\mathbb{S}_{P_3P}^*[P_0]}{\rho_0(P)C_0(P,P,P_0)} = \frac{\mathbb{S}_{PP_3}^*[P_0]}{\rho_0(P_3)C_0(P_3,P_3,P_0)}.
\end{equation}
Doing so, we learn that the Virasoro TQFT partition function on the `regulated' three-boundary wormhole $M_3^{\O_0}$ is given by:
\begin{equation}\label{eq:Z_micro}
    Z_\text{Vir}(M_3^{\O_0};P_{1,2,3}) = \frac{1}{\rho_0(P_3)C_0(P_3,P_3,P_0)} \int_0^\infty \dd P \,\frac{\sker{P_1}{P}{P_0}\sker{P_2}{P}{\bbi}\mathbb{S}_{PP_3}^*[P_0]}{\sker{\bbi}{P}{\bbi}}\,.
\end{equation}
 Remarkably, the right-hand side is proportional to the fusion density $\N_{P_0}[P_1,P_2,P_3]$ by virtue of the Virasoro-Verlinde formula \eqref{eq:main_formula}! The proportionality factor is precisely the norm of the torus one-point conformal block in the Hilbert space $\mathcal{H}_{1,1}$ (see \eqref{eq:inner_product}). 
 
 Rewriting the answer in terms of $\widehat{\N}$ in the Racah-Wigner normalization, and multiplying by the anti-holomorphic counterpart, we obtain the microcanonical partition function:
\begin{equation}\label{eq:microcan}
    Z_\text{grav}(M_3^{\O_0};P_{1,2,3},\bar P_{1,2,3}) = \frac{1}{\sqrt{\mathsf{C}_{101}\mathsf{C}_{303}}} \,\widehat{\N}_{P_0}[P_1,P_2,P_3]\,\widehat{\N}_{\bar P_0}[\bar P_1,\bar P_2,\bar P_3].
\end{equation}
Here we defined the normalization factors $\mathsf{C}_{123} \coloneqq C_0(P_1,P_2,P_3)C_0(\bar P_1,\bar P_2,\bar P_3)$.\footnote{Interestingly, these are precisely the `vertex factors' of \cite{Collier:2024mgv}, corresponding to the vertices of the tetrahedron to which the $P_0$ edge is attached; recall eq.\! \eqref{eq:tetrahedron_diagram}.} The result \eqref{eq:microcan} should be seen as the analogue of the compact Chern-Simons result \eqref{eq:CS_simple}, now for the non-compact gauge group $PSL(2,\mathbb{R})\times PSL(2,\mathbb{R})$. 

To calculate the gravity path integral with asymptotic boundaries, one has to integrate the microcanonical partition function against the torus 0- and 1-point conformal blocks, and sum over the boundary modular group. But before doing so, let us discuss the limit $P_0 \to \bbi$ of the result \eqref{eq:microcan}.

\paragraph{Removing the regulator and RMT.} The appearance of the normalization factors in \eqref{eq:microcan} is interesting, because it allows us to take the limit that the external operator $\O_0$ is taken to the identity. In this limit, there is no longer the semiclassical interpretation as a conical defect geometry, but in principle the Virasoro TQFT framework accommodates arbitrarily light matter fields. In particular, our computation above shows that the limit exists and is equal to 1:
\begin{equation}
    \lim_{P_0\to \frac{iQ}{2}} Z_\text{Vir}(M_3^{\O_0};P_{1,2,3}) = \lim_{P_0\to \frac{iQ}{2}} \frac{\fker{P_0}{P_2}{P_3}{P_1}{P_1}{P_3}}{\fker{\bbi}{P_2}{P_3}{P_1}{P_1}{P_3}} = 1.
\end{equation}
We would like to stress that this is a non-trivial result: 
equation \eqref{eq:Z_micro} is of the form $\frac{\infty}{\infty}$ when $P_0=\bbi$, since both the $P$-integral and the normalization factor diverge; however the Virasoro-Verlinde formula shows (using e.g.\! \eqref{eq:fusion_dens}) that the limit exists and is equal to 1. In particular, the result is the same for any (real) value of $P_{1,2,3}$. 

This simple answer is reminiscent of `square root edge' double-scaled random matrix models\footnote{By this we mean that the spectral curve has the form $(x(z),y(z)) = (z^2, \sum_{n=0}^\infty a_n z^{2n+1}$). These models are known to be dual to 2d topological gravity in a general background \cite{Okuyama:2019xbv}. Examples include the Kontsevich-Witten model \cite{Eynard:2014zxa}, JT gravity \cite{Saad:2019lba} and the Virasoro minimal string \cite{Collier:2023cyw}.}, which have a universal form for the leading three-point correlator of resolvents, $\omega_{0,3}(z_1,z_2,z_3)$, which after Laplace transforming in $z_{1,2,3}$ gives $V_{0,3}(\ell_1,\ell_2,\ell_3) = 1$. The connection to random matrix theory can be understood more precisely via the Virasoro minimal string \cite{Collier:2023cyw}, which was shown to $a)$ be dual to a random matrix model with a square root edge and $b)$ describe a chiral half of 3d gravity on manifolds of the form $\Sigma_{g,n}\times S^1$. Indeed, we have for $P_i \in \mathbb{R}_+$:
\begin{equation}
    \mathsf{V}^{(b)}_{0,3}(P_1,P_2,P_3) = Z_{\text{Vir}}(M_3;P_{1,2,3}) = 1,
\end{equation}
where we used the notation of \cite{Collier:2023cyw} for the quantum volumes.\footnote{In general, the Virasoro minimal string is not identical to Virasoro TQFT on $\Sigma_{g,n}\times S^1$, because the former is quotiented by the 2d mapping class group; but this quotient is trivial for $\Sigma_{0,3}$.}

\paragraph{The canonical partition function.}

So far, we computed the partition function on $M_3^{\O_0}$ with finite-size boundaries with specified holonomies $P_{1,2,3}$ along meridians of the boundary tori. A semiclassical $(b\to 0)$ argument due to Verlinde \cite{Verlinde:1989ua} relates the $P_i$ to lengths $\ell_i$ of the boundaries of $\Sigma_{0,3}$ through $P = \frac{\ell}{4\pi b}$. To get asymptotic boundaries, with fixed conformal moduli $\tau_{1,2,3}$, we have to integrate \eqref{eq:microcan} against the appropriate `trumpets', which include the contribution from boundary gravitons. 

For the torus without puncture, the trumpet evaluates to the Virasoro character $\mathcal{F}_{1,0}(P;\tau)\coloneqq \chi_P(\tau)$, as was shown in \cite{Cotler:2020ugk}. For the once-punctured torus, the relevant trumpet factor is the one-point torus conformal block \cite{Collier_2023}. So before summing over modular images, we can write the `seed' canonical partition function at fixed $\tau_{1,2,3}$ as:
\begin{multline}\label{eq:Z_seed}
    Z^{(\text{seed})}_{\text{grav}}(M_3^{\O_0};\tau_{1,2,3},\bar\tau_{1,2,3}) = \Bigg |\int_0^\infty \dd P_1\dd P_2\dd P_3\,\rho_{1,1}(P_1;P_0) \,\rho_{1,1}(P_3;P_0) \\ \times Z_\text{Vir}(M_3^{\O_0};P_{1,2,3}) \,\mathcal{F}_{1,1}(P_1;P_0; \tau_1)\,\mathcal{F}_{1,0}(P_2,\tau_2)\,\mathcal{F}^*_{1,1}(P_3;P_0; \tau_3)\,\Bigg|^2\,.
\end{multline}
Here we used the notation that $|\cdot|^2$ signifies the product of the holomorphic piece with its anti-holomorphic counterpart $P_i\to \bar P_i, \tau_i \to \bar\tau_i$. The integrand in \eqref{eq:Z_seed} includes the appropriate measure factors for the integrals over the one-point blocks, coming from the normalization of the inner product \eqref{eq:inner_product}, $
    \rho_{1,1}(P_i;P_0) \coloneqq \rho_0(P_i)C_0(P_i,P_i,P_0).$
We also took a flat measure in $P_2$ for the integral over the Virasoro character.\footnote{This should be contrasted to the measure that was used to compute the two-boundary `double trumpet' in 3d gravity \cite{Cotler:2020ugk}, which is flat in $\Delta_2$. Our choice of measure aligns with \cite{Collier_2023}, and, as we will show, reproduces the prediction from the ETH-like CFT ensemble of \cite{Belin:2020hea, Chandra:2022bqq,deBoer:2024kat}.} Lastly, we multiplied by $\mathcal{F}_{1,1}^*(\tau_3) = \mathcal{F}_{1,1}(-\tau_3)$, since one of the torus boundaries has the opposite orientation compared to the other two, as can be seen from \eqref{eq:regulated_manifold}. 

Using the Virasoro-Verlinde formula \eqref{eq:Z_micro}, we can simplify equation \eqref{eq:Z_seed} to:
\begin{multline}\label{eq:Z_seed2}
    Z^{(\text{seed})}_{\text{grav}}(M_3^{\O_0};\tau_{1,2,3},\bar \tau_{1,2,3}) \\= \Bigg |\int_0^\infty \dd P \,C_0(P,P,P_0)\,\mathcal{F}_{1,1}(P;P_0; -\tfrac{1}{\tau_1})\,\mathcal{F}_{1,0}(P,-\tfrac{1}{\tau_2})\,\mathcal{F}^*_{1,1}(P;P_0; -\tfrac{1}{\tau_3})\,\Bigg|^2\,.
\end{multline}
 To arrive at this answer, we swapped the arguments of $\mathbb{S}_{P_1P}[P_0]$ using \eqref{eq:S_swap}, and then recognized the $P_{1,2,3}$ integrals as S-transforms of the blocks. Notice the similarity between \eqref{eq:Z_seed2} and the Chern-Simons computation \eqref{eq:CS_result}: the blocks are in the S-dual channel and they are all paired with the same momentum $P$.

\paragraph{Modular completion.} One immediate consequence of \eqref{eq:Z_seed2} is that the seed partition function has the following property under modular T-transformations:\footnote{This property is a consequence of the \emph{bulk} mapping class group: a Dehn twist along a cycle parallel to the $S^1$ direction in $\Sigma_{0,3}\times S^1$ is equivalent to a T-transformation on the $b$-cycle of any of the three boundaries. This can be seen by deforming a $b$-cycle on the outer torus in \eqref{eq:regulated_manifold} through the bulk to a $b$-cycle on either of the grey inner tori.}\vspace{1mm}  
\begin{equation}\label{eq:T-invariance}
    Z^{(\text{seed})}_{\text{grav}}(\mathsf{T}^n\tau_1, \tau_2, \tau_3) = Z^{(\text{seed})}_{\text{grav}}(\tau_1, \mathsf{T}^n\tau_2, \tau_3) = Z^{(\text{seed})}_{\text{grav}}(\tau_1, \tau_2, (\mathsf{T}^*)^{n}\tau_3).\vspace{1mm}
\end{equation}
Here we defined $\mathsf{T}^n = S^{-1}T^nS$, with $S,T$ the standard generators of $\mathrm{PSL}(2,\mathbb{Z})$. We also suppressed the dependence on $\bar\tau_i$ in the notation for clarity. Put differently, two factors of $\mathbb{Z}$ are redundant in the sum over modular images. If we sum over the remaining $\mathbb{Z}$ subgroup, we get a partition function that has a full $\mathbb{Z}^3$ invariance:\vspace{1mm}
\begin{equation}\label{eq:Dehn_sum}
    \widetilde{Z}^{(\text{seed})}_{\text{grav}}(\tau_1, \tau_2, \tau_3) \coloneqq \sum_{n\in \mathbb{Z}}Z^{(\text{seed})}_{\text{grav}}(\mathsf{T}^n\tau_1, \tau_2, \tau_3).
\end{equation}
This sum over $\mathbb{Z}$ implements the spin quantization of the states in the integral over $P,\bar P$ in \eqref{eq:Z_seed2}. Namely, applying the $\mathsf{T}^n$ transform in \eqref{eq:Z_seed2} brings a factor of $\mathbb{T}_P^n = \e^{-2\pi i n \Delta_P}$ inside the $P$ integral, and a factor of $\e^{2\pi i n \bar \Delta_P}$ inside the $\bar P$ integral. Then, the sum over $n\in\mathbb{Z}$ can be done using the Poisson summation formula, giving
\begin{equation}
    \sum_{n\in \mathbb{Z}} \mathbb{T}_P^n (\mathbb{T}^*_{\bar P})^n = \sum_{n\in \mathbb{Z}} \e^{-2\pi i n (P^2-\bar P^2)} = \sum_{n\in\mathbb{Z}}\delta(J-n) \eqqcolon \delta_{\mathbb{Z}}(J),
\end{equation}
where we recognized $J = P^2-\bar P^2$. In the integral for $\widetilde{Z}^{\text{(seed)}}$ that arises in this way, one can then change variables from $P,\bar P$ to $J$ and $E = \frac{c}{12} + P^2+\bar P^2$, so that the Dirac comb will turn the $J$-integral into a sum over integer $J$'s. The remaining $E$-integral is strictly above the black hole threshold $E\geq \frac{c}{12}$, precisely as expected for a Euclidean wormhole contribution \cite{Schlenker:2022dyo}. 

To obtain a fully modular invariant object, we follow the prescription of \cite{Collier_2023} and sum over the remaining boundary mapping class group. As the seed is already invariant under $\mathbb{Z}^3$, the modular completion is a sum over $G = \left(\mathrm{PSL}(2,\mathbb{Z})/\mathbb{Z}\right)^3$:
\begin{equation}\label{eq:Z_3bdy_final}
    Z_{\text{grav}}(M_3^{\O_0};\tau_{1,2,3},\bar \tau_{1,2,3}) = \sum_{(\gamma_1,\gamma_2,\gamma_3) \in G} \widetilde Z^{(\text{seed})}_{\text{grav}}(\gamma_1\cdot\tau_1,\gamma_2\cdot \tau_2,\gamma_3\cdot \tau_3).
\end{equation}
One can view the modular images as coming from bulk 3-manifolds that are topologically equivalent to $M_3^{\O_0}$, but where each `trumpet' is glued with a relative modular transformation $\gamma_i$.\footnote{More precisely, the meridian curve $[m]$ on a geodesic boundary of the `microcanonical' wormhole (the convex core) is identified with a curve $p[m]+q[l]$ on the geodesic boundary of the trumpet, where $p$ and $q$ are coprime positive integers.} 

Notice that the structure of the modular sum differs from that of the two-boundary torus wormholes encountered in e.g.\! \cite{Cotler:2020ugk}. For the two-boundary case, there is only a single (relative) modular sum, since the modular transformation on one boundary can be `undone' by an opposite modular transformation on the other boundary. For the three-boundary case, this is no longer possible, since there is a topological obstruction to deforming a general large diffeomorphism of a given boundary through the bulk to any of the other boundaries (with the exception of the subgroup of T-transforms that we identified in \eqref{eq:T-invariance}).    

\paragraph{Match to the large-$c$ OPE ensemble.}
Since $M_3^{\O_0}$ is a Euclidean wormhole between two punctured tori and one non-punctured torus, it is natural to expect that $Z_{\text{grav}}(M_3^{\O_0})$ computes the connected correlation function:
\begin{equation}\label{eq:cubic_moment}
\overline{\expval{\O_0}_{\tau_1,\bar\tau_1}\expval{\bbi}_{\tau_2,\bar\tau_2}\expval{\O_0}_{\tau_3,\bar\tau_3}}^{\,\text{conn.}}
\end{equation}
in the statistical ensemble of OPE coefficients introduced in \cite{Chandra:2022bqq}. However, if we take the strictly Gaussian \emph{ansatz} of \cite{Chandra:2022bqq}, this moment would factorize as $Z(\tau_2)\, \overline{\expval{\O_0}_{\tau_1}\!\!\expval{\O_0}_{\tau_3}}$, so the cubic cumulant vanishes. Indeed, it was argued in \cite{deBoer:2024kat} that non-Gaussian higher statistical moments are \emph{necessary} to explain multiboundary wormholes in 3d gravity.
These non-Gaussian moments are further suppressed in $\e^{-S}$ (where $S$ is the Cardy entropy) compared to the Gaussian contraction, but they are still important for the consistency of the OPE ensemble \cite{Belin:2021ryy}. 

Interestingly, requiring that the cubic moment \eqref{eq:cubic_moment} matches to our three-boundary wormhole computation gives a non-trivial prediction for the statistical correlation between the density of states and two OPE coefficients:
\begin{equation}\label{eq:cubic_moment2}
    \overline{\rho(P_2,\bar P_2)\,C_{1\O 1}C_{3\O 3}}^{\,\text{conn.}} = \left|\frac{\delta(P_1-P_3)}{\rho_0(P_1)}\frac{\delta(P_2-P_3)}{\rho_0(P_2)} \,C_0(P_3,P_3,P_\O)\right|^2\, + \dots
\end{equation}
The dots signify other subleading contractions that we will discuss momentarily. 

The cubic moment $\overline{\rho CC}$ has not appeared in the literature before, and it is interesting for the following reason. When we expand the product of expectation values \eqref{eq:cubic_moment} in conformal blocks, we get
\begin{multline}\label{eq:block_expansion}
\overline{\expval{\O}_{\tau_1}\expval{\bbi}_{\tau_2}\expval{\O}_{\tau_3}} = \int \prod_{j=1}^3\dd \Delta_{j}\dd \bar \Delta_{j} \,\overline{\rho(\Delta_1,\bar \Delta_1)\rho(\Delta_2,\bar \Delta_2)\rho(\Delta_3,\bar \Delta_3)C_{1\O 1}C_{3\O3}} \\ \times \left|\mathcal{F}_{1,1}(\Delta_1;\Delta_\O; -\tfrac{1}{\tau_1})\,\mathcal{F}_{1,0}(\Delta_2,-\tfrac{1}{\tau_2})\,\mathcal{F}^*_{1,1}(\Delta_3;\Delta_\O; -\tfrac{1}{\tau_3})\right|^2.
\end{multline}
We expanded the blocks in the S-dual channel, which is allowed since the left-hand side of \eqref{eq:block_expansion} is modular invariant in $\tau_1,\tau_2$ and $\tau_3$. At this point, we have not made any approximation, and we defined $\rho(\Delta,\bar\Delta) = \sum_{h,\bar h \in \mathcal{S}} \delta(h-\Delta)\delta(\bar h-\bar \Delta)$, where $\mathcal{S}$ is the exact primary spectrum of the CFT. 

Now, if $\overline{(\bullet)}$ denotes the average in any consistent statistical ensemble of $\rho$ and $C$'s, then all lower moments should appear as terms in the average. We can organize their contributions as:
\begin{center}
\begin{tabular}{ l l }
 Fully factorized contraction: & $\bar{\rho}\,\bar{\rho}\,\bar{\rho} \,\overline{C^2}$  \\[0.5em] 
 Single contractions: & $\bar\rho\,\overline{\rho^2}\,\overline{C^2}$ and $\bar{\rho}\,\bar{\rho}\,\overline{\rho C^2}$.  \\[0.5em]  
 Double contractions: & $\overline{\rho^2}\,\overline{\rho C^2}$ and $\bar{\rho}\,\overline{\rho^2 C^2}$. \\[0.5em]
 Fully connected contraction: & $\overline{\rho^3 C^2}$.
\end{tabular}
\end{center}
The contractions that involve higher moments of $\rho$ arise from spectral correlations, which are believed to arise due to \emph{off-shell} Euclidean wormholes in the gravitational path integral \cite{toappear}. Such spectral correlations can usually be disregarded at early enough times where the Eigenstate Thermalization Hypothesis applies (for ergodic systems, spectral correlations become important after the Thouless time \cite{Altland:2020ccq}). So if we ignore all higher spectral correlations, then the only non-trivial contraction pattern other than the fully factorized contraction is
\begin{equation}\label{eq:coarse_grained}
    \bar{\rho}\,\bar{\rho}\,\overline{\rho C^2}^{\,\text{conn.}} \approx \rho_0 \,\rho_0 \,\overline{\rho C^2}^{\,\text{conn.}},
\end{equation}
which is precisely the type of cubic contraction that we have found in \eqref{eq:cubic_moment}. This contraction sits at an interesting intermediate point in the statistical description of CFT data, where we do not take into account eigenvalue repulsion, but we do get non-trivial correlations between $\rho$ and $C$.  

In \eqref{eq:coarse_grained}, we approximated the coarse-grained density of states $\bar\rho$ by the smooth Cardy density $\rho_0(\Delta_1,\bar \Delta)\dd \Delta \dd \bar\Delta \coloneqq \delta_{\mathbb{Z}}(J)\rho_0(P)\rho_0(\bar P) \dd P \dd \bar P$.\footnote{This is a slight refinement of the OPE ensemble of \cite{Chandra:2022bqq}, taking into account spin quantization.} This approximation is justified for CFT's with a sparse light spectrum, a gap of the order $c \gg 1$, and a dense spectrum of states above the black hole threshold \cite{Hartman:2014oaa}. If we combine this with the cubic moment \eqref{eq:cubic_moment2}, then we can approximate the CFT average \eqref{eq:cubic_moment} by:\footnote{To arrive at this answer, we approximated the CFT spectrum by the identity $\bbi$ plus a continuum of black hole states with $\Delta,\bar\Delta \geq \frac{c-1}{24}$. We also used that $C_{\bbi \O \bbi} = 0$, meaning that the contributions from $\Delta_{1,3} = 0$ to \eqref{eq:block_expansion} vanish. Finally, we used that the contribution from $\Delta_2 = 0$ drops out of the connected average.}
\begin{multline}\label{eq:CFT_ensemble_3}
\overline{\expval{\O}_{\tau_1}\expval{\bbi}_{\tau_2}\expval{\O}_{\tau_3}}^{\,\text{conn.}} \approx  \int_0^\infty \dd P \dd \bar P \,\delta_{\mathbb{Z}}(J)\, C_0(P,P,P_\O)C_0(\bar P,\bar P,\bar P_\O)\\ \times\left|\mathcal{F}_{1,1}(P;P_\O; -\tfrac{1}{\tau_1})\,\mathcal{F}_{1,0}(P,-\tfrac{1}{\tau_2})\,\mathcal{F}^*_{1,1}(P;P_\O; -\tfrac{1}{\tau_3})\right|^2 .
\end{multline}
This precisely matches to the answer obtained in equations \eqref{eq:Z_seed2} and \eqref{eq:Dehn_sum} for the 3d gravity partition function on the three-boundary wormhole $M_3^{\O_0}$. Put differently, if one assumes that an averaged version of AdS$_3$/CFT$_2$ exists for pure gravity, then consistency requires a statistical correlation for $\overline{\rho C^2}$ of the leading form \eqref{eq:cubic_moment2}. 

We can also give an independent CFT motivation for the cubic moment $\eqref{eq:cubic_moment2}$ by appealing to the concept of \emph{typicality}. For sufficiently chaotic quantum systems, it is expected that in the dense part of the spectrum, there is approximate unitary invariance in microcanonical energy bands. This general expectation is at the core of the Eigenstate Thermalization Hypothesis \cite{Foini:2018sdb}. A small amount of unitary averaging then fixes the index structure of the OPE coefficients in \eqref{eq:cubic_moment} to the `diagonal' contraction:
\begin{equation}
\begin{tikzpicture}[x=0.75pt,y=0.75pt,yscale=-1,xscale=1,baseline={([yshift=-.5ex]current bounding box.center)}]
\draw [thin]   (152.26,160.08) -- (168.91,160.08) ;
\draw [thin]   (152.26,160.59) -- (152.26,153.21) ;
\draw [thin]   (168.49,160.5) -- (168.49,153.12) ;
\draw [thin]   (183.48,159.9) -- (200.13,159.9) ;
\draw [thin]   (183.48,160.41) -- (183.48,153.03) ;
\draw [thin]   (199.71,160.32) -- (199.71,152.94) ;
\draw [thin]   (137.4,168.71) -- (215,168.94) ;
\draw [thin]   (137.29,153.42) -- (137.4,169.23) ;
\draw [thin]   (214.58,153.07) -- (214.67,168.88) ;
\draw [thin]   (122.17,130) -- (219,130) ;
\draw (121,132.4) node [anchor=north west][inner sep=0.75pt]    {$C_{1\O1} C_{2\bbi 2} C_{3\O3}$};
\end{tikzpicture}
\end{equation}
which sets $\Delta_1=\Delta_2=\Delta_3$.
It was moreover shown in \cite{deBoer:2024kat} that the cubic moment is suppressed by a factor $e^{-S}$ of compared to the Gaussian contraction. Our proposal for \eqref{eq:cubic_moment2} satisfies both conditions: the selection rule $\delta_{13}$ is imposed by the continuous delta function $\frac{\delta(P_1-P_3)}{\rho_0(P_1)}$, and there is an extra factor of $1/\rho_0(P_2,\bar P_2) \sim e^{-S_{\text{Cardy}}}$ compared to the Gaussian moment. Finally, the precise functional form of \eqref{eq:cubic_moment2} can be obtained from imposing crossing symmetry on the CFT's torus three-point function. Indeed, taking $\O_1 = \O_3$ and $\O_2 = \bbi$ in equation (2.44) of \cite{deBoer:2024kat} we retrieve \eqref{eq:cubic_moment2}. 

\paragraph{Corrections from modular images.}
Let us now briefly discuss the corrections to the cubic moment \eqref{eq:cubic_moment2}. From equation \eqref{eq:CFT_ensemble_3} we can see that the `diagonal' approximation does not preserve modular invariance in $\tau_{1,2,3}$, $\bar\tau_{1,2,3}$. This is a general feature of the large-$c$ OPE ensemble for wormhole amplitudes --- for a similar discussion, see section 3.2 in \cite{deBoer:2024kat}. We can restore modular invariance by adding to the OPE contraction \eqref{eq:cubic_moment2} all the corrections from the modular images in the triple Poincar\'e sum \eqref{eq:Z_3bdy_final}. For example, the contributions
\begin{equation}\label{eq:cubic_moment3}
    \overline{\rho(P_2,\bar P_2)\,C_{1\O 1}C_{3\O 3}}^{\,\text{conn.}} \supset \left|\frac{\delta(P_2-P_3)}{\rho_0(P_2)} \frac{\sker{P_3}{P_1}{P_\O}}{\rho_0(P_1)}\,C_0(P_3,P_3,P_\O)\right|^2\, + (1\leftrightarrow 3)
\end{equation}
give rise to the modular images $\tau_1 \to S\cdot\tau_1$ and $\tau_3 \to S\cdot\tau_3$. Note that thanks to the symmetry relation \eqref{eq:S_swap}, the second term in the above expression only differs from the first by the substitution $\delta(P_2-P_3)\mapsto \delta(P_2-P_1)$. 

We can in principle add all modular images in this way. However, it is expected that at high enough temperatures and large $c$ the diagonal contribution \eqref{eq:CFT_ensemble_3} dominates. A heuristic argument goes as follows. A simple analysis shows that $C_0(P,P,P_\O)$ is peaked at $\Delta_P^* \approx \Delta_\O^2/\Delta_{t}$, where $\Delta_{t} \coloneqq \frac{c-1}{24}$ is the black hole threshold. This value of $\Delta_P^*$ is very small for any sub-threshold $\Delta_\O$, when $c$ is large. For larger values of $\Delta_P$, the $C_0$ function decays exponentially. So the largest contribution to the modular sum will come from those terms where the $P$-integral is dominated by a saddle point close to $\Delta_{P}^*$. At high temperatures $\mathrm{Im}(\tau_i)\to 0$, this is the case when all the conformal blocks are in the S-dual channel. This selects \eqref{eq:CFT_ensemble_3} as the dominant contribution.

\paragraph{The off-shell 3-boundary wormhole.}
 We have seen that the microcanonical partition function on $\Sigma_{0,3}\times S^1$ is finite in the limit that the matter insertion $\mathcal{W}_0$ is removed. However, the canonical partition function with fixed boundary moduli diverges, as can be seen from eq.\! \eqref{eq:Z_seed2}: the function $C_0(P,P,P_\O)$ has a simple pole at $P_\O = \frac{iQ}{2}$. The divergence comes from the factors of $\rho_{1,1}$ in \eqref{eq:Z_seed}, in exactly the same way that the two-boundary punctured torus wormhole partition function diverges as $\O_0 \to \bbi$ \cite{Yan:2023rjh}. This divergence signals the fact that the 3-boundary torus wormhole without matter is off-shell: an analysis similar to that of Cotler and Jensen \cite{Cotler:2020ugk} should be done to determine the correct wormhole partition function in 3d gravity, which would contribute to the ``RMT$_2$'' statistics of $\overline{\rho^3}$ \cite{DiUbaldo:2023qli, Haehl:2023mhf}. We leave this as a problem for future work.

\section{Comments on general non-rational CFTs}\label{sec:universal}

The main formula \eqref{eq:main_formula} (or its variant \eqref{eq:intro_6j}) can be viewed as a mathematical identity between the $\mathbb{S}$ and $\mathbb{F}$ kernels which have independent and concrete definitions as meromorphic functions \cite{Ponsot:1999uf, Teschner:2003at}. Our assertion so far, however, is that it represents something more profound. Crucially, these crossing kernels are the unique solutions\footnote{\label{foot:cleq1kernels}As it was explained in \cite{Eberhardt:2023mrq} the uniqueness argument relies crucially on the assumptions of meromorphicity in the $P$ variables and continuity in $b\in (0,1]$. It is useful to contrast this with the situation at $c\leq 1$ where it was shown in \cite{Ribault:2023vqs} that the analogous unique meromorphic and continuous in $b\in i\mathbb{R}_{(0,1]}$ solutions of the shift relations lead to \textit{non-physical} crossing kernels (i.e. kernels that do not implement the fusion of $c\leq1$ blocks). It is still an open question to date to determine the \textit{physical} $\mathbb{S},\mathbb{F}$ kernels for $c\leq1$ in the same way that \cite{Ponsot:1999uf,Teschner:2003at} did for $c\geq 25$.} that satisfy the Moore-Seiberg consistency relations for $c\geq 25$ 2d CFTs \cite{Eberhardt:2023mrq}, which makes them universal quantities in any theory with these values of the central charge. Even more, as we have seen, the derivation of \eqref{eq:main_formula} from both the Virasoro TQFT perspective (c.f. Section \ref{sec:VTQFT_proof}) and the (Liouville) CFT perspective (c.f. App. \ref{sec:CFTproof}) proceeded through a natural extension of the steps that led to the derivation of the original Verlinde formula in Chern-Simons theory \cite{Witten:1988hf} and RCFTs \cite{Moore:1988uz, Moore:1988qv}, respectively. As such, it is reasonable to posit that our main formula encodes some structure related to the fusion coefficients in non-rational 2d CFTs with just Virasoro symmetry which we wish to make precise in the current section. By `non-rational' here we basically refer to theories with an infinite, yet \textit{discrete}, spectra of representations with respect to the chiral and anti-chiral Virasoro algebra. 

To set the stage, we first review some general facts about the fusion coefficients in section \ref{sec:basicsfusion}. We then return to the Virasoro-Verlinde formula and ask whether it defines a fusion product. In the limit where the `regulator' $\mathcal{O}_0$ becomes the identity operator, we argue that our formula connects with a particular fusion product considered previously by J.Teschner in \cite{Teschner:2008qh}, which we newly analyze in section \ref{sec:analytic_Liouville_fp}. We are then led to conjecture, in section \ref{sec:conjectureBoot}, a natural generalization of our main formula to non-rational theories which properly accounts for the discreteness of the allowed representations. Given such formula, we show that there is a universal contribution to the fusion numbers of such theories that is captured by two copies of the Virasoro-Verlinde formula. We also discuss some novel bootstrap implications that arise from the conjectured formula.

\subsection{Fusion coefficients in CFTs with only Virasoro symmetry}\label{sec:basicsfusion}
Let us review briefly what we mean by fusion coefficients in 2d CFTs with just Virasoro symmetry. By definition the fusion coefficients $N_{ij}\text{}^k\in \mathbb{Z}_{\geq 0}$ are commonly associated with the \textit{chiral} algebra of the theory (in this case either $\text{Vir}_{(\text{left})}$ or $\text{Vir}_{(\text{right})}$) and encode the multiplicities in which a given representation $R_k$ of, say, $ \text{Vir}_{(\text{left})}$ appears in the fusion of two other representations $R_i,R_j$. As it is known from the plethora of examples in RCFTs, in a diagonal theory --- meaning a theory with only scalar operators --- the total \textit{fusion number} associated to the full left-right representations factorizes as $N_{\text{tot.}} = N_{\text{left}}\times N_{\text{right}}$, where $N_{\text{left}}\equiv N_{ij}\text{}^k$ and $N_{\text{right}}\equiv N_{\bar{i} \bar{j}}\text{}^{\bar{k}}$ (corresponding to the representations $R_{\bar{i}},R_{\bar{j}},R_{\bar{k}}\in \text{Vir}_{(\text{right})}$). On the other hand, in a non-diagonal theory, the corresponding fusion rules must take into account the asymmetric left-right pairing of the Virasoro primaries, and hence will not be chiral in general. In this case, one simply defines\footnote{\label{foot:commentRCFTVerlinde}In the case of non-diagonal RCFTs, one usually proceeds further by defining \textit{extended-symmetry characters} which re-organize e.g. the non-diagonal partition function of the theory to a diagonal modular invariant in those extended characters. The associated \textit{extended fusion numbers} then factorize into left and right fusion coefficients, just like in the usual diagonal case, which are in turn described by an `extended' Verlinde formula.} a single fusion number as $N_{\text{(tot.)}}=N_{(i,\bar{i}),(j,\bar{j})}\text{}^{(k,\bar{k})}$. In summary,
\begin{equation*}
\bea
\text{diagonal theories:} \ \ \ \ \ \ \ N_{\text{tot.}} &= N_{ij}\text{}^k\times N_{\bar{i} \bar{j}}\text{}^{\bar{k}}\ \ \ \ \ \in\{0,1\} \\ 
\text{non-diagonal theories:} \ \ \ \ \ \ \ N_{\text{tot.}} &= N_{(i,\bar{i}),(j,\bar{j})}\text{}^{(k,\bar{k})}\ \ \ \ \in\{0,1\} \ .
\eea
\end{equation*}
The condition that the fusion number of $(R_i,R_{\bar{i}}), (R_j,R_{\bar{j}})\in \text{Vir}_{\text{(left})}\times \text{Vir}_{(\text{right})}$ contains the representation $(R_k,R_{\bar{k}})$ is equivalent with the associated three-point structure constant $C_{(i,\bar{i}),(j,\bar{j})}\text{}^{(k,\bar{k})}$ being non-trivial. In particular, the statement is that the fusion number coincides with the \textit{dimension} of the space of three-point structure constants involving the aforementioned representations.

In the case where the chiral symmetry algebra is just Virasoro the fusion coefficients can take values $N_{ij}\text{}^{k}=\{0,1\}$. In other words, absent any extra symmetry, given three representations of the Virasoro algebra the three-point structure constants are either unique or non-existent. This essentially relies on the structure of the Virasoro Ward identities which imply that three-point functions of descendant fields are completely determined from the three-point functions of the corresponding primary fields (see e.g. \cite{Ribault:2014hia}).\footnote{The situation is different in e.g. RCFTs with extended symmetry or theories with $\mathcal{W}_n$ symmetry algebras where the fusion coefficients can take values larger than one (see e.g. \cite{Bouwknegt:1992wg, DiFrancesco:1997nk}).} Based on that, it is clear that the fusion numbers are in general determined in part by symmetry. In contrast, the theory-specific information is encoded on the actual \textit{support} of the fusion numbers, which in turn depends on the type of representations of the Virasoro algebra that we are studying. Here we are primarily concerned with unitary (or `positive energy') representations characterized by central charge $c>0$ and chiral conformal dimensions $\Delta,\bar{\Delta}\geq 0$, thereby setting aside several important classes of theories such as non-unitary minimal models or logarithmic CFTs.\footnote{For a discussion on the analog of the Verlinde formula in logarithmic CFTs see \cite{Ridout:2014yfa}.} The role of the type of the representation is pivotal in the fusion rules as can be seen already from the well-known case of unitary Virasoro minimal models with $c<1$: the corresponding highest-weight representations are reducible due to the presence of null states which results into a finite number of primary fields propagating in the fusion rules \cite{Belavin:1984vu}.  On the other hand, for $c>1$ it is known that the only degenerate representation occurs for the identity field with $\Delta=\bar{\Delta}=0$, whereas all other highest-weight representations with $\Delta,\bar{\Delta}>0$ are non-degenerate. We will restrict attention to exactly those, in particular for $c\geq 25$ (or $0<b\leq 1$). It is in this sense that we will refer to the fusion numbers/rules of non-rational theories, where by `non-rational' we specifically mean theories with $c\geq25$ and an infinite number of unitary highest-weight representations labeled by \textit{discrete} $\Delta,\bar{\Delta}>0$, together with a \textit{unique} degenerate identity module with $\Delta_0=\bar{\Delta_0}=0$. As usual, in terms of the Liouville momentum labels $P,\bar{P}$ it will be convenient to `split' the representations into `light' (sometimes also called `complementary series') which occur when $P,\bar{P}\in i\mathbb{R}_{(0,\frac{Q}{2}]}$ or $0\leq \Delta,\bar{\Delta}<\frac{c-1}{24}$, and `heavy' (sometimes also called `principal series') which occur when $P,\bar{P}\in \mathbb{R}$ or $\Delta,\bar{\Delta}\geq \frac{c-1}{24}$.
\subsection{The analytic Liouville fusion product}\label{sec:analytic_Liouville_fp}

After these preparations let us return to the Virasoro-Verlinde formula, in particular its form in the $6j$ normalization \eqref{eq:VV_6j}. We repeat it here in an equivalent form\footnote{We have renamed $P_{1,2,3}\equiv P_{i,k,j}, P_0\equiv z$ and we divided both sides by the normalization factors $\sqrt{C_0(z,P_i,P_i)C_0(z,P_j,P_j)}$; recall equation \eqref{eq:boundedness}.}: 
\be\label{eq:formula_revisitL}
\N^{\text{(L)}}_{ij}(P_k|z)= 
\int_0^\infty \dd \mu(P) \left(\frac{\hat{\mathbb{S}}_{P_iP}[z]}{\sqrt{C_0(z,P_i,P_i)}}\right)
\frac{\mathbb{S}_{P_kP}[\bbi]}{\mathbb{S}_{\bbi P}[\bbi]}
\left(\frac{\hat{\mathbb{S}}_{PP_j}^*[z]}{\sqrt{C_0(z,P_j,P_j)}}\right)
\ee
where
\be\label{eq:N6jfull}
\N^{\text{(L)}}_{ij}(P_k|z) := \left(\rho_0(P_k)C_0(P_i,P_j,P_k)\right)^{-1}\,\fker{z,}{P_k}{P_j}{P_i}{P_i}{P_j} \ .
\ee
Normalizing both sides in this manner helps circumvent the various singularities described in section \ref{sec:properties} and further reveals the physical meaning of the formula, as we now discuss.

Given the striking resemblance of \eqref{eq:formula_revisitL} with the ordinary Verlinde formula it is tempting to suggest that $\N^{\text{(L)}}_{ij}(P_k|z)$ defines a particular fusion product within the unitary highest-weight representations of the Virasoro algebra with $c\geq 25$. For arbitrary $z$, this is not quite true for the simple reason that 
$\N^{\text{(L)}}_{ij}(P_k|z)$ can take arbitrary values other than $\{0,1\}$ (which are the expected values, as we discussed, of the fusion coefficients in a theory with just Virasoro symmetry). The situation, on the other hand, is better in the limit $z\rightarrow \frac{iQ}{2}$. Indeed, we see right away that\footnote{From the LHS of \eqref{eq:formula_revisitL}, this is obvious. From the RHS, this result is attained from a cancellation between a pinching singularity of the integral and a zero coming from an overall factor of $C_0(z,P_j,P_j)^{-1}$, when $z\rightarrow \frac{iQ}{2}$.  } 
$$\lim_{z\rightarrow \frac{iQ}{2}}\N^{\text{(L)}}_{ij}(P_k|z)=1 .
$$
The interpretation of this `1' should be handled with caution though, as its derivation implicitly makes some assumption about the \textit{support} of the representations involved. In particular, this quick calculation obscures any \textit{absence} of particular representations in a given fusion (i.e. cases where the fusion coefficient is zero).

To be more rigorous, for two unitary representations $R_{P_i}$ and $R_{P_j}$ of, say, Vir$\text{}_{\text{(left)}}$ we can define a formal \textit{fusion product} $\boxtimes: (R,R)\rightarrow R$ inspired by \eqref{eq:formula_revisitL} as follows:
\be\label{eq:fusion_product}
\bea
R_{P_i} \boxtimes R_{P_j} \simeq \int_{\mathbb{U}}^{\oplus} \dd \mu(P) \ R_{P} \otimes \text{Hom}^{\text{(L)}}\left(R_{P_i} \boxtimes R_{P_j};R_{P}\right) . 
\eea
\ee
Here $\simeq$ means unitary equivalence, and $\mathbb{U}$ is some measure set of unitary representations of \text{Vir}$\text{}_{\text{(left)}}$ (of given $c\geq 25$) that we need to specify. Given the original connection of our formula with Virasoro TQFT and Liouville theory, it is natural to consider $\mathbb{U}=\mathbb{R}_+$, i.e. the (chiral) continuous spectrum of normalizable states in the Hilbert space of Liouville theory. The measure then is $\dd\mu(P)=\dd P \rho_0(P)$ which is the standard continuous Plancherel measure in the space of representations of the Virasoro algebra. Finally, $\text{Hom}^{\text{(L)}}\left(R_{P_i} \boxtimes R_{P_j};R_{P}\right)$ is the space of homomorphisms $R_{P_i}\boxtimes R_{P_j}\rightarrow R_P$, which is isomorphic to the Hilbert space of (chiral) conformal blocks on the three-punctured sphere (or, more precisely, of the so-called `intertwiners'\footnote{ The role of the intertwiners (or `intertwining operators') was originally emphasized in \cite{Moore:1988uz} (see also \cite{Gaberdiel:1993td,Gaberdiel:1993mt}) to stress the fact that the fusion product $\boxtimes$ is \textit{not} a tensor product (usually denoted by $\otimes$). The tensor product of two representations is a representation that has as central charge the \textit{sum} of the two initial central charges. In the fusion product we are interested in obtaining a representation with the same central charge.}) and has dimension
\be
\text{dim}\left[\text{Hom}^{\text{(L)}}\left(R_{P_i} \boxtimes R_{P_j};R_{P}\right)\right]= \lim_{z\to \frac{iQ}{2}}\N^{\text{(L)}}_{ij}(P|z) = 1 \ .
\ee
As emphasized in section \ref{sec:properties}, for arbitrary $z\in i\mathbb{R}_{(0,\frac{Q}{2})}$ the quantity $\N^{\text{(L)}}_{ij}(P|z)$ is a positive definite number and can thus be regarded in our setup as a \textit{regularized dimension} for the space of three-point conformal blocks on the sphere. This is, however, physically relevant for the Virasoro case only in the limit $z\rightarrow \frac{iQ}{2}$, where $\N^{\text{(L)}}_{ij}(P|z)\rightarrow 1$. Similarly, on the space of three-point conformal blocks one can formally define a \textit{metric} which is proportional to the physical (chiral) structure constants $C_{ij}\text{}^P$. From \eqref{eq:formula_revisitL}, it is suggested that there is a \textit{regularized metric} which reads $C_0(P_i,P_j,P)\N^{\text{(L)}}_{ij}(P|z)=\rho_0(P)^{-1}\fker{z,}{P}{P_j}{P_i}{P_i}{P_j}$. However this, again, only makes sense as a physical structure constant in the limit $z\rightarrow \frac{iQ}{2}$, where it reduces to $C_0(P_i,P_j,P)$.

The fusion product \eqref{eq:fusion_product} with the aforementioned choices has interestingly appeared before in \cite{Teschner:2008qh}\footnote{See section 7.2 thereof. The function $D(\alpha_3,\alpha_2,\alpha_1)$ of \cite{Teschner:2008qh} is up to trivial factors equal to $C_0$.}. 
 It enjoys various nice properties that we now discuss. First, the identity selection rule is automatically satisfied since the metric in $\text{Hom}^{\text{(L)}}$ obeys $\lim_{P_i\rightarrow P_{\bbi}}C_0(P_i,P_j,P)=\rho_0(P)^{-1}\delta(P-P_j)$. For general values of $P_i,P_j$, the metric is permutation-symmetric and hence the fusion product \eqref{eq:fusion_product} is \textit{commutative}. In addition, for $\mathbb{U}=\mathbb{R}_+$ the \textit{associativity} property holds. This follows directly from crossing symmetry of four-point functions in Liouville theory and the existence of the fusion kernel $\mathbb{F}$. Indeed, we compute
\be\label{eq:associativity_fusionprod}
\bea
\left(R_{P_1} \boxtimes R_{P_2}\right) \boxtimes R_{P_3} &\simeq \int_{\mathbb{U}}^{\oplus} \dd\mu (P) \ R_{P} \otimes \text{Hom}^{\text{(s)}}\left(R_{P_1} \boxtimes R_{P_2}\boxtimes R_{P_3};R_P
\right)  \\
R_{P_1} \boxtimes \left(R_{P_2} \boxtimes R_{P_3}\right) &\simeq \int_{\mathbb{U}}^{\oplus} \dd\mu (P) \ R_{P} \otimes \text{Hom}^{\text{(t)}}\left(R_{P_1} \boxtimes R_{P_2}\boxtimes R_{P_3};R_P
\right) 
\eea
\ee
where $\text{Hom}^{\text{(s,t)}}$ are isomorphic to the (infinite-dimensional) spaces of the $s,t-$channels four-point Liouville conformal blocks, with \textit{metrics} given by $C_0(P_1,P_2,P_s)C_0(P_s,P_3,P)$ and $C_0(P_3,P_2,P_t)C_0(P_t,P_1,P)$, respectively. The unitary equivalence of the two expressions in \eqref{eq:associativity_fusionprod} for $\mathbb{U}=\mathbb{R}_+$ is therefore guaranteed from crossing symmetry in Liouville theory, or equivalently, from the existence of a unitary fusion kernel $\mathbb{F}$ that maps $t$-channel Liouville conformal blocks to $s$-channel ones \cite{Teschner:2008qh}. We will refer to the fusion product \eqref{eq:fusion_product} with $\dd\mu(P)=\dd P\rho_0(P)$, $\mathbb{U}=\mathbb{R}_+$ and metric in the space $\text{Hom}^{\text{(L)}}$ given by $C_0$, as the \textit{analytic Liouville fusion product}. 

In order to understand further the support of its measure set $\mathbb{U}$ we will analyze the OPE between chiral vertex operators $V_{P_i},V_{P_j}$. In Liouville theory, the standard form of the OPE reads   
\be\label{eq:OPE_Liouville}
V_{P_i}(x,\bar{x})V_{P_j}(0) \sim \int_{\mathbb{R}_+} \frac{\dd P}{B(P)} \   C_{\text{DOZZ}}(P_i, P_j, P) \ (x\bar{x})^{\Delta-\Delta_i-\Delta_j} \left[V_P\right](0) \ .
\ee
Here $B(P)$ is a given two-point function normalization, $C_{\text{DOZZ}}(P_1,P_2,P_3)$ is the `DOZZ formula' compatible with this two-point function, and $\left[V_P\right]$ denotes the chain operator that contains all the contributions from the holomorphic descendants of the primary $V_P$. It can be shown using standard methods that if we normalize our chiral vertex operators such that $B(P)=\rho_0^{-1}(P)$, then $C_{\text{DOZZ}}$ coincides with the function $C_0$ (see appendix B.1 of \cite{Collier_2023}). We then have
\be\label{eq:OPE_LiouvC0norm}
\bea
\hat{V}_{P_i}(x,\bar{x})\hat{V}_{P_j}(0) &\sim \int_{\mathbb{R}_+} \dd P \rho_0(P) C_{0}(P_i, P_j, P) \ (x\bar{x})^{\Delta-\Delta_i-\Delta_j} [\hat{V}_P](0)
\eea
\ee
which is exactly the map of the fusion product \eqref{eq:fusion_product} to the corresponding space of chiral vertex operators. 

As it was first observed in the original work of \cite{Zamolodchikov:1995aa}, the Liouville OPE can have either \textit{continuous} or \textit{discrete} support depending on the values of $P_i,P_j$. The discrete series can arise when we analytically continue the momenta $P_i,P_j$ \textit{outside} the standard Liouville regime $P\in\mathbb{R}_+$, in which case several poles of the integrand can cross the contour. To maintain analyticity in the parameters we need to deform the contour and pick the residues of these contributions. 
For the case of the normalization \eqref{eq:OPE_LiouvC0norm} these poles are determined basically by the analytic structure of $C_0(P_i,P_j,P)$ (since $\rho_0(P)$ is analytic). There are \textit{four} distinct cases regarding the support of the OPE \eqref{eq:OPE_LiouvC0norm} --- and hence of \eqref{eq:fusion_product} --- as dictated by the analytic structure of $C_0$. We summarize them in Table \ref{tab:supportdmmuij}.

\begin{table}[h]
\centering
\begin{tabular}{|l|l|}
    \hline
    $P_i,P_j$ & support of \eqref{eq:fusion_product} \\ \hline
    heavy, heavy & continuous \\
    heavy, light & continuous \\
    light, light; with $\Im(P_i+P_j)\leq \frac{Q}{2}$ & continuous \\
    light, light; with $\Im(P_i+P_j)> \frac{Q}{2}$ & discrete $+$ continuous \\ \hline
\end{tabular}
\caption{\footnotesize{Four distinct cases of two unitary highest-weight representations of the chiral Virasoro algebra labeled by $P_i,P_j$ and the support of their corresponding fusion product \eqref{eq:fusion_product} as dictated by the OPE \eqref{eq:OPE_LiouvC0norm}. The characterization as `light' or `heavy' corresponds to conformal dimensions $0\leq \Delta<\frac{c-1}{24}$ or $\Delta\geq \frac{c-1}{24}$, respectively.}} 
    \label{tab:supportdmmuij}
\end{table}  

In the first three cases, where we have at least one `heavy' representation or two `light' ones such that $\Im(P_i+P_j)\leq \frac{Q}{2}$, it can be seen that the poles of $C_0(P_i,P_j,P)$ do not cross the contour $P\in\mathbb{R}_+$. Therefore the corresponding OPE has the form \eqref{eq:OPE_LiouvC0norm}, i.e. with a \textit{continuous} support. On the other hand, for two light representations with $\Im(P_i+P_j)> \frac{Q}{2}$ a finite set of poles of the measure can cross the contour $P\in\mathbb{R}_+$ and therefore, by analyticity, contribute an additional \textit{discrete} sum in the OPE. In particular, it can be seen that these poles occur at pure imaginary $P=P_m$ such that $\Im(P_m)>0$ (thus corresponding to `light' unitary representations):
\be\label{eq:VMFT}
P_m = P_i + P_j - \frac{iQ}{2} - imb , \ \ \ \  \text{with} \ \ m\in \mathbb{Z}_{\geq0} \ : \ \Im(P_m)>0
\ee
along with all their reflections. The corresponding OPE can then be written compactly in this case as\footnote{It is worth noting that from the analysis of $C_0(P_i,P_j,P)$ (given in \eqref{eq:C_0_function}) one can see that these discrete contributions exist for $m\geq 1$ only for $c>25$ (or $0<b<1$).
}
\be\label{eq:discreteOPE}
\bea
\hat{V}_{P_i}(x,\bar{x})\hat{V}_{P_j}(0) \sim \int_{\Gamma} \dd P \ \mathcal{C}^{(L)}_{ij}(P) \ (x\bar{x})^{\Delta-\Delta_i-\Delta_j} [\hat{V}_P](0)
\eea
\ee
where $\Gamma$ is an appropriate contour that passes through the \textit{extended}, but still unitary, regime of $P$ (namely, $i\mathbb{R}_{(0,\frac{Q}{2}]}\cup \mathbb{R}_{\geq0}$), and
\be\label{eq:OPEliouvilledensity}
\bea
\mathcal{C}^{(L)}_{ij}(P):=\sum_m -2\pi i \underset{\ P=P_m}{\text{Res}}\big[\rho_0(P) C_{0}(P_i, P_j, P)\big]\delta(P - P_m) + \rho_0(P) C_{0}(P_i, P_j, P)\Theta(P).
\eea
\ee
This is the `analytic' OPE spectral density, where $\Theta$ denotes the standard Heaviside step function.
It is therefore clear from the OPE analysis that, by analyticity, the measure set $\mathbb{U}$ in the fusion product \eqref{eq:fusion_product} can have both \textit{discrete} and \textit{continuous} support depending on the type of (unitary) representations that undergo fusion. As a result, we can write the analytic Liouville fusion product formally as
\be\label{eq:analytic_Liouville_fusion}
\bea
R_{P_i} \boxtimes R_{P_j} \simeq&
\bigoplus_{\substack{P = P_m}}  R_{P}\otimes \text{Hom}^{\text{(discr.)}}\left(R_{P_i} \boxtimes R_{P_j};R_{P}\right) \ \\
&\quad \quad \quad \oplus \ \int_{\mathbb{R}_+}^{\oplus} \dd \mu(P) \ R_{P} \otimes \text{Hom}^{\text{(L)}}\left(R_{P_i} \boxtimes R_{P_j};R_{P}\right) 
\eea
\ee
with the understanding that the discrete series of representations is only present in the case $\Im(P_i+P_j)> \frac{Q}{2}$. The metric in the one-dimensional Hilbert space denoted as $\text{Hom}^{\text{(discr.)}}$ is proportional to $\underset{\  P=P_m}{\text{Res}}\big[\rho_0(P) C_{0}(P_i, P_j, P)\big]$. 

This concludes our definition of the analytic Liouville fusion product. It is worth noting that the contributions \eqref{eq:VMFT}, \eqref{eq:discreteOPE} were studied in \cite{Collier:2018exn} (see also \cite{Kusuki:2018wpa}) within the context of the lightcone bootstrap of four-point functions in unitary 2d CFTs with 
$c>1$ and a twist gap above the vacuum. The corresponding spectrum was dubbed \textit{Virasoro Mean Field Theory} and it was shown to govern universally the support of the $s$-channel OPE spectral density of four-point functions for pairwise-identical operators at large spin. Crucially, such theories are characterized by a discrete set of non-diagonal left-right representations of the Virasoro algebra and, as such, are not directly related with the diagonal case of Liouville theory that we discussed here. We will explore the extent to which we can describe such theories via a Verlinde formula in the next section.

\subsection{A conjecture for a novel bootstrap approach}\label{sec:conjectureBoot}
In a general unitary $c>1$ CFT with just Virasoro symmetry, a unique $\mathfrak{sl}(2,\mathbb{C})$-invariant vacuum state, and a discrete spectrum of infinitely-many primary operators at each spin sector, there is a corresponding set of modules $\{\mathcal{M}_{(\bbi,\bbi)}, \mathcal{M}_{(i,\bar{i})},\mathcal{M}_{(j,\bar{j})},\cdots\}$ which includes all the pairings of left-right Virasoro primaries with conformal weights $\Delta_i,\bar{\Delta}_i$ along with their descendants. Since such a theory contains primary operators at arbitrarily large spins \cite{Collier:2017shs}, we in general have modules with $\Delta_i\neq \bar{\Delta}_i$. This differs substantially from the `diagonal' case of Liouville theory that we studied above. It is reasonable, however, to expect that there still exists an associative and commutative fusion ring structure on the set of allowed modules of the theory: 
\be
\mathcal{M}_{I}\boxtimes \mathcal{M}_{J} \simeq \bigoplus_{\substack{K\in \mathcal{S}_{IJ}}} \mathcal{N}_{IJ}\text{}^K \mathcal{M}_{K}
\ee
where we denoted for brevity $I\equiv (i,\bar{i}), J\equiv (j,\bar{j{}})$ etc. In a theory with just Virasoro symmetry, the fusion numbers $\mathcal{N}_{IJ}\text{}^K$ are expected to take values in $\{0,1\}$. 

The original Verlinde formula \cite{Verlinde:1988sn} as well as the Virasoro-Verlinde formula \eqref{eq:formula_revisitL} are inherently `chiral' equations, in the sense that they describe the chiral part of the symmetry algebra. For a non-diagonal theory, and absent any other extended symmetry beyond Virasoro, it is intriguing to conjecture the existence of an analogous `generalized Virasoro-Verlinde formula' that connects in a similar way the fusion numbers of left-right modules of such theories with the CFT data on the torus\footnote{To the best of our knowledge, there is no analogous construction of a `generalized Verlinde formula' in non-diagonal RCFTs that yields the left-right fusion numbers as a sum over the full spectrum weighted by the left and right copies of the modular kernel (c.f. discussion in footnote \ref{foot:commentRCFTVerlinde}). Nevertheless, we will proceed with the conjecture presented here for reasons that we will argue below.}, namely
\be\label{eq:conjecturalVV}
\bea
\mathcal{N}_{IJ}\text{}^K(z,\bar{z})\stackrel{!}{=} \int_{\mathcal{C}} \dd P \dd \bar{P} \rho(P,\bar{P}) \left(\frac{\hat{\mathbb{S}}_{P_iP}[z]\hat{\mathbb{S}}_{\bar{P}_i\bar{P}}[\bar{z}]}{\sqrt{\mathsf{C}_{zii}}}\right)\frac{\mathbb{S}_{P_kP}[\bbi]\mathbb{S}_{\bar{P}_k\bar{P}}[\bbi]}{\mathbb{S}_{\bbi P}[\bbi]\mathbb{S}_{\bbi \bar{P}}
[\bbi]}\left(\frac{\hat{\mathbb{S}}_{PP_j}^*[z]\hat{\mathbb{S}}_{\bar{P} \bar{P}_j}^*[\bar{z}]}{\sqrt{\mathsf{C}_{zjj}}}\right).
\eea
\ee
The function $\mathcal{N}_{IJ}\text{}^K(z,\bar{z})$ on the LHS represents a `deformed' (for arbitrary values of $z,\bar{z}$) fusion number for the modules $\mathcal{M}_{I},\mathcal{M}_{J}, \mathcal{M}_{K}$, and is expected to take values in $\{0,1\}$ in the limit $z,\bar{z}\rightarrow \frac{iQ}{2}$. In less abstract terms, we view the LHS as defined mostly by the integral on the RHS which we now explain. The dependence on $z,\bar{z}$ on the RHS enters universally through the arguments of the modular S-kernels and the functions $\mathsf{C}$ (related to $C_0$). The behaviour of the integral is fundamentally determined by the density $\rho$, which we take it to be the modular-invariant density of \textit{primary} states in the $P,\bar{P}$ variables that enters in the usual partition function of the theory on the torus
\be
Z(\tau,\bar{\tau}) = \int_{\mathcal{C}} \dd P \dd \bar{P} \rho(P,\bar{P}) \frac{q^{P^2}}{\eta(\tau)} \frac{\bar{q}^{\bar{P}^2}}{\eta(-\bar{\tau})} \ , \ \ \ \ \ q=e^{2\pi i \tau},\,\bar{q}=e^{-2\pi i \bar{\tau}} \ .
\ee
Here $\mathcal{C}$ is an appropriate contour in the $P,\bar{P}$ plane that includes the support of $\rho$. In a given non-rational 2d CFT with a discrete spectrum, $\rho$ is a generalized function and can be thought of as a sum of delta functions supported at integer spins and a discrete set of scaling dimensions at each spin sector. 

We will now show that, assuming the existence of \eqref{eq:conjecturalVV} and relying solely on known universal statements about the primary density of states $\rho$ on the RHS,
the `deformed fusion numbers' $\mathcal{N}_{IJ}\text{}^K(z,\bar{z})$ are captured universally by two copies of the Virasoro-Verlinde formula \eqref{eq:formula_revisitL} which are, crucially, non-identical. Indeed, in a modular-invariant theory with $c>1$ and a twist gap above the vacuum, the density of states at large energies or large spin is famously governed by the Cardy formula \cite{Cardy:1986ie} (established rigorously and extended in \cite{Collier:2016cls,Afkhami-Jeddi:2017idc, Maxfield:2019hdt,Benjamin:2019stq,Mukhametzhanov:2019pzy,Mukhametzhanov:2020swe,Pal:2019zzr, Pal:2022vqc, Pal:2023cgk}). In the $(P,\bar{P})$ variables the \textit{averaged} density reads\footnote{The notation $a\approx b$ here means that $\frac{a}{b}\rightarrow 1$ in the limit of interest.}
\be\label{eq:Cardyrho}
\overline{\rho}(P,\bar{P})\approx  \rho_{\text{Cardy}}(P,\bar{P})\equiv \rho_0(P)\rho_0(\bar{P}) \ \ \ \ \ \ \ \ \text{as} \ \ \ P,\bar{P}\rightarrow \infty, \ \text{or} \ P\rightarrow \infty, \bar{P} \ \text{fixed},
\ee
and further corrections coming from non-trivial primaries above the vacuum are exponentially suppressed in at least one of $P\sim\sqrt{\Delta}$ or $\bar{P}\sim\sqrt{\bar{\Delta}}$. 

From the perspective of \eqref{eq:conjecturalVV} this means that we could in principle study the behavior of the `deformed fusion numbers' on the LHS as dictated from the internal high-energy part of the spectrum (one can think of these contributions as ``high-energy loops''), by approximating the measure on the RHS as
\be\label{eq:highenergy_approx}
\int_{\mathcal{C}} \dd P \dd \bar{P} \rho(P,\bar{P})\longrightarrow \int_0^{\infty}\dd\mu(P)\dd\mu(\bar{P}).
\ee 
Accordingly, this defines an \textit{averaged} `deformed fusion number' that we denote as $\overline{\mathcal{N}_{IJ}\text{}^K}(z,\bar{z})$. It is not at all clear a priori that by doing this crude approximation we will end up with something physical on the LHS (i.e. a set of fusion numbers) after taking the limits $z,\bar{z}\rightarrow \frac{iQ}{2}$. Nonetheless, given the existence of our original formula \eqref{eq:formula_revisitL} we find
\be\label{eq:genVVapprox}
\bea
  \overline{\mathcal{N}_{IJ}\text{}^K}(z,\bar{z})& \approx \N^{\text{(L)}}_{ij}(P_k|z) \times \N^{\text{(L)}}_{\bar{i}\bar{j}}(\bar{P_k}|\bar{z})\\
&= \left(\rho_{\text{Cardy}}(P_k,\bar{P}_k) \ \mathsf{C}_{ijk}\right)^{-1}\fker{z,}{P_k}{P_j}{P_i}{P_i}{P_j}\fker{\bar{z},}{\bar{P}_k}{\bar{P}_j}{\bar{P}_i}{\bar{P}_i}{\bar{P}_j}.
\eea
\ee
The corrections to \eqref{eq:genVVapprox} are suppressed in the approximation \eqref{eq:highenergy_approx}, and are generically controlled by corrections to the Cardy formula coming from non-vacuum primaries (as computed within \eqref{eq:conjecturalVV}). Quite remarkably, this appears to be a physical result for the fusion numbers. Indeed, the limit
$$
\lim_{z,\bar{z}\rightarrow\frac{iQ}{2}}\overline{\mathcal{N}_{IJ}\text{}^K}(z,\bar{z})=1
$$ recovers the correct dimension of the space of conformal blocks in the three-punctured sphere for a theory with just Virasoro symmetry. Furthermore, we see that the expression factorizes into left and right `fusion coefficients' which -- contrary to the diagonal case of Liouville theory -- have in principle different left and right labels for all the indices. This shows the precise way in which the Virasoro-Verlinde formula captures the fusion numbers in general non-rational and non-diagonal theories, subject to the conjecture \eqref{eq:conjecturalVV}. The rigorous implications of this factorized structure deserve further understanding, especially in relation to the analytic bootstrap results of \cite{Collier:2018exn, Kusuki:2018wpa, Collier_2020}, where we believe there is a highly suggestive connection. We leave these explorations to future work. 

A key point to emphasize is that the only input in this derivation (subject to the conjecture \eqref{eq:conjecturalVV}) was the universality of the density of states at large scaling dimension or large spin, as given by the Cardy formula. From this, we were able to obtain universal information about the fusion numbers. This kind of relation is precisely at the heart of the Verlinde formula on general grounds: it connects in a non-trivial way the spectrum of the theory on $S^1$ with the space of three-point conformal blocks on the sphere. Conversely, further exploring this connection for non-rational two-dimensional CFTs could serve as a promising approach and a novel means for constraining the spectral density $\rho$ itself, using universal features of the fusion coefficients.

An immediate bootstrap constraint of this form is the following. Within the space of all modular-invariant spectral densities $\rho(P,\bar{P})$, not all of them will possibly produce a \textit{positive-definite} and \textit{integral} distribution $\mathcal{N}_{IJ}\text{}^K$ in the limit $z,\bar{z}\rightarrow iQ/2$ of \eqref{eq:conjecturalVV}. Even more, in theories with just Virasoro symmetry, the allowed spectral densities should be able to capture the fusion numbers which take values in $\{0,1\}$. This is a concrete and novel way to distinguish between theories with just Virasoro symmetry and theories with extended symmetry algebras. The possible existence of a `generalized Verlinde formula' such as \eqref{eq:conjecturalVV}, therefore, defines an intriguing classification problem for allowed distributions $\rho$ (with either discrete or continuous support). We think of this approach as equivalent and/or complementary to the usual modular bootstrap approach \cite{Hellerman:2009bu, Collier:2016cls, Afkhami-Jeddi:2019zci, Hartman:2019pcd, Dey:2024nje}. The existence of a Verlinde formula is a unique and fascinating feature of two-dimensional theories that we believe should be an integral part of the bootstrap toolkit for exploring the rich structure of these theories. It is critical, of course, to understand whether similar relations exist for CFTs in higher dimensions.

\clearpage

\section{Discussion}

In this work, we derived a new formula in non-rational 2d conformal field theories with $c\geq 25$ and only Virasoro symmetry that we view as the analog of the Verlinde formula for the Virasoro algebra. Remarkably,  \eqref{eq:intro_6j} assembles all the basic Virasoro crossing kernels in one formula: the identity S-kernel, the modular S-kernel, the one-point S-kernel and the Virasoro $6j$ symbol (with pairwise identical external arguments). Moreover, we have seen that the regularized `fusion density' $\N_{P_0}[P_1,P_2,P_3]$ elegantly packages the fusion rules of both degenerate and non-degenerate representations of the Virasoro algebra in one meromorphic function.

Our derivation relied on the internal consistency of the Virasoro TQFT, which was shown to admit a well-defined inner product (for stable surfaces and $c\geq 25$) in \cite{Collier_2023, Collier:2024mgv}. We can think of the TQFT as `encoding' the non-rational Moore-Seiberg consistency conditions into topological properties of 3-manifolds. Apart from showcasing the utility of the Virasoro TQFT formalism, we showed that the Virasoro-Verlinde formula has interesting applications in a variety of theories that are governed by Virasoro symmetry. Moreover, it offers intriguing implications for generic non-rational 2d CFTs. 

In this discussion, we will pose some open questions and indicate future directions.

\paragraph*{The one-point S-kernel diagonalizes the $6j$ symbol.}
As already remarked in the introduction, our formula has the structural form of a diagonalization $S D S^\dagger$ of the $6j$ symbol (or equivalently, the fusion kernel) with pairwise identical external operators. Since the formula follows from basic consistency conditions required in any CFT$_2$ (as shown in appendix \ref{sec:CFTproof}), there should be a version of our formula for arbitrary chiral algebras, i.e.
\begin{equation}\label{eq:6j_generalized}
    \begin{Bmatrix}
  h_3 & h_3 & h_0 \\
  h_1 & h_1 & h_2 
 \end{Bmatrix}_{6j} = \int_{\mathcal{C}} \dd \mu(h)\,\frac{\hat{S}_{h_1h}[h_0]\,S_{h_2h}[\bbi]\,\hat{S}_{hh_3}^*[h_0]}{S_{\bbi h}[\bbi]},
\end{equation}
for some appropriate measure $\mu(h)$. In particular, for RCFT's $\mu(h)$ is a finite sum of delta functions supported on the spectrum of the RCFT, and the right-hand side becomes a product of matrices (for fixed $h_0$).\footnote{For example, for $SU(2)$ Chern-Simons theory at level $k$ one can explicitly check that the matrix product on the right-hand side reproduces the $6j$ symbol of the $SU(2)_k$ quantum group.} For non-rational theories with non-compact representations -- such as the coset WZNW models on $SL(2,\mathbb{R})/U(1)$ or $H^+_3 = SL(2,\mathbb{C})/SU(2)$, or theories with $\mathcal{W}_n$ symmetry -- 
the measure $\mu(h)$ will be the Plancherel measure on the representation space. Furthermore, we expect that \eqref{eq:6j_generalized} will generalize straightforwardly to supersymmetric CFT's, such as $\mathcal{N}=1$ or $\mathcal{N}=2$ super-Liouville theory\footnote{Recently, a supersymmetric generalization of the Virasoro TQFT was given in \cite{Bhattacharyya:2024vnw}, which uses the known fusion and one-point S-kernels of $\mathcal{N}=1$ super-Liouville CFT. This means that the TQFT argument of section \ref{sec:VTQFT_proof} directly carries over to the $\mathcal{N}=1$ case.} \cite{Distler:1989nt,Kutasov:1991pv}. 

Finally, in the realm of CFTs whose chiral algebra is just Virasoro, there is still an interesting open problem of understanding the modular and fusion kernels at $c\leq 1$ in great generality (for a recent proposal, see \cite{Roussillon:2024wmr}), or even an inner product in the space of $c\leq 1$ conformal blocks (in the sense of \cite{Collier_2023}). Understanding this would ultimately lead to a $c\leq 1$ analog of the formula we presented, where an intricate connection with the so-called `timelike' Liouville theory is anticipated, in a similar way to how our formula is connected with the ordinary (spacelike) Liouville theory for $c>1$. 

\paragraph*{Large central charge and the Schwarzian limit.} It is interesting to study systematically various large central charge limits of our formula. As a notable example, it is known that the Virasoro $6j$ symbol reduces to the classical $6j$ symbol associated to $\mathfrak{sl}(2,\mathbb{R})$ in the so-called \emph{Schwarzian limit} \cite{Mertens:2017mtv}. From the bulk point of view, the Schwarzian limit can be understood as the dimensional reduction of Chern-Simons theory to $SL^+(2,\mathbb{R})$ BF theory, which is equivalent to JT with heavy matter on the disk \cite{Blommaert:2018iqz}. Explicitly, the Schwarzian limit is a combination of the semiclassical limit $b\to 0$ (i.e. $c\to\infty$) with a scaling of the conformal weights to the threshold $P\to 0$ \cite{Ghosh:2019rcj}. Keeping the ratio $k \coloneqq P/b$ fixed, the classical $6j$ symbol is retrieved as the limit:
\begin{equation}
    \lim_{b\to 0}b^3 \begin{Bmatrix}
  P_3 & P_3 & P_0 \\
  P_1 & P_1 & P_2 
 \end{Bmatrix}_{6j} = \begin{Bmatrix}
  k_3 & k_3 & k_0 \\
  k_1 & k_1 & k_2 
 \end{Bmatrix}_{\mathfrak{sl}(2,\mathbb{R})} \,.
\end{equation}
An interesting question arises when we take this Schwarzian limit on both sides of the Virasoro-Verlinde formula. We know that in this limit, $\sker{\bbi}{P}{\bbi}$ reduces to the Schwarzian density of states, while the classical $6j$ symbol computes the Schwarzian OTOC \cite{Mertens:2017mtv}. However, it is not so clear what the interpretation of the one-point S-kernel $\sker{k_1}{k}{k_0}$ is in the Schwarzian limit. Perhaps it can be understood in terms of boundary states in Liouville theory, as in section \ref{sec:open_closed}, by using the connection between Liouville CFT and the Schwarzian theory \cite{Mertens:2018fds}. This leads us to the next point.

\paragraph{Boundary states in Liouville CFT.}
In rational BCFT, the standard Cardy boundary states are written as a superposition of Ishibashi states, weighted by the ratio of modular S-matrices $\frac{S_{kh}}{\sqrt{S_{\bbi h}}}$ \cite{Cardy:1989ir}. In this case, Cardy's boundary states are consistent with the open-closed duality of the annulus partition function thanks to the Verlinde formula \cite{Cardy:1991tv}. Similarly, the FZZT and ZZ brane boundary states in Liouville theory can be written as a continuous superposition:
\begin{equation}\label{eq:bdy_state}
    \ket{P_a} = \int_0^\infty \dd P\, \e^{i\delta(P)} \frac{\sker{P_a}{P}{\bbi}}{\sqrt{\sker{\bbi}{P}{\bbi}}} \,|P\rangle\!\rangle,
\end{equation}
where the Ishibashi state is defined by the condition $(L_n-\bar L_{-n})|P\rangle\!\rangle = 0$. For the FZZT boundary condition, $\sker{P_a}{P}{\bbi}$ is the non-degenerate modular S-kernel \eqref{eq:torus_mod_S} \cite{Teschner:2000md}, while for the ZZ boundary state it is the degenerate modular S-kernel \eqref{eq:degenerateS} with $P_a=P_{\langle m,n\rangle}$ \cite{Zamolodchikov:2001ah}. Given the fact that we showed in section \ref{sec:open_closed} that the open-closed duality of the annulus boundary one-point function in Liouville is guaranteed by the Virasoro-Verlinde formula, it is natural to conjecture that we can consistently define a  `punctured boundary state' of the form:
\begin{equation}\label{eq:punc_bdy}
  \ket{P_a;P_0} = \int_0^\infty\dd P\,e^{i\delta(P)}  \frac{\sker{P_a}{P}{P_0}}{\sqrt{\sker{\bbi}{P}{\bbi}}} \,|P;P_0\rangle\!\rangle,
\end{equation}
where the phase factor is the same as in \eqref{eq:bdy_state} (namely the square root of the bulk reflection amplitude) and $|P;P_0\rangle\!\rangle$ is now a `punctured Ishibashi state' that includes the contribution from the descendants of $\O_0$. Since the open-closed duality we derived in section \ref{sec:open_closed} was independent of the structure constant $g^{(\mathsf{ab})}_{P}$, and since eq.\! \eqref{eq:bdy_3pt} is the generic form of a boundary 3-point OPE coefficient in BCFT, it is plausible that a punctured boundary state of the form \eqref{eq:punc_bdy} may be defined in greater generality.

\paragraph*{Open and closed Verlinde lines.} The quantization of the Teichm\"uller space of Riemann surfaces is known to be intimately connected to Liouville CFT and Virasoro modular geometry. Two natural observables in quantum Teichm\"uller theory are the closed Verlinde loop operator \cite{Verlinde:1989ua, Alday:2009fs, Drukker:2009id, Drukker:2010jp} and the open Verlinde line operator \cite{Gaiotto:2014lma}. Pictorially, the loop operator acts on a line labeled by $P$ as:
\begin{equation}\label{eq:intro_open_closed}
    \begin{tikzpicture}[x=0.75pt,y=0.75pt,yscale=-1.3,xscale=1.3,baseline={([yshift=-.5ex]current bounding box.center)}] 
\draw   (100.1,105) -- (119.11,105) ;
\draw  [draw opacity=0] (146.06,109.65) .. controls (145.08,121.23) and (140.5,130) .. (135,130) .. controls (128.79,130) and (123.75,118.81) .. (123.75,105) .. controls (123.75,91.19) and (128.79,80) .. (135,80) .. controls (140.68,80) and (145.38,89.35) .. (146.14,101.5) -- (135,105) -- cycle ; \draw   (146.06,109.65) .. controls (145.08,121.23) and (140.5,130) .. (135,130) .. controls (128.79,130) and (123.75,118.81) .. (123.75,105) .. controls (123.75,91.19) and (128.79,80) .. (135,80) .. controls (140.68,80) and (145.38,89.35) .. (146.14,101.5) ;  
\draw    (128.11,105) -- (168.75,105) ;
\draw (155,93.4) node [anchor=north west][inner sep=0.75pt]  [font=\scriptsize]  {$P$};
\draw (112,82.4) node [anchor=north west][inner sep=0.75pt]  [font=\scriptsize]  {$P_2$};
\end{tikzpicture} \;=\, \frac{\mathbb{S}_{P_2P}[\bbi]}{\mathbb{S}_{\bbi P}[\bbi]}\,\,
\begin{tikzpicture}[x=0.75pt,y=0.75pt,yscale=-1.2,xscale=1.2,baseline={([yshift=-1.7ex]current bounding box.center)}]
\draw    (99.9,105) -- (168.75,105) ;
\draw (126.2,92.2) node [anchor=north west][inner sep=0.75pt]  [font=\scriptsize]  {$P$};
\end{tikzpicture}\,,
\end{equation}
which for $P_2 = P_{\langle 2,1\rangle}$ (semiclassically) computes the geodesic length of the cycle dual to $P$. The open Verlinde line operator acts on three-point vertex as:
\begin{equation}
    \begin{tikzpicture}[x=0.75pt,y=0.75pt,yscale=-1,xscale=1,baseline={([yshift=-.5ex]current bounding box.center)}]
\draw    (215.27,169.62) -- (251.71,169.62) ;
\draw [shift={(215.27,169.62)}, rotate = 0] [color={rgb, 255:red, 0; green, 0; blue, 0 }  ][fill={rgb, 255:red, 0; green, 0; blue, 0 }  ]      (0, 0) circle [x radius= 1.34, y radius= 1.34]   ;
\draw    (199.12,136.75) -- (215.27,169.62) ;
\draw    (199.12,202.48) -- (215.27,169.62) ;
\draw [color={rgb, 255:red, 208; green, 2; blue, 27 }  ,draw opacity=1 ]   (207.2,153.18) .. controls (221.43,144) and (233.71,153.14) .. (233.49,169.62) ;
\draw [shift={(233.49,169.62)}, rotate = 90.78] [color={rgb, 255:red, 208; green, 2; blue, 27 }  ,draw opacity=1 ][fill={rgb, 255:red, 208; green, 2; blue, 27 }  ,fill opacity=1 ]      (0, 0) circle [x radius= 1.34, y radius= 1.34]   ;
\draw [shift={(207.2,153.18)}, rotate = 327.16] [color={rgb, 255:red, 208; green, 2; blue, 27 }  ,draw opacity=1 ][fill={rgb, 255:red, 208; green, 2; blue, 27 }  ,fill opacity=1 ]      (0, 0) circle [x radius= 1.34, y radius= 1.34]   ;
\draw (259.1,165.34) node [anchor=north west][inner sep=0.75pt]  [font=\scriptsize]  {$P_{3}$};
\draw (180.78,126.72) node [anchor=north west][inner sep=0.75pt]  [font=\scriptsize]  {$P_{1}$};
\draw (181.18,199.43) node [anchor=north west][inner sep=0.75pt]  [font=\scriptsize]  {$P_{2}$};
\draw (229.06,141.3) node [anchor=north west][inner sep=0.75pt]  [font=\scriptsize]  {$P_0$};
\end{tikzpicture} =  \frac{1}{\sqrt{C_{101}C_{303}}}\begin{Bmatrix}
  P_3 & P_3 & P_0 \\
  P_1 & P_1 & P_2 
 \end{Bmatrix}_{6j} \quad\begin{tikzpicture}[x=0.75pt,y=0.75pt,yscale=-1,xscale=1,baseline={([yshift=-.5ex]current bounding box.center)}]
\draw    (215.27,169.62) -- (251.71,169.62) ;
\draw [shift={(215.27,169.62)}, rotate = 0] [color={rgb, 255:red, 0; green, 0; blue, 0 }  ][fill={rgb, 255:red, 0; green, 0; blue, 0 }  ]      (0, 0) circle [x radius= 1.34, y radius= 1.34]   ;
\draw    (199.12,136.75) -- (215.27,169.62) ;
\draw    (199.12,202.48) -- (215.27,169.62) ;
\draw (259.1,165.34) node [anchor=north west][inner sep=0.75pt]  [font=\scriptsize]  {$P_{3}$};
\draw (180.78,126.72) node [anchor=north west][inner sep=0.75pt]  [font=\scriptsize]  {$P_{1}$};
\draw (181.18,199.43) node [anchor=north west][inner sep=0.75pt]  [font=\scriptsize]  {$P_{2}$};
\end{tikzpicture}\,
\end{equation}
where $C_{101} \coloneqq C_0(P_1,P_0,P_1)$. The Virasoro-Verlinde formula \eqref{eq:VV_6j} can now be read as saying that the open and closed loop/line operators are related by a `change of basis', implemented by the (unitary) one-point S-kernel. This can be made more explicit by introducing a complete basis $\ket{P}$ associated to an annulus (i.e. a tubular neighborhood of the line $P$ in \eqref{eq:intro_open_closed}). We choose this basis such that it diagonalizes the loop operator:
\begin{equation}
    \mathcal{L}_{(2)}^{\text{closed}} \ket{P} = \frac{\mathbb{S}_{P_2P}[\bbi]}{\mathbb{S}_{\bbi P}[\bbi]}\ket{P}.
\end{equation}
In quantum Teichm\"uller theory, this basis simultaneously diagonalizes the Dehn twist operator \cite{Kashaev:2000ku, Teschner:2003em}. The one-point S-kernel is written as a matrix element in this basis, 
$
    \widehat{\mathbb{S}}_{PP'}[P_0] = \bra{P'}\mathsf{S}_{(0)} \ket{P}
$ \cite{Teschner:2013tqy}.
Using the completeness relation $1 = \int_0^\infty \dd \mu(P) \ket{P}\bra{P}$, we can then rewrite the Virasoro-Verlinde formula as the operator identity
\begin{equation}\label{eq:VV_operator}
   \mathcal{L}^{\text{open}}_{(0,2)}  =  \mathsf{S}_{(0)}\mathcal{L}^{\text{closed}}_{(2)} \mathsf{S}_{(0)}^\dagger
\end{equation}
where we define $\mathcal{L}^{\text{open}}_{(0,2)}$ through its matrix elements
\begin{equation}
    \bra{P_1}\mathcal{L}^{\text{open}}_{(0,2)} \ket{P_3} = \begin{Bmatrix}
  P_3 & P_3 & P_0 \\
  P_1 & P_1 & P_2 
 \end{Bmatrix}_{6j}.
\end{equation}
Indeed, taking the matrix element $\bra{P_1}\cdots\ket{P_3}$ on both sides of \eqref{eq:VV_operator} retrieves our formula \eqref{eq:VV_6j}. It would be interesting to gain a deeper understanding of the significance of the `open-closed' operator equality \eqref{eq:VV_operator} in quantum Teichm\"uller theory.

\paragraph{Mixed correlations in the large-$c$ ensemble.} In section \ref{sec:3boundary}, we used the Virasoro-Verlinde formula to compute a three-boundary wormhole, which led to a new formula for the cubic connected average $\overline{\rho C^2}$ in the large-$c$ ensemble of OPE data. It would be interesting to show whether our result can be derived directly from the matrix-tensor model of \cite{Belin:2023efa,Jafferis:2024jkb}. Since the matrix/tensor model obeys the Moore-Seiberg consistency conditions in the triple scaling limit, it seems that in principle the result \eqref{eq:cubic_moment2} should agree with ours (although showing this may be technically hard). 

A related question is whether our prediction for the mixed correlation agrees with the \emph{state averaging ansatz} proposed in \cite{deBoer:2023vsm}\footnote{We thank Jan de Boer for this suggestion.}. There, the framework of state averaging was introduced as an alternative for ensemble averaging over OPE coefficients. The conceptual advantage is that state averaging is well-defined in a single CFT$_2$, where the $C_{ijk}$ are non-random. Moreover, the ensemble of mixed states, which was derived using a principle of maximum entropy,  was shown to agree with Euclidean wormhole computations in 3d gravity \cite{deBoer:2023vsm}. In the present case, the three-boundary wormhole computed in section \ref{sec:3boundary} would contribute to the connected average 
\begin{equation}
    \overline{\Tr(\rho \O)\Tr(\rho)\Tr(\rho \O)}
\end{equation}
where $\rho$ now denotes a density matrix drawn from an ensemble of random mixed states. The state averaging perspective has the advantage of treating the three factors inside the average on the same footing: instead of a mixed correlation between $\rho(E_2,J_2)$ and $C_{1\O1}C_{3\O3}$, the average is simply over a density matrix  $\rho\otimes \rho\otimes \rho$ on the tripled CFT Hilbert space. For the state averaging ansatz, the connected contribution to the average $\overline{\rho\otimes \rho\otimes \rho}$ was shown to be of the `cyclic' type, which precisely agrees with the diagonal pattern of index contractions of our result \eqref{eq:cubic_moment2}. It would be interesting to know what would be the low energy input in the maximum ignorance ensemble of states that would also reproduce the precise smooth functional form of the three-boundary wormhole \eqref{eq:Z_seed2}.  

\paragraph{Reconstruction fantasies.}\hspace{-4mm}\footnote{We have kindly borrowed the title of this paragraph from the name of a section in \cite{Moore:1988uz} which has greatly influenced the present work.} The problem of finding examples of non-rational two-dimensional theories with just Virasoro symmetry beyond Liouville theory remains tantalizing, as it continues to challenge our understanding of two-dimensional CFTs (see \cite{Antunes:2022vtb} for recent progress constructing such candidate examples). In section \ref{sec:conjectureBoot}, we conjectured the existence of a generalized Virasoro-Verlinde formula, which we believe merits further exploration and could provide essential insights into addressing this longstanding challenge. One possibly interesting concrete problem to study is the observation that the RHS of \eqref{eq:conjecturalVV} obeys a \textit{shift relation} in the variables $z,\bar{z}$ as dictated from the analogous shift relations obeyed by the modular kernel and the function $C_0$ (see e.g. \cite{Eberhardt:2023mrq}). This directly translates into shift relations for the conjectured `deformed fusion number' $\mathcal{N}_{IJ}\text{}^K(z,\bar{z})$ on the LHS. It will be interesting to systematically study (and classify) the solutions of such shift relations and investigate their physical interpretation.

\section*{Acknowledgements}

We would like to thank 
Jan de Boer, Scott Collier,
Lorenz Eberhardt,
Diego Liška,
Dalimil Mazáč,
Eric Perlmutter,
Sylvain Ribault,
Erik Verlinde,
and Jakub Vošmera
for stimulating discussions. 
BP is supported by the European
Research Council under the European Unions Seventh Framework Programme (FP7/2007-2013),
ERC Grant agreement ADG 834878. IT is supported by the ERC Starting Grant 853507.

\appendix

\addtocontents{toc}{\protect\setcounter{tocdepth}{1}}

\section{Basics of crossing kernels and Virasoro TQFT}\label{app:prelim}

In this appendix we give a brief introduction to the Virasoro crossing kernels in CFT$_2$, and how they appear in the 3d Virasoro TQFT introduced in \cite{Collier_2023}. For a more in-depth analysis of the Virasoro crossing kernels, we refer to the excellent notes of \cite{Eberhardt:2023mrq}. 

As usual, we adopt the Liouville parametrization for the the central charge $c$ and the conformal weights $\Delta,\bar\Delta$ of highest-weight representations of the Virasoro algebra:
\begin{equation}\label{eq:Liouville_par}
    \Delta = \frac{c-1}{24} + P^2, \quad \bar \Delta = \frac{c-1}{24} + \bar{P}^2, \quad  c = 1 + 6 Q^2  , \quad Q = b + b^{-1} \ .
\end{equation}
The corresponding scaling dimensions $E$ (on the plane) and spin $J$ are given by: 
\begin{equation}
    E=\frac{c-1}{12}+P^2+\bar{P}^2,\quad J=P^2-\bar{P}^2.
\end{equation}
Note the reflection symmetry $P\rightarrow-P$ of the Liouville parametrization \eqref{eq:Liouville_par}. Unless explicitly stated, we are interested in the regime $c\geq25$, which means $b\in(0,1]$. We also demand $\Delta,\bar \Delta\geq 0$, so we take $P,\bar{P}\in i\mathbb{R}_{(0,\frac{Q}{2}]}\cup \R_{+}$ modulo reflections. The vacuum state corresponds to pure imaginary Liouville momentum $P=\bar P = \frac{iQ}{2}$, which we denote by ``$P=\bbi$''.

For any compact unitary CFT the spectrum of primary conformal dimensions is discrete and the spin is quantized. Nonetheless, the space of non-degenerate representations of the full Virasoro algebra $\mathrm{Vir}\times\overline{\mathrm{Vir}}$ is described by continuous parameters $P,\bar P$. Since the Virasoro conformal blocks factorize into holomorphic and anti-holomorphic blocks, the Virasoro crossing kernels are also holomorphically factorized. As our paper is mostly concerned with the representation theoretic aspects of the Virasoro algebra, we will mostly ignore the dependence on the anti- holomorphic sector. However, it is understood that the conformal blocks and crossing kernels discussed below all have an anti-holomorphic counterpart.

\subsection{Crossing kernels}

The simplest example of a crossing kernel is the modular transformation that maps the non-degenerate Virasoro character on the torus $\chi_P(\tau)$ to its S-dual character $\chi_{P'}(-\tfrac{1}{\tau})$. It is simply the Fourier kernel:
\begin{equation}\label{eq:torus_mod_S}
    \mathbb{S}_{P'P}[\bbi] = 2\sqrt{2}\cos(4\pi PP').
\end{equation}
Pictorially, the modular S-transform maps the $a$-cycle of the torus to the $b$-cycle:
\begin{equation}
   \begin{tikzpicture}[x=0.75pt,y=0.75pt,yscale=-1,xscale=1,baseline={([yshift=-.5ex]current bounding box.center)}]
\draw  [draw opacity=0] (295.6,139.69) .. controls (291.92,145.6) and (285.1,149.57) .. (277.29,149.57) .. controls (269.37,149.57) and (262.46,145.48) .. (258.82,139.42) -- (277.29,129.57) -- cycle ; \draw   (295.6,139.69) .. controls (291.92,145.6) and (285.1,149.57) .. (277.29,149.57) .. controls (269.37,149.57) and (262.46,145.48) .. (258.82,139.42) ;  
\draw  [draw opacity=0] (261.5,142.47) .. controls (264.74,138.08) and (270.69,135.14) .. (277.49,135.14) .. controls (284.11,135.14) and (289.92,137.93) .. (293.22,142.12) -- (277.49,150.14) -- cycle ; \draw   (261.5,142.47) .. controls (264.74,138.08) and (270.69,135.14) .. (277.49,135.14) .. controls (284.11,135.14) and (289.92,137.93) .. (293.22,142.12) ;  
\draw  [draw opacity=0] (278.35,166.61) .. controls (278.07,166.7) and (277.79,166.75) .. (277.49,166.75) .. controls (274.87,166.75) and (272.75,163.03) .. (272.75,158.45) .. controls (272.75,153.86) and (274.87,150.14) .. (277.49,150.14) .. controls (278.43,150.14) and (279.31,150.62) .. (280.04,151.44) -- (277.49,158.45) -- cycle ; \draw  [color={rgb, 255:red, 208; green, 2; blue, 27 }  ,draw opacity=1 ] (278.35,166.61) .. controls (278.07,166.7) and (277.79,166.75) .. (277.49,166.75) .. controls (274.87,166.75) and (272.75,163.03) .. (272.75,158.45) .. controls (272.75,153.86) and (274.87,150.14) .. (277.49,150.14) .. controls (278.43,150.14) and (279.31,150.62) .. (280.04,151.44) ;  
\draw  [draw opacity=0][dash pattern={on 1.5pt off 1.5pt on 1.5pt off 1.5pt}] (280.89,152.66) .. controls (281.73,154.15) and (282.24,156.19) .. (282.24,158.45) .. controls (282.24,162.45) and (280.62,165.79) .. (278.47,166.58) -- (277.49,158.45) -- cycle ; \draw  [color={rgb, 255:red, 208; green, 2; blue, 27 }  ,draw opacity=1 ][dash pattern={on 1.5pt off 1.5pt on 1.5pt off 1.5pt}] (280.89,152.66) .. controls (281.73,154.15) and (282.24,156.19) .. (282.24,158.45) .. controls (282.24,162.45) and (280.62,165.79) .. (278.47,166.58) ;  
\draw   (243.94,142.66) .. controls (243.94,128.99) and (258.88,117.91) .. (277.32,117.91) .. controls (295.76,117.91) and (310.71,128.99) .. (310.71,142.66) .. controls (310.71,156.33) and (295.76,167.42) .. (277.32,167.42) .. controls (258.88,167.42) and (243.94,156.33) .. (243.94,142.66) -- cycle ;
\draw (257.34,150.88) node [anchor=north west][inner sep=0.75pt]  [font=\scriptsize,color={rgb, 255:red, 208; green, 2; blue, 27 }  ,opacity=1 ]  {$P'$};
\end{tikzpicture} \,= \int_0^\infty \dd P\,\mathbb{S}_{P'P}[\bbi]\,\,\begin{tikzpicture}[x=0.75pt,y=0.75pt,yscale=-1,xscale=1,baseline={([yshift=-.5ex]current bounding box.center)}]
\draw  [draw opacity=0] (275.6,119.69) .. controls (271.92,125.6) and (265.1,129.57) .. (257.29,129.57) .. controls (249.37,129.57) and (242.46,125.48) .. (238.82,119.42) -- (257.29,109.57) -- cycle ; \draw   (275.6,119.69) .. controls (271.92,125.6) and (265.1,129.57) .. (257.29,129.57) .. controls (249.37,129.57) and (242.46,125.48) .. (238.82,119.42) ;  
\draw  [draw opacity=0] (241.5,122.47) .. controls (244.74,118.08) and (250.69,115.14) .. (257.49,115.14) .. controls (264.11,115.14) and (269.92,117.93) .. (273.22,122.12) -- (257.49,130.14) -- cycle ; \draw   (241.5,122.47) .. controls (244.74,118.08) and (250.69,115.14) .. (257.49,115.14) .. controls (264.11,115.14) and (269.92,117.93) .. (273.22,122.12) ;  
\draw  [color={rgb, 255:red, 208; green, 2; blue, 27 }  ,draw opacity=1 ] (232.68,122.22) .. controls (232.68,111.19) and (243.72,102.25) .. (257.32,102.25) .. controls (270.93,102.25) and (281.97,111.19) .. (281.97,122.22) .. controls (281.97,133.25) and (270.93,142.19) .. (257.32,142.19) .. controls (243.72,142.19) and (232.68,133.25) .. (232.68,122.22) -- cycle ;
\draw   (223.57,122.22) .. controls (223.57,107.83) and (238.69,96.17) .. (257.32,96.17) .. controls (275.96,96.17) and (291.07,107.83) .. (291.07,122.22) .. controls (291.07,136.61) and (275.96,148.27) .. (257.32,148.27) .. controls (238.69,148.27) and (223.57,136.61) .. (223.57,122.22) -- cycle ;
\draw (253.71,131.26) node [anchor=north west][inner sep=0.75pt]  [font=\scriptsize,color={rgb, 255:red, 208; green, 2; blue, 27 }  ,opacity=1 ]  {$P$};
\end{tikzpicture}\,.
\end{equation}

We can also study degenerate representations of the Virasoro algebra, which occur at the special values
   \be 
   P_{\langle m,n\rangle} = \frac{ibm}{2}+ \frac{in}{2b}.
   \ee
To construct the corresponding degenerate characters, one needs to subtract the null-states in a systematic way. The S-transform of these characters is then implemented by the degenerate modular S-kernel:
\begin{equation}\label{eq:degenerateS}
    \mathbb{S}_{{\langle m,n\rangle} P}[\bbi] = 4\sqrt{2}\sinh(2\pi b^{-1}  mP)\sinh(2\pi b \, n  P).
\end{equation}
The special case $m=n=1$ corresponds to the identity module, $P_{\langle 1,1\rangle} = \bbi$. The identity S-kernel defines the universal Cardy density of states via $\rho_{\text{Cardy}}(P,\bar P) = \rho_0(P)\rho_0(\bar P)$, where 
\begin{equation}\label{eq:rho_0}
    \rho_0(P) \coloneqq \sker{\bbi}{P}{\bbi} = 4\sqrt{2}\sinh(2\pi b P)\sinh(2\pi b^{-1}  P).
\end{equation}
This density of states indeed exhibits the usual Cardy growth $\rho_{\text{Cardy}} \sim \e^{2\pi Q (P+\bar P)}$ at high energies \cite{Collier_2020}.

More complicated crossing kernels exist for conformal blocks on punctured Riemann surfaces with negative Euler characteristic. The two basic building blocks are:
\be\bea\label{eq:crossing_kernels}
    \text{Torus one-point:} \quad \quad  \tau ^{\Delta_0}\mathcal{F}_{P_1}[P_0;\tau] &= \int_0^\infty dP_2 \  \mathbb{S}_{P_1P_2}[P_0] \ \mathcal{F}_{P_2}[P_0;-\tfrac{1}{\tau}] \ , \\[1em]
    \text{Sphere four-point:} \quad \quad \quad \quad   \mathcal{V}_{P_s}[P_i;z] &= \int_0^\infty dP_t \  \fker{P_s}{P_t}{P_1}{P_2}{P_3}{P_4} \ \mathcal{V}_{P_t}[P_i;1-z] \ .
\eea\ee
The torus one-point blocks are normalized as $\mathcal{F}_{P_1}[P_0;\tau]=q^{\Delta_{1}-\frac{c}{24}}\left(1+O(q)\right)$ 
with nome $q\equiv \e^{2\pi i \tau}$ and $\tau\in\mathbb{H}_+$, and the sphere four-point block is normalized as 
$\mathcal{V}_{P_s}[P_i;z] = z^{\Delta_s - \Delta_1 - \Delta_2}(1+O(z))$. We call $\mathbb{S}$ the \emph{one-point S-kernel} and $\mathbb{F}$ the \emph{fusion kernel}. 

Pictorially, the one-point S-kernel implements the transformation between two canonical pair-of-pants decompositions of the one-holed torus:
\begin{equation}
   \begin{tikzpicture}[x=0.75pt,y=0.75pt,yscale=-1,xscale=1,baseline={([yshift=-.5ex]current bounding box.center)}]
\draw  [draw opacity=0] (286.51,134.45) .. controls (280.84,142.26) and (269.89,147.55) .. (257.32,147.55) .. controls (238.92,147.55) and (224,136.21) .. (224,122.22) .. controls (224,108.23) and (238.92,96.89) .. (257.32,96.89) .. controls (268.86,96.89) and (279.03,101.34) .. (285.01,108.11) -- (257.32,122.22) -- cycle ; \draw   (286.51,134.45) .. controls (280.84,142.26) and (269.89,147.55) .. (257.32,147.55) .. controls (238.92,147.55) and (224,136.21) .. (224,122.22) .. controls (224,108.23) and (238.92,96.89) .. (257.32,96.89) .. controls (268.86,96.89) and (279.03,101.34) .. (285.01,108.11) ;  
\draw  [draw opacity=0] (275.6,119.69) .. controls (271.92,125.6) and (265.1,129.57) .. (257.29,129.57) .. controls (249.37,129.57) and (242.46,125.48) .. (238.82,119.42) -- (257.29,109.57) -- cycle ; \draw   (275.6,119.69) .. controls (271.92,125.6) and (265.1,129.57) .. (257.29,129.57) .. controls (249.37,129.57) and (242.46,125.48) .. (238.82,119.42) ;  
\draw  [draw opacity=0] (241.5,122.47) .. controls (244.74,118.08) and (250.69,115.14) .. (257.49,115.14) .. controls (264.11,115.14) and (269.92,117.93) .. (273.22,122.12) -- (257.49,130.14) -- cycle ; \draw   (241.5,122.47) .. controls (244.74,118.08) and (250.69,115.14) .. (257.49,115.14) .. controls (264.11,115.14) and (269.92,117.93) .. (273.22,122.12) ;  
\draw   (293.44,122.13) .. controls (293.44,118.88) and (294.46,116.25) .. (295.72,116.25) .. controls (296.98,116.25) and (298,118.88) .. (298,122.13) .. controls (298,125.37) and (296.98,128) .. (295.72,128) .. controls (294.46,128) and (293.44,125.37) .. (293.44,122.13) -- cycle ;
\draw  [draw opacity=0] (258.35,146.61) .. controls (258.07,146.7) and (257.79,146.75) .. (257.49,146.75) .. controls (254.87,146.75) and (252.75,143.03) .. (252.75,138.45) .. controls (252.75,133.86) and (254.87,130.14) .. (257.49,130.14) .. controls (258.43,130.14) and (259.31,130.62) .. (260.04,131.44) -- (257.49,138.45) -- cycle ; \draw  [color={rgb, 255:red, 208; green, 2; blue, 27 }  ,draw opacity=1 ] (258.35,146.61) .. controls (258.07,146.7) and (257.79,146.75) .. (257.49,146.75) .. controls (254.87,146.75) and (252.75,143.03) .. (252.75,138.45) .. controls (252.75,133.86) and (254.87,130.14) .. (257.49,130.14) .. controls (258.43,130.14) and (259.31,130.62) .. (260.04,131.44) ;  
\draw    (284.87,107.93) .. controls (289.03,112.42) and (291.85,115.75) .. (295.72,116.25) ;
\draw    (286.4,134.6) .. controls (288.77,132.08) and (291.47,128.92) .. (295.72,128) ;
\draw  [draw opacity=0][dash pattern={on 1.5pt off 1.5pt on 1.5pt off 1.5pt}] (260.89,132.66) .. controls (261.73,134.15) and (262.24,136.19) .. (262.24,138.45) .. controls (262.24,142.45) and (260.62,145.79) .. (258.47,146.58) -- (257.49,138.45) -- cycle ; \draw  [color={rgb, 255:red, 208; green, 2; blue, 27 }  ,draw opacity=1 ][dash pattern={on 1.5pt off 1.5pt on 1.5pt off 1.5pt}] (260.89,132.66) .. controls (261.73,134.15) and (262.24,136.19) .. (262.24,138.45) .. controls (262.24,142.45) and (260.62,145.79) .. (258.47,146.58) ;  
\draw (298.71,114.69) node [anchor=north west][inner sep=0.75pt]  [font=\small]  {$P_{0}$};
\draw (237.34,131.88) node [anchor=north west][inner sep=0.75pt]  [font=\scriptsize,color={rgb, 255:red, 208; green, 2; blue, 27 }  ,opacity=1 ]  {$P_{1}$};
\end{tikzpicture} = \int_0^\infty \dd P\,\mathbb{S}_{P_1P}[P_0]\; \begin{tikzpicture}[x=0.75pt,y=0.75pt,yscale=-1,xscale=1,baseline={([yshift=-.5ex]current bounding box.center)}]
\draw  [draw opacity=0] (266.51,114.45) .. controls (260.84,122.26) and (249.89,127.55) .. (237.32,127.55) .. controls (218.92,127.55) and (204,116.21) .. (204,102.22) .. controls (204,88.23) and (218.92,76.89) .. (237.32,76.89) .. controls (248.86,76.89) and (259.03,81.34) .. (265.01,88.11) -- (237.32,102.22) -- cycle ; \draw   (266.51,114.45) .. controls (260.84,122.26) and (249.89,127.55) .. (237.32,127.55) .. controls (218.92,127.55) and (204,116.21) .. (204,102.22) .. controls (204,88.23) and (218.92,76.89) .. (237.32,76.89) .. controls (248.86,76.89) and (259.03,81.34) .. (265.01,88.11) ;  
\draw  [draw opacity=0] (255.6,99.69) .. controls (251.92,105.6) and (245.1,109.57) .. (237.29,109.57) .. controls (229.37,109.57) and (222.46,105.48) .. (218.82,99.42) -- (237.29,89.57) -- cycle ; \draw   (255.6,99.69) .. controls (251.92,105.6) and (245.1,109.57) .. (237.29,109.57) .. controls (229.37,109.57) and (222.46,105.48) .. (218.82,99.42) ;  
\draw  [draw opacity=0] (221.5,102.47) .. controls (224.74,98.08) and (230.69,95.14) .. (237.49,95.14) .. controls (244.11,95.14) and (249.92,97.93) .. (253.22,102.12) -- (237.49,110.14) -- cycle ; \draw   (221.5,102.47) .. controls (224.74,98.08) and (230.69,95.14) .. (237.49,95.14) .. controls (244.11,95.14) and (249.92,97.93) .. (253.22,102.12) ;  
\draw   (273.44,102.13) .. controls (273.44,98.88) and (274.46,96.25) .. (275.72,96.25) .. controls (276.98,96.25) and (278,98.88) .. (278,102.13) .. controls (278,105.37) and (276.98,108) .. (275.72,108) .. controls (274.46,108) and (273.44,105.37) .. (273.44,102.13) -- cycle ;
\draw  [color={rgb, 255:red, 208; green, 2; blue, 27 }  ,draw opacity=1 ] (212.68,102.22) .. controls (212.68,91.19) and (223.72,82.25) .. (237.32,82.25) .. controls (250.93,82.25) and (261.97,91.19) .. (261.97,102.22) .. controls (261.97,113.25) and (250.93,122.19) .. (237.32,122.19) .. controls (223.72,122.19) and (212.68,113.25) .. (212.68,102.22) -- cycle ;
\draw    (264.87,87.93) .. controls (269.03,92.42) and (271.85,95.75) .. (275.72,96.25) ;
\draw    (266.4,114.6) .. controls (268.77,112.08) and (271.47,108.92) .. (275.72,108) ;
\draw (279.71,94.69) node [anchor=north west][inner sep=0.75pt]  [font=\small]  {$P_{0}$};
\draw (233.71,111.26) node [anchor=north west][inner sep=0.75pt]  [font=\scriptsize,color={rgb, 255:red, 208; green, 2; blue, 27 }  ,opacity=1 ]  {$P$};
\end{tikzpicture}.
\end{equation}
Similarly, the fusion kernel implements the change of pair-of-pants decomposition of the four-holed sphere from the $s$-channel to the $t$-channel :
    \begin{equation}
    \begin{tikzpicture}[x=0.75pt,y=0.75pt,yscale=-0.5,xscale=0.5,baseline={([yshift=-0.5ex]current bounding box.center)}]
\draw  [draw opacity=0]  (225.66,257.5) .. controls (214.69,257.25) and (205.83,233.69) .. (205.83,204.67) .. controls (205.83,175.65) and (214.69,152.1) .. (225.65,151.84) -- (225.83,204.67) -- cycle ; \draw  [color={rgb, 255:red, 208; green, 2; blue, 27 }  ,draw opacity=1 ]  (225.66,257.5) .. controls (214.69,257.25) and (205.83,233.69) .. (205.83,204.67) .. controls (205.83,175.65) and (214.69,152.1) .. (225.65,151.84) ;  
\draw  [draw opacity=0][dash pattern={on 1.5pt off 1.5pt on 1.5pt off 1.5pt}]  (226.88,151.9) .. controls (237.44,153.35) and (245.83,176.42) .. (245.83,204.67) .. controls (245.83,233.85) and (236.88,257.5) .. (225.83,257.5) .. controls (225.73,257.5) and (225.62,257.5) .. (225.51,257.49) -- (225.83,204.67) -- cycle ; \draw  [color={rgb, 255:red, 208; green, 2; blue, 27 }  ,draw opacity=1 ][dash pattern={on 1.5pt off 1.5pt on 1.5pt off 1.5pt}]  (226.88,151.9) .. controls (237.44,153.35) and (245.83,176.42) .. (245.83,204.67) .. controls (245.83,233.85) and (236.88,257.5) .. (225.83,257.5) .. controls (225.73,257.5) and (225.62,257.5) .. (225.51,257.49) ;  
\draw   (159.9,169.9) .. controls (154.43,164.43) and (156.38,153.62) .. (164.26,145.74) .. controls (172.14,137.86) and (182.95,135.91) .. (188.42,141.38) .. controls (193.89,146.85) and (191.93,157.66) .. (184.06,165.54) .. controls (176.18,173.41) and (165.37,175.37) .. (159.9,169.9) -- cycle ;
\draw   (261.38,141.58) .. controls (266.85,136.11) and (277.66,138.07) .. (285.54,145.94) .. controls (293.41,153.82) and (295.37,164.63) .. (289.9,170.1) .. controls (284.43,175.57) and (273.62,173.62) .. (265.74,165.74) .. controls (257.86,157.86) and (255.91,147.05) .. (261.38,141.58) -- cycle ;
\draw   (188.71,268.05) .. controls (183.24,273.52) and (172.42,271.56) .. (164.55,263.69) .. controls (156.67,255.81) and (154.72,245) .. (160.19,239.53) .. controls (165.65,234.06) and (176.47,236.01) .. (184.34,243.89) .. controls (192.22,251.77) and (194.17,262.58) .. (188.71,268.05) -- cycle ;
\draw   (290.19,239.33) .. controls (295.65,244.8) and (293.7,255.61) .. (285.82,263.49) .. controls (277.95,271.36) and (267.13,273.32) .. (261.67,267.85) .. controls (256.2,262.38) and (258.15,251.56) .. (266.03,243.69) .. controls (273.9,235.81) and (284.72,233.86) .. (290.19,239.33) -- cycle ;
\draw    (188.42,141.38) .. controls (207,154.25) and (243,155.25) .. (261.38,141.58) ;
\draw    (291.64,167.64) .. controls (278.77,186.22) and (277.62,222.41) .. (291.29,240.79) ;
\draw    (261.67,267.85) .. controls (243.08,254.98) and (207.08,254.38) .. (188.71,268.05) ;
\draw    (158.58,241.46) .. controls (171.46,222.88) and (172.46,186.88) .. (158.79,168.5) ;
\draw (210.5,194.3) node [anchor=north west][inner sep=0.75pt]  [font=\scriptsize]  {$P_s$};
\draw (161.5,145.9) node [anchor=north west][inner sep=0.75pt]  [font=\tiny]  {$P_{2}$};
\draw (263,244.4) node [anchor=north west][inner sep=0.75pt]  [font=\tiny]  {$P_{4}$};
\draw (265.5,147.4) node [anchor=north west][inner sep=0.75pt]  [font=\tiny]  {$P_{1}$};
\draw (164,244.4) node [anchor=north west][inner sep=0.75pt]  [font=\tiny]  {$P_{3}$};
\end{tikzpicture} \,\,=\,\, \int_0^\infty \dd P_t\,\fker{P_s}{P_t}{P_1}{P_2}{P_3}{P_4} \,\,\begin{tikzpicture}[x=0.75pt,y=0.75pt,yscale=-0.5,xscale=0.5,baseline={([yshift=-0.5ex]current bounding box.center)}]
\draw  [draw opacity=0]  (261,184.85) .. controls (260.72,195.81) and (235.83,204.67) .. (205.17,204.67) .. controls (174.51,204.67) and (149.63,195.82) .. (149.34,184.86) -- (205.17,184.67) -- cycle ; \draw  [color={rgb, 255:red, 208; green, 2; blue, 27 }  ,draw opacity=1 ]  (261,184.85) .. controls (260.72,195.81) and (235.83,204.67) .. (205.17,204.67) .. controls (174.51,204.67) and (149.63,195.82) .. (149.34,184.86) ; 
\draw  [draw opacity=0][dash pattern={on 1.5pt off 1.5pt on 1.5pt off 1.5pt}]  (149.42,183.56) .. controls (151.02,173.03) and (175.37,164.67) .. (205.17,164.67) .. controls (236,164.67) and (261,173.62) .. (261,184.67) .. controls (261,184.78) and (261,184.89) .. (260.99,185.01) -- (205.17,184.67) -- cycle ; \draw  [color={rgb, 255:red, 208; green, 2; blue, 27 }  ,draw opacity=1 ][dash pattern={on 1.5pt off 1.5pt on 1.5pt off 1.5pt}]  (149.42,183.56) .. controls (151.02,173.03) and (175.37,164.67) .. (205.17,164.67) .. controls (236,164.67) and (261,173.62) .. (261,184.67) .. controls (261,184.78) and (261,184.89) .. (260.99,185.01) ;  
\draw   (139.9,149.9) .. controls (134.43,144.43) and (136.38,133.62) .. (144.26,125.74) .. controls (152.14,117.86) and (162.95,115.91) .. (168.42,121.38) .. controls (173.89,126.85) and (171.93,137.66) .. (164.06,145.54) .. controls (156.18,153.41) and (145.37,155.37) .. (139.9,149.9) -- cycle ;
\draw   (241.38,121.58) .. controls (246.85,116.11) and (257.66,118.07) .. (265.54,125.94) .. controls (273.41,133.82) and (275.37,144.63) .. (269.9,150.1) .. controls (264.43,155.57) and (253.62,153.62) .. (245.74,145.74) .. controls (237.86,137.86) and (235.91,127.05) .. (241.38,121.58) -- cycle ;
\draw   (168.71,248.05) .. controls (163.24,253.52) and (152.42,251.56) .. (144.55,243.69) .. controls (136.67,235.81) and (134.72,225) .. (140.19,219.53) .. controls (145.65,214.06) and (156.47,216.01) .. (164.34,223.89) .. controls (172.22,231.77) and (174.17,242.58) .. (168.71,248.05) -- cycle ;
\draw   (270.19,219.33) .. controls (275.65,224.8) and (273.7,235.61) .. (265.82,243.49) .. controls (257.95,251.36) and (247.13,253.32) .. (241.67,247.85) .. controls (236.2,242.38) and (238.15,231.56) .. (246.03,223.69) .. controls (253.9,215.81) and (264.72,213.86) .. (270.19,219.33) -- cycle ;
\draw    (168.42,121.38) .. controls (187,134.25) and (223,135.25) .. (241.38,121.58) ;
\draw    (271.64,147.64) .. controls (258.77,166.22) and (257.62,202.41) .. (271.29,220.79) ;
\draw    (241.67,247.85) .. controls (223.08,234.98) and (187.08,234.38) .. (168.71,248.05) ;
\draw    (138.58,221.46) .. controls (151.46,202.88) and (152.46,166.88) .. (138.79,148.5) ;
\draw (198.5,178.9) node [anchor=north west][inner sep=0.75pt]  [font=\scriptsize]  {$P_{t}$};
\draw (141,125.9) node [anchor=north west][inner sep=0.75pt]  [font=\tiny]  {$P_{2}$};
\draw (244,224.4) node [anchor=north west][inner sep=0.75pt]  [font=\tiny]  {$P_{4}$};
\draw (244.5,127.4) node [anchor=north west][inner sep=0.75pt]  [font=\tiny]  {$P_{1}$};
\draw (143,224.4) node [anchor=north west][inner sep=0.75pt]  [font=\tiny]  {$P_{3}$};
\end{tikzpicture}.
\end{equation}

Combining the one-point S-kernel and fusion kernel with the braiding phase 
\begin{equation}
    \mathbb{B}_{P_1}^{P_2P_3} = \e^{\pi i (\Delta_1-\Delta_2-\Delta_3)}
\end{equation}
leads to a non-rational generalization of the Moore-Seiberg construction for rational theories \cite{Moore:1988qv,Teschner:2005bz}. In other words, any element $\gamma$ of the mapping class group of given punctured Riemann surface $\Sigma_{g,n}$ can be represented as an integral kernel $\mathbb{K}^\gamma$ acting on the corresponding conformal block $\mathcal{F}_{\Sigma_{g,n}}$. The generators $\mathbb{S}$, $\mathbb{F}$ and $\mathbb{B}$ satisfy a continuous version of the pentagon identity and hexagon identity, as well as fundamental identities on the 0-, 1- and 2-punctured torus \cite{Eberhardt:2023mrq}. The kernels are also invertible, in a distributional sense:
\be
\bea
    \int_0^\infty \dd P_t\,\fker{P_s}{P_t}{P_1}{P_2}{P_3}{P_4}\fker{P_t}{P_s'}{P_2}{P_3}{P_4}{P_1} = \delta(P_s-P_s') \\[1em]
    \int_0^\infty \dd P_2 \,\mathbb{S}_{P_1P_2}[P_0]
\,\mathbb{S}^*_{P_2P_3}[P_0] = \delta(P_1-P_3).
\eea
\ee
 Remarkably, these crossing kernels are known in closed form, as originally derived in \cite{Ponsot:1999uf,Ponsot:2000mt}
. In the next appendix \ref{app:kernels}, we write down their explicit form. They involve the use of the special function $S_b(z)$, known as the \emph{double sine function}, which is closely related to Faddeev's quantum dilogarithm $\Phi_b(z)$ \cite{Faddeev:1993rs} and the Barnes double gamma function $\Gamma_b(z)$. 

The one-point S-kernel reduces in the limit $P_0\to\bbi$ to the modular S-kernel $\sker{P'}{P}{\bbi}$ given in \eqref{eq:torus_mod_S}. The fusion kernel also has a well-defined limit when the first internal index goes to the identity and the external operators are pairwise-identical:
\begin{equation}\label{eq:C0}
  \fker{\bbi}{P_1}{P_3}{P_2}{P_2}{P_3} = \rho_0(P_1)C_0(P_1,P_2,P_3).\vspace{1mm}
\end{equation}
The density of states $\rho_0(P)$ was defined in \eqref{eq:rho_0} and $C_0$ is a totally symmetric function (see equation \eqref{eq:C_0_function}) that controls the leading asymptotic growth of high energy OPE coefficients $|C_{123}|^2$ in any non-rational $c>1$ CFT$_2$ with only Virasoro symmetry \cite{Collier_2020}.

Lastly, we note here the relation between the Virasoro $6j$ symbol and the fusion kernel \cite{Eberhardt:2023mrq}:
\begin{equation}\label{eq:6j_fusion}
    \begin{Bmatrix}
  P_4 & P_1 & P_s \\
  P_2 & P_3 & P_t 
 \end{Bmatrix}_{6j} = \frac{1}{\rho_0(P_t)} \sqrt{\frac{C_0(P_1,P_4,P_s)C_0(P_2,P_3,P_s)}{C_0(P_1,P_2,P_t)C_0(P_3,P_4,P_t)}} \,\fker{P_s}{P_t}{P_1}{P_2}{P_3}{P_4}.
\end{equation}
This relation has been used in the main text to connect the fusion density to the Virasoro $6j$ symbol. The normalization  \eqref{eq:6j_fusion} of the fusion kernel is called the `Racah-Wigner normalization', as the $6j$ symbol plays the role of the Racah-Wigner coefficient for a continuous series representation of the quantum group $\mathcal{U}_q(\mathfrak{sl}(2,\mathbb{R}))$ \cite{Ponsot:2000mt}.

\subsection{Integral form}\label{app:kernels}

Here we write down the closed-form expressions for the crossing kernels used in the main text. They are expressed in terms of the Barnes double gamma function $\Gamma_b(z)$, and the double sine function $S_b(z) = \Gamma_b(z)/\Gamma_b(Q-z)$. See \cite{Eberhardt:2023mrq} for a detailed derivation of these formulas.

The one-point S-kernel is given by an integral over a ratio of double sine functions:
\begin{equation}\label{eq:one-point_S}
    \mathbb{S}_{P_1P_2}[P_0] = \e^{\frac{i\pi}{2}\Delta_0} \,N_b(P_1,P_2)\,\rho_0(P_2) \int_{\mathbb{R}} \dd x\, \frac{S_b(\frac{\mu}{2} \pm i P_2\pm i x)}{S_b(\mu)}\,\e^{4\pi i P_1x}.
\end{equation}
We defined $\mu \coloneqq \frac{Q}{2}+iP_0$, and the normalization factor $N_b$ is a ratio of Barnes double gamma functions:
\begin{equation}
    N_b(P_1,P_2) \coloneqq \frac{\Gamma_b(Q\pm 2iP_1)\Gamma_b(\tfrac{Q}{2}-iP_0\pm 2i P_2)}{\Gamma_b(Q\pm 2i P_2)\Gamma_b(\tfrac{Q}{2}-iP_0\pm 2i P_1)}.
\end{equation}
The notation $\pm$ denotes a product over all possible choices of sign, for example $f(a\pm b) = f(a+b)f(a-b)$. The integral representation converges for $P_{0,1,2}\in \mathbb{R}$ and can be analytically continued to a meromorphic function for complex arguments. 

The fusion kernel is also an integral over a ratio of double sine functions:\footnote{Note that this formula differs from equation (2.10) in \cite{Collier:2018exn} by the exchange of $s\leftrightarrow t$. Compared to \cite{Collier:2018exn} our conventions are $\mathbb{F}_{P_s P_t}^{\text{here}} \sbmatrix{P_2 & P_1 \\ P_3 & P_4} = \mathbb{F}_{P_t P_s}^{\text{there}} \sbmatrix{P_2 & P_1 \\ P_3 & P_4}$. Our convention aligns with \cite{Eberhardt:2023mrq}.}
\begin{equation}\label{eq:F_kernel_integral}
    \fker{P_s}{P_t}{P_1}{P_2}{P_3}{P_4} = M_b(P_i;P_s,P_t) \int_{\frac{Q}{4}+i\mathbb{R}}\frac{dz}{i}\prod_{k=1}^4\frac{S_b(z+u_k)}{S_b(z+v_k)},
\end{equation}
where the normalization factor is given by
\begin{equation}
    M_b(P_i;P_s,P_t) = \frac{\Gamma_b(Q\pm 2iP_s)}{\Gamma_b(\pm 2iP_t)} \frac{\Gamma_b(\frac{Q}{2}\pm iP_t\pm iP_3-iP_4)\Gamma_b(\frac{Q}{2}\pm iP_t \pm i P_1+iP_2)}{\Gamma_b(\frac{Q}{2}\pm iP_s \pm iP_1-iP_4)\Gamma_b(\frac{Q}{2}\pm iP_s \pm iP_3+iP_2)},
\end{equation}
and we defined the variables
\begin{equation}
    \bea 
    u_1 &= iP_1-iP_4,  &&v_1 =  \tfrac{Q}{2}-iP_t + iP_2-iP_4\\
    u_2 & = -iP_1-iP_4,  &&v_2= \tfrac{Q}{2}+iP_t + iP_2-iP_4 \\
    u_3 &= i P_2+iP_3,  &&v_3= \tfrac{Q}{2}+iP_s\\
    u_4 &= iP_2-iP_3, &&v_4= \tfrac{Q}{2} -iP_s.
    \eea
\end{equation}

Lastly, the explicit form of the $C_0$ function is given by:
\begin{equation}\label{eq:C_0_function}
    C_0(P_1,P_2,P_3) = \frac{\Gamma_b(2Q)\Gamma_b(\frac{Q}{2}\pm i P_1 \pm i P_2 \pm i P_3)}{\sqrt{2}\Gamma_b(Q)^3\prod_{k=1}^3\Gamma_b(Q\pm 2i P_k)}.
\end{equation}
This totally symmetric function is proportional to the identity fusion kernel via \eqref{eq:C0}.

\subsection{Virasoro TQFT}

The Virasoro TQFT was introduced in \cite{Collier_2023} as a tool to quantize pure AdS$_3$ gravity on a given (fixed) hyperbolic 3-manifold. The phase space of 3d gravity on a codimension-one slice $\Sigma$ is the product of Teichm\"uller spaces $\mathcal{T}_\Sigma\times \overline{\mathcal{T}}_\Sigma$, and on each copy of Teichm\"uller space one considers the line bundle of Virasoro conformal blocks \cite{Verlinde:1989ua,Verlinde:1989hv}. In the quantum theory, wavefunctions (i.e.\! sections of this line bundle) are products of left- and right-moving conformal blocks $\mathcal{F}_\Sigma\,\overline{\mathcal{F}}_\Sigma$. If $\Sigma$ is a stable surface with genus $g$ and $n$ punctures, the Hilbert space of (Liouville) conformal blocks $\mathcal{H}_{\text{Vir}}(\Sigma_{g,n})$ is equipped with an inner product \cite{Collier_2023}:
\begin{equation}\label{eq:inner_product}
    \braket{\mathcal{F}^{\mathcal{C}}_{\Sigma_{g,n}}(\mathbf{P},\mathbf{P}_\O)}{\mathcal{F}^{\,\mathcal{C}}_{\Sigma_{g,n}}(\mathbf{P}',\mathbf{P}_\O)} = \frac{\delta^{3g-3+n}(\mathbf{P}-\mathbf{P}')}{\rho_{g,n}(\mathbf{P};\mathbf{P}_{\O})}.
\end{equation}
Here $\mathcal{C}$ denotes a particular choice of pair-of-pants decomposition of the surface $\Sigma_{g,n}$ (also called an \emph{OPE channel}) and the `internal momenta' $\mathbf{P} = (P_1,\dots,P_{3g-3+n})$ are associated to the set of non-intersecting cycles defining the pair-of-pants decomposition. The `external momenta' $\mathbf{P}_\O = (P_{\O_1},\dots, P_{\O_n})$ label the $n$ punctures. 

The normalization factor appearing in the inner product \eqref{eq:inner_product} is given by\vspace{1mm}
\begin{equation}
    \rho_{g,n}(\mathbf{P};\mathbf{P}_{\O}) = \prod_{a} \rho_0(P_a)\prod_{(i,j,k)}C_0(P_i,P_j,P_k),
\end{equation}
where the index $a$ runs over all embedded cylinders (`cuffs') and the triples $(i,j,k)$ label the three-holed spheres (`pair-of-pants'), in the channel $\mathcal{C}$. In other words, if we represent the pair-of-pants decomposition as a trivalent graph, each edge is weighted by the $\rho_0$ function \eqref{eq:rho_0} and each vertex by the $C_0$ function \eqref{eq:C_0_function}.

The Hilbert space $\mathcal{H}_{\text{Vir}}(\Sigma_{g,n})$ carries a projective unitary representation of the mapping class group of $\Sigma_{g,n}$, through the action of the crossing kernels $\mathbb{F},\mathbb{S}$ and $\mathbb{B}$ discussed above. The action of the Moore-Seiberg generators together with the inner product \eqref{eq:inner_product} defines a 3d TQFT, called the Virasoro TQFT. As usual in TQFT, the 3d partition function $Z_{\text{Vir}}$ on a 3-manifold $M$ with boundary $\Sigma$ creates a state in the Hilbert space $\mathcal{H}_{\text{Vir}}(\Sigma)$:
\begin{equation}
Z_{\text{Vir}}\left(\hspace{-3mm}\begin{tikzpicture}[x=0.75pt,y=0.75pt,yscale=-0.7,xscale=0.7,baseline={([yshift=-0.5ex]current bounding box.center)}]
\draw    (230,146.67) .. controls (230,129.6) and (242.24,120) .. (260,120) .. controls (277.76,120) and (290,129.87) .. (290,146.67) .. controls (290,163.47) and (278,167.2) .. (278,180) .. controls (278,192.8) and (290.24,197.07) .. (290,213.33) .. controls (289.76,229.6) and (277.76,240) .. (260,240) .. controls (242.24,240) and (230,230.93) .. (230,213.33) .. controls (230,195.73) and (241.76,193.87) .. (242,180) .. controls (242.24,166.13) and (230,164) .. (230,146.67) -- cycle ;
\draw    (265.78,162.42) .. controls (251.78,164.02) and (250.58,142.02) .. (264.18,141.62) ;
\draw    (260.38,161.52) .. controls (266.18,160.62) and (267.78,143.22) .. (259.18,143.02) ;
\draw    (266.18,218.82) .. controls (252.18,220.42) and (250.98,198.42) .. (264.58,198.02) ;
\draw    (260.78,217.92) .. controls (266.58,217.02) and (268.18,199.62) .. (259.58,199.42) ;
\draw    (248.38,238.46) .. controls (167.98,223.66) and (167.7,137.7) .. (248.7,121.5) ;
\draw (202.4,170.6) node [anchor=north west][inner sep=0.75pt]    {$M$};
\draw (252.8,171.2) node [anchor=north west][inner sep=0.75pt]    {$\Sigma $};
\end{tikzpicture}\hspace{2mm}\right) \in \mathcal{H}_{\text{Vir}}(\Sigma).\vspace{1mm}
\end{equation}

Boundaries can be glued (with opposite orientation) using the inner product \eqref{eq:inner_product}. In this way, one can build the TQFT partition function on any hyperbolic 3-manifold by slicing along stable surfaces. Multiplying the holomorphic partition function by the anti-holomorphic counterpart, and dividing out by the mapping class group, one obtains the 3d gravity partition function as in \eqref{eq:grav_part} \cite{Collier_2023}.

\subsection{TQFT identities}\label{app:TQFT_identities}

Here we collect and prove some identities in Virasoro TQFT that are needed in the main text. These can be understood either as diagrammatic relations between configurations of bulk Wilson lines, or as non-perturbative identities satisfied by conformal blocks on (punctured) surfaces surrounding the Wilson lines.  

\subsection*{Unlinking formula}
We will start with a useful formula that applies when two bulk Wilson lines are linked:
\be
\begin{tikzpicture}[x=0.75pt,y=0.75pt,yscale=-1,xscale=1,baseline={([yshift=-0.5ex]current bounding box.center)}]
\draw  [dash pattern={on 1.5pt off 1.5pt on 1.5pt off 1.5pt}] (90,128.6) .. controls (90,100.16) and (113.06,77.1) .. (141.5,77.1) .. controls (169.94,77.1) and (193,100.16) .. (193,128.6) .. controls (193,157.04) and (169.94,180.1) .. (141.5,180.1) .. controls (113.06,180.1) and (90,157.04) .. (90,128.6) -- cycle ;
\draw  [draw opacity=0][dash pattern={on 1.5pt off 1.5pt on 1.5pt off 1.5pt}] (193,127.41) .. controls (192.79,134.58) and (169.81,140.38) .. (141.5,140.38) .. controls (113.19,140.38) and (90.21,134.58) .. (90,127.41) -- (141.5,127.31) -- cycle ; \draw  [dash pattern={on 1.5pt off 1.5pt on 1.5pt off 1.5pt}] (193,127.41) .. controls (192.79,134.58) and (169.81,140.38) .. (141.5,140.38) .. controls (113.19,140.38) and (90.21,134.58) .. (90,127.41) ;  
\draw  [draw opacity=0] (115.39,99.82) .. controls (135.47,100.16) and (151.63,113.57) .. (151.63,130.07) .. controls (151.63,136.81) and (148.93,143.04) .. (144.37,148.06) -- (114.63,130.07) -- cycle ; \draw   (115.39,99.82) .. controls (135.47,100.16) and (151.63,113.57) .. (151.63,130.07) .. controls (151.63,136.81) and (148.93,143.04) .. (144.37,148.06) ;  
\draw  [draw opacity=0] (144.24,106.82) .. controls (150.09,102.58) and (157.57,100.03) .. (165.71,100) -- (165.84,130.25) -- cycle ; \draw   (144.24,106.82) .. controls (150.09,102.58) and (157.57,100.03) .. (165.71,100) ;  
\draw  [draw opacity=0] (136.98,154.17) .. controls (130.96,157.91) and (123.49,160.18) .. (115.39,160.31) -- (114.63,130.07) -- cycle ; \draw   (136.98,154.17) .. controls (130.96,157.91) and (123.49,160.18) .. (115.39,160.31) ;  
\draw  [draw opacity=0] (165.69,160.5) .. controls (146.9,160.43) and (131.69,146.91) .. (131.69,130.25) .. controls (131.69,123.46) and (134.21,117.2) .. (138.47,112.15) -- (165.84,130.25) -- cycle ; \draw   (165.69,160.5) .. controls (146.9,160.43) and (131.69,146.91) .. (131.69,130.25) .. controls (131.69,123.46) and (134.21,117.2) .. (138.47,112.15) ;  
\draw   (165.69,160.5) .. controls (165.69,160.04) and (166.06,159.67) .. (166.52,159.67) .. controls (166.98,159.67) and (167.35,160.04) .. (167.35,160.5) .. controls (167.35,160.96) and (166.98,161.33) .. (166.52,161.33) .. controls (166.06,161.33) and (165.69,160.96) .. (165.69,160.5) -- cycle ;
\draw   (113.93,160.32) .. controls (113.93,159.86) and (114.31,159.49) .. (114.76,159.49) .. controls (115.22,159.49) and (115.59,159.86) .. (115.59,160.32) .. controls (115.59,160.78) and (115.22,161.15) .. (114.76,161.15) .. controls (114.31,161.15) and (113.93,160.78) .. (113.93,160.32) -- cycle ;
\draw   (165.71,100) .. controls (165.71,99.54) and (166.08,99.17) .. (166.54,99.17) .. controls (167,99.17) and (167.37,99.54) .. (167.37,100) .. controls (167.37,100.46) and (167,100.83) .. (166.54,100.83) .. controls (166.08,100.83) and (165.71,100.46) .. (165.71,100) -- cycle ;
\draw   (114.11,99.82) .. controls (114.11,99.36) and (114.48,98.99) .. (114.94,98.99) .. controls (115.39,98.99) and (115.76,99.36) .. (115.76,99.82) .. controls (115.76,100.28) and (115.39,100.65) .. (114.94,100.65) .. controls (114.48,100.65) and (114.11,100.28) .. (114.11,99.82) -- cycle ;
\draw (102.86,101.64) node [anchor=north west][inner sep=0.75pt]  [font=\small]  {$P_{1}$};
\draw (163.61,101.64) node [anchor=north west][inner sep=0.75pt]  [font=\small]  {$P_{2}$};
\end{tikzpicture}
\ee 
Inserting an identity line and applying fusion and braiding moves, we obtain:
\begin{align}
\begin{tikzpicture}[x=0.75pt,y=0.75pt,yscale=-1,xscale=1,baseline={([yshift=-0.5ex]current bounding box.center)}]
\draw  [draw opacity=0] (135.39,119.82) .. controls (155.47,120.16) and (171.63,133.57) .. (171.63,150.07) .. controls (171.63,156.81) and (168.93,163.04) .. (164.37,168.06) -- (134.63,150.07) -- cycle ; \draw   (135.39,119.82) .. controls (155.47,120.16) and (171.63,133.57) .. (171.63,150.07) .. controls (171.63,156.81) and (168.93,163.04) .. (164.37,168.06) ;  
\draw  [draw opacity=0] (164.24,126.82) .. controls (170.09,122.58) and (177.57,120.03) .. (185.71,120) -- (185.84,150.25) -- cycle ; \draw   (164.24,126.82) .. controls (170.09,122.58) and (177.57,120.03) .. (185.71,120) ;  
\draw  [draw opacity=0] (156.98,174.17) .. controls (150.96,177.91) and (143.49,180.18) .. (135.39,180.31) -- (134.63,150.07) -- cycle ; \draw   (156.98,174.17) .. controls (150.96,177.91) and (143.49,180.18) .. (135.39,180.31) ;  
\draw  [draw opacity=0] (185.69,180.5) .. controls (166.9,180.43) and (151.69,166.91) .. (151.69,150.25) .. controls (151.69,143.46) and (154.21,137.2) .. (158.47,132.15) -- (185.84,150.25) -- cycle ; \draw   (185.69,180.5) .. controls (166.9,180.43) and (151.69,166.91) .. (151.69,150.25) .. controls (151.69,143.46) and (154.21,137.2) .. (158.47,132.15) ;  
\draw   (185.69,180.5) .. controls (185.69,180.04) and (186.06,179.67) .. (186.52,179.67) .. controls (186.98,179.67) and (187.35,180.04) .. (187.35,180.5) .. controls (187.35,180.96) and (186.98,181.33) .. (186.52,181.33) .. controls (186.06,181.33) and (185.69,180.96) .. (185.69,180.5) -- cycle ;
\draw   (133.93,180.32) .. controls (133.93,179.86) and (134.31,179.49) .. (134.76,179.49) .. controls (135.22,179.49) and (135.59,179.86) .. (135.59,180.32) .. controls (135.59,180.78) and (135.22,181.15) .. (134.76,181.15) .. controls (134.31,181.15) and (133.93,180.78) .. (133.93,180.32) -- cycle ;
\draw   (185.71,120) .. controls (185.71,119.54) and (186.08,119.17) .. (186.54,119.17) .. controls (187,119.17) and (187.37,119.54) .. (187.37,120) .. controls (187.37,120.46) and (187,120.83) .. (186.54,120.83) .. controls (186.08,120.83) and (185.71,120.46) .. (185.71,120) -- cycle ;
\draw   (134.11,119.82) .. controls (134.11,119.36) and (134.48,118.99) .. (134.94,118.99) .. controls (135.39,118.99) and (135.76,119.36) .. (135.76,119.82) .. controls (135.76,120.28) and (135.39,120.65) .. (134.94,120.65) .. controls (134.48,120.65) and (134.11,120.28) .. (134.11,119.82) -- cycle ;
\draw  [dash pattern={on 1.5pt off 1.5pt on 1.5pt off 1.5pt}]  (152.59,150.29) -- (172.24,150.29) ;
\draw (122.86,121.64) node [anchor=north west][inner sep=0.75pt]  [font=\small]  {$P_{1}$};
\draw (183.61,121.64) node [anchor=north west][inner sep=0.75pt]  [font=\small]  {$P_{2}$};
\draw (156.59,136.34) node [anchor=north west][inner sep=0.75pt]  [font=\small]  {$\mathbb{1}$};
\end{tikzpicture} &= \int_0^\infty \dd P\,\fker{\bbi}{P}{P_2}{P_1}{P_1}{P_2}\begin{tikzpicture}[x=0.75pt,y=0.75pt,yscale=-1,xscale=1,baseline={([yshift=-0.5ex]current bounding box.center)}]
\draw  [draw opacity=0] (155.39,139.82) .. controls (166.13,140) and (175.75,143.92) .. (182.38,150.06) -- (154.63,170.07) -- cycle ; \draw   (155.39,139.82) .. controls (166.13,140) and (175.75,143.92) .. (182.38,150.06) ;  
\draw  [draw opacity=0] (184.24,146.82) .. controls (190.09,142.58) and (197.57,140.03) .. (205.71,140) -- (205.84,170.25) -- cycle ; \draw   (184.24,146.82) .. controls (190.09,142.58) and (197.57,140.03) .. (205.71,140) ;  
\draw  [draw opacity=0] (177.63,193.76) .. controls (171.5,197.75) and (163.79,200.17) .. (155.39,200.31) -- (154.63,170.07) -- cycle ; \draw   (177.63,193.76) .. controls (171.5,197.75) and (163.79,200.17) .. (155.39,200.31) ;  
\draw   (205.69,200.5) .. controls (205.69,200.04) and (206.06,199.67) .. (206.52,199.67) .. controls (206.98,199.67) and (207.35,200.04) .. (207.35,200.5) .. controls (207.35,200.96) and (206.98,201.33) .. (206.52,201.33) .. controls (206.06,201.33) and (205.69,200.96) .. (205.69,200.5) -- cycle ;
\draw   (153.93,200.32) .. controls (153.93,199.86) and (154.31,199.49) .. (154.76,199.49) .. controls (155.22,199.49) and (155.59,199.86) .. (155.59,200.32) .. controls (155.59,200.78) and (155.22,201.15) .. (154.76,201.15) .. controls (154.31,201.15) and (153.93,200.78) .. (153.93,200.32) -- cycle ;
\draw   (205.71,140) .. controls (205.71,139.54) and (206.08,139.17) .. (206.54,139.17) .. controls (207,139.17) and (207.37,139.54) .. (207.37,140) .. controls (207.37,140.46) and (207,140.83) .. (206.54,140.83) .. controls (206.08,140.83) and (205.71,140.46) .. (205.71,140) -- cycle ;
\draw   (154.11,139.82) .. controls (154.11,139.36) and (154.48,138.99) .. (154.94,138.99) .. controls (155.39,138.99) and (155.76,139.36) .. (155.76,139.82) .. controls (155.76,140.28) and (155.39,140.65) .. (154.94,140.65) .. controls (154.48,140.65) and (154.11,140.28) .. (154.11,139.82) -- cycle ;
\draw    (182.38,150.06) .. controls (190.8,159.27) and (173.47,160.47) .. (178.47,152.15) ;
\draw  [draw opacity=0] (205.29,200.5) .. controls (195.79,200.36) and (187.23,196.79) .. (181.13,191.13) -- (205.84,170.25) -- cycle ; \draw   (205.29,200.5) .. controls (195.79,200.36) and (187.23,196.79) .. (181.13,191.13) ;  
\draw    (181.13,191.13) .. controls (169.19,180.94) and (190.46,179.62) .. (184,188.54) ;
\draw    (181,157.96) -- (181.08,182.71) ;
\draw (142.86,134.89) node [anchor=north west][inner sep=0.75pt]  [font=\footnotesize]  {$1$};
\draw (165.24,163) node [anchor=north west][inner sep=0.75pt]  [font=\small]  {$P$};
\draw (144.06,193.69) node [anchor=north west][inner sep=0.75pt]  [font=\footnotesize]  {$1$};
\draw (208.86,193.69) node [anchor=north west][inner sep=0.75pt]  [font=\footnotesize]  {$2$};
\draw (208.86,134.29) node [anchor=north west][inner sep=0.75pt]  [font=\footnotesize]  {$2$};
\end{tikzpicture} \\[1em]
&= \int_0^\infty \dd P\,\fker{\bbi}{P}{P_2}{P_1}{P_1}{P_2} (\mathbb{B}_P^{P_1P_2})^2\begin{tikzpicture}[x=0.75pt,y=0.75pt,yscale=-1,xscale=1,baseline={([yshift=-0.5ex]current bounding box.center)}]
\draw   (225.69,220.5) .. controls (225.69,220.04) and (226.06,219.67) .. (226.52,219.67) .. controls (226.98,219.67) and (227.35,220.04) .. (227.35,220.5) .. controls (227.35,220.96) and (226.98,221.33) .. (226.52,221.33) .. controls (226.06,221.33) and (225.69,220.96) .. (225.69,220.5) -- cycle ;
\draw   (173.93,220.32) .. controls (173.93,219.86) and (174.31,219.49) .. (174.76,219.49) .. controls (175.22,219.49) and (175.59,219.86) .. (175.59,220.32) .. controls (175.59,220.78) and (175.22,221.15) .. (174.76,221.15) .. controls (174.31,221.15) and (173.93,220.78) .. (173.93,220.32) -- cycle ;
\draw   (225.71,160) .. controls (225.71,159.54) and (226.08,159.17) .. (226.54,159.17) .. controls (227,159.17) and (227.37,159.54) .. (227.37,160) .. controls (227.37,160.46) and (227,160.83) .. (226.54,160.83) .. controls (226.08,160.83) and (225.71,160.46) .. (225.71,160) -- cycle ;
\draw   (174.11,159.82) .. controls (174.11,159.36) and (174.48,158.99) .. (174.94,158.99) .. controls (175.39,158.99) and (175.76,159.36) .. (175.76,159.82) .. controls (175.76,160.28) and (175.39,160.65) .. (174.94,160.65) .. controls (174.48,160.65) and (174.11,160.28) .. (174.11,159.82) -- cycle ;
\draw    (201,177.96) -- (201.08,202.71) ;
\draw    (175.74,160.48) -- (201,177.96) ;
\draw    (201,177.96) -- (225.65,160.74) ;
\draw    (201.08,202.71) -- (225.74,219.96) ;
\draw    (175.57,219.7) -- (201.08,202.71) ;
\draw (162.86,154.89) node [anchor=north west][inner sep=0.75pt]  [font=\footnotesize]  {$1$};
\draw (185.24,183) node [anchor=north west][inner sep=0.75pt]  [font=\small]  {$P$};
\draw (164.06,213.29) node [anchor=north west][inner sep=0.75pt]  [font=\footnotesize]  {$1$};
\draw (228.86,213.69) node [anchor=north west][inner sep=0.75pt]  [font=\footnotesize]  {$2$};
\draw (229.86,154.29) node [anchor=north west][inner sep=0.75pt]  [font=\footnotesize]  {$2$};
\end{tikzpicture} \\[1em]
&= \int_0^\infty \dd P\dd P'\,\fker{\bbi}{P}{P_2}{P_1}{P_1}{P_2} (\mathbb{B}_P^{P_1P_2})^2\fker{P}{P'}{P_2}{P_2}{P_1}{P_1} \begin{tikzpicture}[x=0.75pt,y=0.75pt,yscale=-1,xscale=1,baseline={([yshift=-0.5ex]current bounding box.center)}]
\draw   (255.38,240.81) .. controls (255.38,240.35) and (255.75,239.98) .. (256.21,239.98) .. controls (256.67,239.98) and (257.04,240.35) .. (257.04,240.81) .. controls (257.04,241.26) and (256.67,241.64) .. (256.21,241.64) .. controls (255.75,241.64) and (255.38,241.26) .. (255.38,240.81) -- cycle ;
\draw   (193.93,240.32) .. controls (193.93,239.86) and (194.31,239.49) .. (194.76,239.49) .. controls (195.22,239.49) and (195.59,239.86) .. (195.59,240.32) .. controls (195.59,240.78) and (195.22,241.15) .. (194.76,241.15) .. controls (194.31,241.15) and (193.93,240.78) .. (193.93,240.32) -- cycle ;
\draw   (255.4,180.31) .. controls (255.4,179.85) and (255.77,179.48) .. (256.23,179.48) .. controls (256.69,179.48) and (257.06,179.85) .. (257.06,180.31) .. controls (257.06,180.77) and (256.69,181.14) .. (256.23,181.14) .. controls (255.77,181.14) and (255.4,180.77) .. (255.4,180.31) -- cycle ;
\draw   (194.11,179.82) .. controls (194.11,179.36) and (194.48,178.99) .. (194.94,178.99) .. controls (195.39,178.99) and (195.76,179.36) .. (195.76,179.82) .. controls (195.76,180.28) and (195.39,180.65) .. (194.94,180.65) .. controls (194.48,180.65) and (194.11,180.28) .. (194.11,179.82) -- cycle ;
\draw    (195.74,180.48) -- (210.38,210.38) ;
\draw    (240.32,210.38) -- (255.34,181.05) ;
\draw    (240.32,210.38) -- (255.43,240.32) ;
\draw    (195.57,240.32) -- (210.38,210.38) ;
\draw    (210.38,210.38) -- (240.32,210.38) ;
\draw (182.86,174.89) node [anchor=north west][inner sep=0.75pt]  [font=\footnotesize]  {$1$};
\draw (218.74,193.75) node [anchor=north west][inner sep=0.75pt]  [font=\small]  {$P'$};
\draw (184.06,233.99) node [anchor=north west][inner sep=0.75pt]  [font=\footnotesize]  {$1$};
\draw (258.56,233.99) node [anchor=north west][inner sep=0.75pt]  [font=\footnotesize]  {$2$};
\draw (259.56,174.89) node [anchor=north west][inner sep=0.75pt]  [font=\footnotesize]  {$2$};
\end{tikzpicture} \\[1em]
&= \int_0^\infty \dd P' \frac{\fker{\bbi}{P'}{P_1}{P_1}{P_1}{P_1}\sker{P_1}{P_2}{P'}}{\sker{\bbi}{P_2}{\bbi}} \,\begin{tikzpicture}[x=0.75pt,y=0.75pt,yscale=-1,xscale=1,baseline={([yshift=-0.5ex]current bounding box.center)}]
\draw   (255.38,240.81) .. controls (255.38,240.35) and (255.75,239.98) .. (256.21,239.98) .. controls (256.67,239.98) and (257.04,240.35) .. (257.04,240.81) .. controls (257.04,241.26) and (256.67,241.64) .. (256.21,241.64) .. controls (255.75,241.64) and (255.38,241.26) .. (255.38,240.81) -- cycle ;
\draw   (193.93,240.32) .. controls (193.93,239.86) and (194.31,239.49) .. (194.76,239.49) .. controls (195.22,239.49) and (195.59,239.86) .. (195.59,240.32) .. controls (195.59,240.78) and (195.22,241.15) .. (194.76,241.15) .. controls (194.31,241.15) and (193.93,240.78) .. (193.93,240.32) -- cycle ;
\draw   (255.4,180.31) .. controls (255.4,179.85) and (255.77,179.48) .. (256.23,179.48) .. controls (256.69,179.48) and (257.06,179.85) .. (257.06,180.31) .. controls (257.06,180.77) and (256.69,181.14) .. (256.23,181.14) .. controls (255.77,181.14) and (255.4,180.77) .. (255.4,180.31) -- cycle ;
\draw   (194.11,179.82) .. controls (194.11,179.36) and (194.48,178.99) .. (194.94,178.99) .. controls (195.39,178.99) and (195.76,179.36) .. (195.76,179.82) .. controls (195.76,180.28) and (195.39,180.65) .. (194.94,180.65) .. controls (194.48,180.65) and (194.11,180.28) .. (194.11,179.82) -- cycle ;
\draw    (195.74,180.48) -- (210.38,210.38) ;
\draw    (240.32,210.38) -- (255.34,181.05) ;
\draw    (240.32,210.38) -- (255.43,240.32) ;
\draw    (195.57,240.32) -- (210.38,210.38) ;
\draw    (210.38,210.38) -- (240.32,210.38) ;
\draw (182.86,174.89) node [anchor=north west][inner sep=0.75pt]  [font=\footnotesize]  {$1$};
\draw (218.74,193.75) node [anchor=north west][inner sep=0.75pt]  [font=\small]  {$P'$};
\draw (184.06,233.99) node [anchor=north west][inner sep=0.75pt]  [font=\footnotesize]  {$1$};
\draw (258.56,233.99) node [anchor=north west][inner sep=0.75pt]  [font=\footnotesize]  {$2$};
\draw (259.56,174.89) node [anchor=north west][inner sep=0.75pt]  [font=\footnotesize]  {$2$};
\end{tikzpicture}.\label{eq:linked_lines}
\end{align} 
In the last equality we used the relation between the S kernel and the fusion kernel, see equation (2.65) in \cite{Eberhardt:2023mrq}. That relation can, in turn, also be derived from a simple TQFT argument, as was shown in section 2.5 of \cite{Collier:2024mgv}. Our formula \eqref{eq:linked_lines} shows that the configuration of linked Wilson lines produces a state in the Hilbert space of conformal blocks on the four-punctured sphere, expressed as a superposition of $s$-channel blocks. 

A very similar formula holds for the linking of two Wilson lines with the opposite orientation of the over- and under-crossings:
\begin{equation}
    \begin{tikzpicture}[x=0.75pt,y=0.75pt,yscale=-1,xscale=1,baseline={([yshift=-0.5ex]current bounding box.center)}]
\draw  [draw opacity=0] (155.39,139.82) .. controls (163.34,139.96) and (170.68,142.14) .. (176.64,145.75) -- (154.63,170.07) -- cycle ; \draw   (155.39,139.82) .. controls (163.34,139.96) and (170.68,142.14) .. (176.64,145.75) ;  
\draw  [draw opacity=0] (177.69,187.38) .. controls (173.9,182.51) and (171.69,176.61) .. (171.69,170.25) .. controls (171.69,153.58) and (186.91,140.06) .. (205.71,140) -- (205.84,170.25) -- cycle ; \draw   (177.69,187.38) .. controls (173.9,182.51) and (171.69,176.61) .. (171.69,170.25) .. controls (171.69,153.58) and (186.91,140.06) .. (205.71,140) ;  
\draw  [draw opacity=0] (185.05,152.85) .. controls (189.19,157.73) and (191.63,163.67) .. (191.63,170.07) .. controls (191.63,186.57) and (175.47,199.98) .. (155.39,200.31) -- (154.63,170.07) -- cycle ; \draw   (185.05,152.85) .. controls (189.19,157.73) and (191.63,163.67) .. (191.63,170.07) .. controls (191.63,186.57) and (175.47,199.98) .. (155.39,200.31) ;  
\draw  [draw opacity=0] (205.69,200.5) .. controls (197.91,200.47) and (190.75,198.14) .. (185.02,194.23) -- (205.84,170.25) -- cycle ; \draw   (205.69,200.5) .. controls (197.91,200.47) and (190.75,198.14) .. (185.02,194.23) ;  
\draw   (205.69,200.5) .. controls (205.69,200.04) and (206.06,199.67) .. (206.52,199.67) .. controls (206.98,199.67) and (207.35,200.04) .. (207.35,200.5) .. controls (207.35,200.96) and (206.98,201.33) .. (206.52,201.33) .. controls (206.06,201.33) and (205.69,200.96) .. (205.69,200.5) -- cycle ;
\draw   (153.93,200.32) .. controls (153.93,199.86) and (154.31,199.49) .. (154.76,199.49) .. controls (155.22,199.49) and (155.59,199.86) .. (155.59,200.32) .. controls (155.59,200.78) and (155.22,201.15) .. (154.76,201.15) .. controls (154.31,201.15) and (153.93,200.78) .. (153.93,200.32) -- cycle ;
\draw   (205.71,140) .. controls (205.71,139.54) and (206.08,139.17) .. (206.54,139.17) .. controls (207,139.17) and (207.37,139.54) .. (207.37,140) .. controls (207.37,140.46) and (207,140.83) .. (206.54,140.83) .. controls (206.08,140.83) and (205.71,140.46) .. (205.71,140) -- cycle ;
\draw   (154.11,139.82) .. controls (154.11,139.36) and (154.48,138.99) .. (154.94,138.99) .. controls (155.39,138.99) and (155.76,139.36) .. (155.76,139.82) .. controls (155.76,140.28) and (155.39,140.65) .. (154.94,140.65) .. controls (154.48,140.65) and (154.11,140.28) .. (154.11,139.82) -- cycle ;
\draw (142.86,141.64) node [anchor=north west][inner sep=0.75pt]  [font=\small]  {$P_{1}$};
\draw (203.61,141.64) node [anchor=north west][inner sep=0.75pt]  [font=\small]  {$P_{2}$};
\end{tikzpicture} = \int_0^\infty \dd P \frac{\fker{\bbi}{P}{P_1}{P_1}{P_1}{P_1}\mathbb{S}^*_{P_1P_2}[P]}{\sker{\bbi}{P_2}{\bbi}} \,\begin{tikzpicture}[x=0.75pt,y=0.75pt,yscale=-1,xscale=1,baseline={([yshift=-0.5ex]current bounding box.center)}]
\draw   (255.38,240.81) .. controls (255.38,240.35) and (255.75,239.98) .. (256.21,239.98) .. controls (256.67,239.98) and (257.04,240.35) .. (257.04,240.81) .. controls (257.04,241.26) and (256.67,241.64) .. (256.21,241.64) .. controls (255.75,241.64) and (255.38,241.26) .. (255.38,240.81) -- cycle ;
\draw   (193.93,240.32) .. controls (193.93,239.86) and (194.31,239.49) .. (194.76,239.49) .. controls (195.22,239.49) and (195.59,239.86) .. (195.59,240.32) .. controls (195.59,240.78) and (195.22,241.15) .. (194.76,241.15) .. controls (194.31,241.15) and (193.93,240.78) .. (193.93,240.32) -- cycle ;
\draw   (255.4,180.31) .. controls (255.4,179.85) and (255.77,179.48) .. (256.23,179.48) .. controls (256.69,179.48) and (257.06,179.85) .. (257.06,180.31) .. controls (257.06,180.77) and (256.69,181.14) .. (256.23,181.14) .. controls (255.77,181.14) and (255.4,180.77) .. (255.4,180.31) -- cycle ;
\draw   (194.11,179.82) .. controls (194.11,179.36) and (194.48,178.99) .. (194.94,178.99) .. controls (195.39,178.99) and (195.76,179.36) .. (195.76,179.82) .. controls (195.76,180.28) and (195.39,180.65) .. (194.94,180.65) .. controls (194.48,180.65) and (194.11,180.28) .. (194.11,179.82) -- cycle ;
\draw    (195.74,180.48) -- (210.38,210.38) ;
\draw    (240.32,210.38) -- (255.34,181.05) ;
\draw    (240.32,210.38) -- (255.43,240.32) ;
\draw    (195.57,240.32) -- (210.38,210.38) ;
\draw    (210.38,210.38) -- (240.32,210.38) ;
\draw (178.06,174.89) node [anchor=north west][inner sep=0.75pt]  [font=\footnotesize]  {$P_1$};
\draw (218.74,193.75) node [anchor=north west][inner sep=0.75pt]  [font=\small]  {$P$};
\draw (178.06,233.99) node [anchor=north west][inner sep=0.75pt]  [font=\footnotesize]  {$P_1$};
\draw (259.56,233.99) node [anchor=north west][inner sep=0.75pt]  [font=\footnotesize]  {$P_2$};
\draw (259.56,174.89) node [anchor=north west][inner sep=0.75pt]  [font=\footnotesize]  {$P_2$};
\end{tikzpicture},\label{eq:unlinking2}
\end{equation}
where the complex conjugate of the S kernel is given by $\mathbb{S}^*_{P_1P_2}[P] = \e^{-i\pi \Delta_P}\mathbb{S}_{P_1P_2}[P]$.
The derivation proceeds in the same steps as above, with the only difference that the braiding phase now comes with a complex conjugation $\mathbb{B}^P_{P_1P_2} = (\mathbb{B}_P^{P_1P_2})^*$. Hence we use the complex conjugate of the identity relating $\mathbb{F}$, $\mathbb{B}$ and $\mathbb{S}$. This conjugation only affects the S-kernel in \eqref{eq:unlinking2}, since $\mathbb{F}$ and $\rho_0$ are real for real arguments $P_1,P_2,P$. 


\subsection*{Wilson bubbles}

A second identity that is used in the main text, is the formula for removing loops of bulk Wilson lines:
    \begin{equation}\label{eq:WilsonBubble}
    \begin{tikzpicture}[x=0.75pt,y=0.75pt,yscale=-1.2,xscale=1.2,baseline={([yshift=-.5ex]current bounding box.center)}] 
\draw  [draw opacity=0] (209.99,90.47) .. controls (209.74,95.77) and (205.37,100) .. (200,100) .. controls (194.48,100) and (190,95.52) .. (190,90) .. controls (190,84.48) and (194.48,80) .. (200,80) .. controls (205.52,80) and (210,84.48) .. (210,90) .. controls (210,90.14) and (210,90.29) .. (209.99,90.43) -- (200,90) -- cycle ; \draw   (209.99,90.47) .. controls (209.74,95.77) and (205.37,100) .. (200,100) .. controls (194.48,100) and (190,95.52) .. (190,90) .. controls (190,84.48) and (194.48,80) .. (200,80) .. controls (205.52,80) and (210,84.48) .. (210,90) .. controls (210,90.14) and (210,90.29) .. (209.99,90.43) ; 
\draw    (170,90) -- (190,90) ;
\draw    (210,90) -- (230,90) ;
\draw (165,78) node [anchor=north west][inner sep=0.75pt]  [font=\scriptsize]  {$P_3$};
\draw (195,69.4) node [anchor=north west][inner sep=0.75pt]  [font=\scriptsize]  {$P_1$};
\draw (195,102.4) node [anchor=north west][inner sep=0.75pt]  [font=\scriptsize]  {$P_2$};
\draw (222.5,78) node [anchor=north west][inner sep=0.75pt]  [font=\scriptsize]  {$P_4$};
\end{tikzpicture} \,= \,\frac{\delta(P_3-P_4)}{\rho_0(P_3)C_0(P_1,P_2,P_3)}\,\,\begin{tikzpicture}[x=0.75pt,y=0.75pt,yscale=-1.2,xscale=1.2,baseline={([yshift=-1.5ex]current bounding box.center)}]
\draw    (190,110) -- (250,110) ;
\draw (215,98.4) node [anchor=north west][inner sep=0.75pt]  [font=\scriptsize]  {$P_4$};
\end{tikzpicture}\,.
\end{equation}
This formula is valid when the Wilson lines are inside a topologically trivial region of the bulk 3-manifold. Namely, in that case we can surround the bubble diagram by a two-punctured sphere, and use the fact that the sphere two-point function is delta-function normalized. The normalization factor in \eqref{eq:WilsonBubble} was derived in \cite{Collier_2023} by considering the degeneration limit of a four-boundary wormhole.

We can use this to evaluate Wilson bubbles with three or more external legs. For example, for three external legs we obtain
\begin{equation}\label{eq:TQFT_vertex_rule}
    \begin{tikzpicture}[x=0.75pt,y=0.75pt,yscale=-1,xscale=1,baseline={([yshift=-.5ex]current bounding box.center)}]
\draw    (215.27,169.62) -- (251.71,169.62) ;
\draw [shift={(215.27,169.62)}, rotate = 0] [color={rgb, 255:red, 0; green, 0; blue, 0 }  ][fill={rgb, 255:red, 0; green, 0; blue, 0 }  ]      (0, 0) circle [x radius= 1.34, y radius= 1.34]   ;
\draw    (199.12,136.75) -- (215.27,169.62) ;
\draw    (199.12,202.48) -- (215.27,169.62) ;
\draw [color={rgb, 255:red, 208; green, 2; blue, 27 }  ,draw opacity=1 ]   (207.2,153.18) .. controls (221.43,144) and (233.71,153.14) .. (233.49,169.62) ;
\draw [shift={(233.49,169.62)}, rotate = 90.78] [color={rgb, 255:red, 208; green, 2; blue, 27 }  ,draw opacity=1 ][fill={rgb, 255:red, 208; green, 2; blue, 27 }  ,fill opacity=1 ]      (0, 0) circle [x radius= 1.34, y radius= 1.34]   ;
\draw [shift={(207.2,153.18)}, rotate = 327.16] [color={rgb, 255:red, 208; green, 2; blue, 27 }  ,draw opacity=1 ][fill={rgb, 255:red, 208; green, 2; blue, 27 }  ,fill opacity=1 ]      (0, 0) circle [x radius= 1.34, y radius= 1.34]   ;
\draw (259.1,165.34) node [anchor=north west][inner sep=0.75pt]  [font=\scriptsize]  {$P_{2}$};
\draw (180.78,126.72) node [anchor=north west][inner sep=0.75pt]  [font=\scriptsize]  {$P_{1}$};
\draw (181.18,199.43) node [anchor=north west][inner sep=0.75pt]  [font=\scriptsize]  {$P_{3}$};
\draw (229.06,141.3) node [anchor=north west][inner sep=0.75pt]  [font=\scriptsize]  {$P$};
\end{tikzpicture} \quad = \quad \frac{\fker{P}{P_3}{P_1}{P_2}{P_2}{P_1}}{\fker{\bbi}{P_3}{P_1}{P_2}{P_2}{P_1}}\quad\begin{tikzpicture}[x=0.75pt,y=0.75pt,yscale=-1,xscale=1,baseline={([yshift=-.5ex]current bounding box.center)}]
\draw    (215.27,169.62) -- (251.71,169.62) ;
\draw [shift={(215.27,169.62)}, rotate = 0] [color={rgb, 255:red, 0; green, 0; blue, 0 }  ][fill={rgb, 255:red, 0; green, 0; blue, 0 }  ]      (0, 0) circle [x radius= 1.34, y radius= 1.34]   ;
\draw    (199.12,136.75) -- (215.27,169.62) ;
\draw    (199.12,202.48) -- (215.27,169.62) ;
\draw (259.1,165.34) node [anchor=north west][inner sep=0.75pt]  [font=\scriptsize]  {$P_{2}$};
\draw (180.78,126.72) node [anchor=north west][inner sep=0.75pt]  [font=\scriptsize]  {$P_{1}$};
\draw (181.18,199.43) node [anchor=north west][inner sep=0.75pt]  [font=\scriptsize]  {$P_{3}$};
\end{tikzpicture}.
\end{equation}
This is easily proven by acting with a fusion transformation on the edge highlighted in red, and then removing the Wilson bubble using \eqref{eq:WilsonBubble}.\footnote{In \cite{Gaiotto:2014lma}, the red edge was called an `open Verlinde line'.} The ratio of F kernels in \eqref{eq:TQFT_vertex_rule} can also be derived from a four-boundary wormhole computation, see equation (3.44) in \cite{Collier_2023}. 
Wilson bubbles with more than three external legs can always be evaluated by inserting identity lines, applying fusion moves, and then reducing to the basic case with two external legs \eqref{eq:WilsonBubble}.

\subsection*{Loop operator}
The most important identity that we need in the main text is the formula for the Verlinde loop operator \cite{Verlinde:1989ua, Alday:2009fs, Drukker:2009id, Drukker:2010jp}. Whenever a Wilson loop encircles another Wilson line, we can use the unlinking formula \eqref{eq:linked_lines} to obtain
\begin{equation}
    \begin{tikzpicture}[x=0.75pt,y=0.75pt,yscale=-1.3,xscale=1.3,baseline={([yshift=-.5ex]current bounding box.center)}] 
\draw    (100.1,105) -- (119.11,105) ;
\draw  [draw opacity=0] (146.06,109.65) .. controls (145.08,121.23) and (140.5,130) .. (135,130) .. controls (128.79,130) and (123.75,118.81) .. (123.75,105) .. controls (123.75,91.19) and (128.79,80) .. (135,80) .. controls (140.68,80) and (145.38,89.35) .. (146.14,101.5) -- (135,105) -- cycle ; \draw   (146.06,109.65) .. controls (145.08,121.23) and (140.5,130) .. (135,130) .. controls (128.79,130) and (123.75,118.81) .. (123.75,105) .. controls (123.75,91.19) and (128.79,80) .. (135,80) .. controls (140.68,80) and (145.38,89.35) .. (146.14,101.5) ;  
\draw    (128.11,105) -- (168.75,105) ;
\draw (155,91.4) node [anchor=north west][inner sep=0.75pt]  [font=\scriptsize]  {$P_{1}$};
\draw (115,82.4) node [anchor=north west][inner sep=0.75pt]  [font=\scriptsize]  {$P$};
\end{tikzpicture}\; =\, \int_0^\infty \dd P_2\, \frac{\fker{\bbi}{P_2}{P}{P}{P}{P}\mathbb{S}_{PP_1}[P_2]}{\sker{\bbi}{P_1}{\bbi}}\,\begin{tikzpicture}[x=0.75pt,y=0.75pt,yscale=-1.3,xscale=1.3,baseline={([yshift=-.5ex]current bounding box.center)}]
\draw  [draw opacity=0] (189.99,70.47) .. controls (189.74,75.77) and (185.37,80) .. (180,80) .. controls (174.48,80) and (170,75.52) .. (170,70) .. controls (170,64.48) and (174.48,60) .. (180,60) .. controls (185.52,60) and (190,64.48) .. (190,70) .. controls (190,70.14) and (190,70.29) .. (189.99,70.43) -- (180,70) -- cycle ; \draw  (189.99,70.47) .. controls (189.74,75.77) and (185.37,80) .. (180,80) .. controls (174.48,80) and (170,75.52) .. (170,70) .. controls (170,64.48) and (174.48,60) .. (180,60) .. controls (185.52,60) and (190,64.48) .. (190,70) .. controls (190,70.14) and (190,70.29) .. (189.99,70.43) ; 
\draw    (150,100) -- (210,100) ;
\draw    (180,80) -- (180,100) ;
\draw (181,85.4) node [anchor=north west][inner sep=0.75pt]  [font=\scriptsize]  {$P_{2}$};
\draw (191,62.4) node [anchor=north west][inner sep=0.75pt]  [font=\scriptsize]  {$P$};
\draw (146,87.4) node [anchor=north west][inner sep=0.75pt]  [font=\scriptsize]  {$P_{1}$};
\draw (206,87.4) node [anchor=north west][inner sep=0.75pt]  [font=\scriptsize]  {$P_{1}$};
\end{tikzpicture}
\end{equation}
We can now connect the $P$ loop on the right hand side to the $P_1$ line with an identity line $\bbi$. Then we remove the Wilson bubble with the TQFT rule \eqref{eq:WilsonBubble}.\footnote{The delta function is to be understood as a real distribution in terms of the conformal weight $\Delta_2 = \frac{c-1}{24}+P_2^2$. Thus $\delta(\Delta_2)$ is only non-zero when $\Delta_2 = 0$.} Using the fact that $\fker{\bbi}{P_2}{P}{P}{P}{P} = \rho_0(P_2)C_0(P,P,P_2)$, we obtain the simple formula:
\begin{equation}\label{eq:TQFTrule}
    \begin{tikzpicture}[x=0.75pt,y=0.75pt,yscale=-1.3,xscale=1.3,baseline={([yshift=-.5ex]current bounding box.center)}] 
\draw   (100.1,105) -- (119.11,105) ;
\draw  [draw opacity=0] (146.06,109.65) .. controls (145.08,121.23) and (140.5,130) .. (135,130) .. controls (128.79,130) and (123.75,118.81) .. (123.75,105) .. controls (123.75,91.19) and (128.79,80) .. (135,80) .. controls (140.68,80) and (145.38,89.35) .. (146.14,101.5) -- (135,105) -- cycle ; \draw   (146.06,109.65) .. controls (145.08,121.23) and (140.5,130) .. (135,130) .. controls (128.79,130) and (123.75,118.81) .. (123.75,105) .. controls (123.75,91.19) and (128.79,80) .. (135,80) .. controls (140.68,80) and (145.38,89.35) .. (146.14,101.5) ;  
\draw    (128.11,105) -- (168.75,105) ;
\draw (155,91.4) node [anchor=north west][inner sep=0.75pt]  [font=\scriptsize]  {$P_1$};
\draw (115,82.4) node [anchor=north west][inner sep=0.75pt]  [font=\scriptsize]  {$P$};
\end{tikzpicture} \;=\, \frac{\mathbb{S}_{PP_1}[\bbi]}{\mathbb{S}_{\bbi P_1}[\bbi]}\,\,
\begin{tikzpicture}[x=0.75pt,y=0.75pt,yscale=-1.2,xscale=1.2,baseline={([yshift=-1.7ex]current bounding box.center)}]
\draw    (99.9,105) -- (168.75,105) ;
\draw (126.2,92.2) node [anchor=north west][inner sep=0.75pt]  [font=\scriptsize]  {$P_1$};
\end{tikzpicture}
\end{equation}
Here $\mathbb{S}_{PP_1}[\bbi]$ is the modular S kernel, which is given by $\mathbb{S}_{PP_1}[\bbi] = 2\sqrt{2}\cos(4\pi PP_1)$ for non-degenerate representations. So the action of the loop operator on a Wilson line is just a multiplication with a ratio of S kernels. If we take the loop $P$ to be in the $\langle 2,1\rangle$ degenerate representation, the ratio in \eqref{eq:TQFTrule} gives the hyperbolic length $2\cosh(\ell/2)$ with $\ell \coloneqq 4\pi bP_1$ \cite{Eberhardt:2023mrq}.

\section{Direct CFT proof}\label{sec:CFTproof}

In this appendix, we give a proof of the main formula \eqref{eq:main_formula}, using only the Moore-Seiberg consistency conditions obeyed by the crossing transformations of conformal blocks.  The proof given in this appendix is essentially a series of manipulations of the $(g,n) = (1,2)$ identity described in \cite{Eberhardt:2023mrq}, and proceeds in the following steps.

\textbf{Step 1.} The starting point is the most general crossing equation for the conformal block on the two-punctured torus (equation $(2.50)$ in \cite{Eberhardt:2023mrq}):
\begin{multline}\label{eq:genus_1_identity}
    \mathbb{S}_{P_1P_2}[P_3] \int_0^\infty \dd P_4\,\fker{P_3}{P_4}{P_0'}{P_2}{P_2}{P_0} \e^{2\pi i(\Delta_4-\Delta_2)}\fker{P_4}{P_5}{P_0'}{P_0}{P_2}{P_2} 
    \\[0.8em] = \int_0^\infty \dd P_6\,\fker{P_3}{P_6}{P_0}{P_1}{P_1}{P_0'}\fker{P_1}{P_5}{P_0'}{P_0}{P_6}{P_6}\e^{\pi i(\Delta_0+\Delta_{0}'-\Delta_5)}\mathbb{S}_{P_6P_2}[P_5]\,.
\end{multline}
From this general formula, we can derive various identities as special cases. First, we consider the degeneration limit $P_3 \to \frac{iQ}{2} \equiv \bbi$. Using the explicit form of the identity crossing kernels $\sker{P_1}{P_2}{\bbi}$ and $\mathbb{F}_{\bbi P_6}$, this gives
\begin{multline}\label{eq:step1}
    \mathbb{S}_{P_1P_2}[\bbi] \int_0^\infty \dd P_4\,\fker{\bbi}{P_4}{P_0'}{P_2}{P_2}{P_0} \e^{2\pi i(\Delta_4-\Delta_2)}\fker{P_4}{P_5}{P_0'}{P_0}{P_2}{P_2} 
    \\[0.8em] = \int_0^\infty \dd P_6\,\fker{\bbi}{P_6}{P_0}{P_1}{P_1}{P_0'}\fker{P_1}{P_5}{P_0'}{P_0}{P_6}{P_6}\e^{\pi i(\Delta_0+\Delta_{0}'-\Delta_5)}\mathbb{S}_{P_6P_2}[P_5]\,.
\end{multline}

\textbf{Step 2.} Next, we return to the general formula \eqref{eq:genus_1_identity}, and specialize to $P_0 \to P_0'$. We further take the \emph{simultaneous} degeneration limits $P_1 \to \bbi$ and $P_3\to \bbi$. Using the fact that $\fker{\bbi}{6}{P_0}{\bbi}{\bbi}{P_0}=\delta(P_6-P_0)$, the $P_6$ integral can be done and we obtain
    \be
        \mathbb{S}_{\bbi P_2}[\bbi]\int_0^\infty \dd P_4\,\fker{\bbi}{P_4}{P_0}{P_2}{P_2}{P_0}\e^{2\pi i(\Delta_4-\Delta_2)}\fker{P_4}{P_5}{P_0}{P_0}{P_2}{P_2} = \e^{\pi i (2\Delta_0-\Delta_5)} \fker{\bbi}{P_5}{P_0}{P_0}{P_0}{P_0}\mathbb{S}_{P_0P_2}[P_5]\,.
    \ee
We can now plug this formula into the left-hand side of \eqref{eq:step1}, still with $P_0=P_0'$, and cancel the phase factors. Doing so gives
\be\label{eq:step2}
        \frac{\mathbb{S}_{P_0P_2}[P_5]\,\mathbb{S}_{P_1P_2}[\bbi]\,\fker{\bbi}{P_5}{P_0}{P_0}{P_0}{P_0}}{
        \mathbb{S}_{\bbi P_2}[\bbi]}= \int_0^\infty \dd P_6\,\fker{\bbi}{P_6}{P_0}{P_1}{P_1}{P_0}\fker{P_1}{P_5}{P_0}{P_0}{P_6}{P_6}\mathbb{S}_{P_6P_2}[P_5].
\ee

\textbf{Step 3.} The denominator on the LHS of \eqref{eq:step2} has a double zero at $P_2 = 0$, coming from the explicit form of the identity S-kernel $\mathbb{S}_{\bbi P_2}[\bbi] = 4\sqrt{2}\sinh(2\pi b P_2)\sinh(2\pi b^{-1}P_2)$. This double zero is matched by a double zero (at $P_2=0$) coming from the modular kernel $\mathbb{S}_{P_0P_2}[P_5]$\footnote{For a discussion about the poles and zeroes of the modular kernel see section 3.8 of \cite{Eberhardt:2023mrq}. In particular, the locations and orders of the \textit{zeroes} of $\mathbb{S}_{P_1P_2}[P_0]$ w.r.t. $P_2$ are as follows: i) \textit{double} zeroes at $P_2=0$, ii) \textit{simple} zeroes at $\pm P_2=\frac{ib}{2}\mathbb{Z}_{>0}$ and $\pm P_2=\frac{ib^{-1}}{2}\mathbb{Z}_{>0}$ and iii) \textit{simple} zeroes at $\pm P_2 = \frac{i}{2}\left(b\mathbb{Z}_{>0}+b^{-1}\mathbb{Z}_{>0}\right)$.}.
Hence we can safely integrate both sides of \eqref{eq:step2} over $P_2\in \R_{+}$ after multiplying with $\mathbb{S}_{P_2P_2'}^*[P_5]$. We then use the idempotency of the S-kernel,
\be
    \int_0^\infty \dd P_2 \,\mathbb{S}_{P_6P_2}[P_5]\,\mathbb{S}_{P_2P_2'}^*[P_5] = \delta(P_6-P_2'),  
\ee
where $\mathbb{S}_{P_2P_2'}^*[P_5] \coloneqq \e^{-i\pi \Delta_5}\mathbb{S}_{P_2P_2'}[P_5]$, to simplify the right-hand side of eq.\! \eqref{eq:step2}. For any $P_0,P_5 \in \R_{+}$, the identity kernel $\fker{\bbi}{P_5}{P_0}{P_0}{P_0}{P_0}$ is non-zero, so we can divide by it, obtaining
\be\label{eq:step3}
    \int_0^\infty \dd P_2\frac{\mathbb{S}_{P_0P_2}[P_5]\,\mathbb{S}_{P_1P_2}[\bbi]\,\mathbb{S}_{P_2P_2'}^*[P_5]}{\mathbb{S}_{\bbi P_2}[\bbi]} = \frac{\fker{\bbi}{P_2'}{P_0}{P_1}{P_1}{P_0}\fker{P_1}{P_5}{P_0}{P_0}{P_2'}{P_2'}}{\fker{\bbi}{P_5}{P_0}{P_0}{P_0}{P_0}}\,.
\ee

\textbf{Step 4.} Finally, we use the tetrahedral symmetry of the fusion kernel,
\be\label{eq:step4}
\fker{\bbi}{P_2'}{P_0}{P_1}{P_1}{P_0}\fker{P_1}{P_5}{P_0}{P_0}{P_2'}{P_2'} = \fker{P_0}{P_2'}{P_2'}{P_5}{P_0}{P_1}\fker{\bbi}{P_5}{P_0}{P_0}{P_0}{P_0} \,.
\ee 
This identity is actually a special case of the \emph{pentagon identity} (see equations $(2.47)$ and $(2.57)$ in \cite{Eberhardt:2023mrq}), which in turn is a consistency condition derived from crossing transformations of the five-punctured sphere conformal block.
Plugging \eqref{eq:step4} into \eqref{eq:step3}, and relabeling variables $(P_0,P_1,P_2,P_2',P_5) \mapsto (P_1,P_2,P,P_3,P_0)$, we arrive at the main formula \eqref{eq:main_formula} of the paper.

We stress that the only ingredients of the proof were the consistency conditions for the torus 2-point function and the sphere 5-point function. This parallels the derivations in \cite{Moore:1988qv, Moore:1989vd, DiFrancesco:639405} for the case of rational CFT. 

\subsection*{Explicit check in the case $c=25, \Delta_0=1$.}
There is a special case where we can verify our main formula \eqref{eq:main_formula} analytically. Namely, for $b=1$ (corresponding to $c=25$) and $P_0=\frac{i(b^{-1}-b)}{2}=0$ all the corresponding Virasoro crossing kernels simplify drastically.  
The regulator momentum $P_0=0$ at $b=1$ corresponds to conformal weight $\Delta_0=1$. The modular kernel with external conformal weight $\Delta_0=1$ takes a very simple form which is actually \textit{independent} of the central charge (see e.g. \cite{Kraus:2016nwo}). According to our conventions \eqref{eq:crossing_kernels} the modular S-kernel reads:
\be
\bea
\mathbb{S}_{PP'}\left[\frac{i(b^{-1}-b)}{2}\right] = 2i\sqrt{2} \,\left(\frac{P'}{P}\right)\sin{(4\pi P P')} .
\eea
\ee
So we have all the ingredients to perform the integral on the RHS of \eqref{eq:main_formula}. We find:
\begin{multline}\label{eq:explresc25RHS}
 \int_0^\infty \dd P\,\frac{\sker{P_1}{P}{P_0=0}\sker{P_2}{P}{\bbi}\mathbb{S}_{PP_3}^*[P_0=0]}{\sker{\bbi}{P}{\bbi}} \Bigg\vert_{b=1}
= \\ \frac{P_3}{2P_1}\Big[ P_{123}\coth{(\pi P_{123})}+P_{1-23}\coth{\left(\pi P_{1-23}\right)}-P_{12-3}\coth{\left(\pi P_{12-3}\right)}-P_{1-2-3}\coth{\left(\pi P_{1-2-3}\right)} \Big],
\end{multline}
where we adopted the notation $P_{i\pm j\pm k}=P_i \pm P_j \pm P_k$ for brevity (e.g. $P_{1-23}=P_1-P_2+P_3$ etc.). This result is formally valid under the condition $\sum_{i=1}^3 |\Im P_i|<1$.

According to the Virasoro-Verlinde formula, the explicit function on the RHS of \eqref{eq:explresc25RHS} should be equal to the fusion kernel 
\begin{equation}
    \fker{P_1}{P_3}{P_3}{\frac{i(b^{-1}-b)}{2}}{P_1}{P_2} \Bigg \vert_{b=1}.
\end{equation}
 One way to check that is indeed the case, is by using the result of \cite{Ribault:2023vqs} where the $c=25$ fusion kernel was given an explicit form in terms of the Barnes$-G$ function for arbitrary internal and external momenta. After studying carefully the $P_0\rightarrow 0$ limit in that formula we are indeed able to reproduce \eqref{eq:explresc25RHS}.\footnote{This limit is subtle because the formula of \cite{Ribault:2023vqs} involves an inverse square root of a determinant that naively diverges in the limit $P_0\rightarrow 0$. However this divergence is compensated by an appropriate zero in the numerator with the net result being exactly \eqref{eq:explresc25RHS}.}

It is worth mentioning a few non-trivial features of the explicit expression \eqref{eq:explresc25RHS} and hence verify the general properties mentioned in Section \ref{sec:properties}. It is straightforward to see that the expression is reflection symmetric in all $P_1,P_2,P_3$, and also positive-definite
\be
\mathcal{N}_{\frac{i(b^{-1}-b)}{2}}\left[P_1,P_2,P_3\right]\text{}\big|_{b=1} \geq 0 \ , \ \ \ \ \forall P_1,P_2,P_3\in\mathbb{R} \ .
\ee
Furthermore, it is \textit{bounded} both as a function of $P_1$ (for fixed $P_2,P_3\in\mathbb{R}$) and as a function $P_2$ (for fixed $P_1,P_3$). In particular,
\be
\lim_{P_1,P_2\rightarrow \pm \infty} \mathcal{N}_{\frac{i(b^{-1}-b)}{2}}\left[P_1,P_2,P_3\right]\text{}\big|_{b=1} = 0 \ .
\ee
As a function of $P_3$ the expression diverges linearly $\mathcal{N}\sim \pm 2P_3$ as $P_3\rightarrow\pm \infty$. This is just an artifact of the particular normalization we are using for the fusion kernel. If we instead use the $6j$ normalization as in \eqref{eq:6jnormN} (c.f. \eqref{eq:relNNhat}) we find: 
\be
\bea
\widehat{\N}_{\frac{i(b^{-1}-b)}{2}}&\left[P_1,P_2,P_3\right]\text{}\big|_{b=1} \\
& \qquad = \frac{P_1}{4\sqrt{2}P_3}\left(\sinh(2\pi P_1)\sinh(2\pi P_3)\right)^{-1} \times \mathcal{N}_{\frac{i(b^{-1}-b)}{2}}\left[P_1,P_2,P_3\right]\text{}\big|_{b=1} \ .
\eea
\ee
It is then obvious that this expression is again positive-definite, and moreover bounded irrespective of the choice of momentum:
\be
\lim_{P_1,P_2,P_3\rightarrow \pm \infty} \widehat{\N}_{\frac{i(b^{-1}-b)}{2}}\left[P_1,P_2,P_3\right]\text{}\big|_{b=1} = 0 \ .
\ee
It is also manifestly symmetric under the exchange $P_1\leftrightarrow P_3$ as expected from the symmetries of the $6j$ symbol.

\bibliographystyle{JHEP}
\bibliography{bibliography}
\end{document}